\documentclass[aps, pra, showpacs,twocolumn]{revtex4}
\usepackage{graphicx}
\usepackage{amsmath}
\usepackage{amsfonts}
\usepackage{amssymb}
\usepackage{epsfig}
\usepackage{booktabs}

\begin{document}

\title{Fidelity approach to quantum phase transitions}
\date{\today }
\author{Shi-Jian Gu}
\email{sjgu@phy.cuhk.edu.hk}
\affiliation{Department of Physics and Institute of Theoretical Physics, The Chinese
University of Hong Kong, Hong Kong, China}

\begin{abstract}
We review briefly the quantum fidelity approach to quantum phase transitions
in a pedagogical manner. We try to relate all established but scattered
results on the leading term of the fidelity into a systematic theoretical
framework, which might provide an alternative paradigm for understanding
quantum critical phenomena. The definition of the fidelity and the scaling
behavior of its leading term, as well as their explicit applications to the
one-dimensional transverse-field Ising model and the Lipkin-Meshkov-Glick
model, are introduced at the graduate-student level. In addition, we survey
also other types of fidelity approach, such as the fidelity per site,
reduced fidelity, thermal-state fidelity, operator fidelity, etc; as well as
relevant works on the fidelity approach to quantum phase transitions
occurring in various many-body systems.
\end{abstract}

\maketitle
\tableofcontents





\section{Introduction}

\subsection{Overview: quantum phase transitions}

Quantum phase transitions \cite{Sachdev} of a quantum many-body system are
characterized by the change in the ground-state properties caused by
modifications in the interactions among the system's constituents. Contrary
to thermal phase transitions where the temperature plays a crucial role,
quantum phase transitions are completely driven by quantum fluctuations and
are incarnated via the non-analytic behavior of the ground-state properties
as the system's Hamiltonian $H(\lambda )$ varies across a transition point $%
\lambda _{c}$.

From the point view of eigenenergy, quantum phase transitions are caused by
the reconstruction of the Hamiltonian's energy spectra, especially of the
low-lying excitation spectra \cite{GSTian2003}. More precisely, the
low-energy spectra can be reconstructed in two qualitatively different ways
around the critical point $\lambda _{c}$, and hence the physical quantities
show different behaviors. The first one is the ground-state level-crossing
in which the first derivative of the ground-state energy with respect to $%
\lambda $ is usually discontinuous at the transition point. Such a
transition is called the first-order phase transition. The second one
corresponds roughly to all other cases in the absence of the ground-state
level-crossing. It is usually a continuous phase transition.

Traditionally, continuous phase transitions can be characterized by the
Landau-Ginzburg-Wilson spontaneous symmetry-breaking theory where the
correlation function of local order parameters plays a crucial role.
Nevertheless, some systems cannot be described in this framework built on
the local order parameter. This might be due to the absence of preexistent
symmetry in the Hamiltonian, such as systems undergoing topological phase
transitions \cite{wen-book} and Beresinskii-Kosterlitz-Thouless phase
transitions \cite{VLBeresinskii,JMKosterlitz73}.

\subsection{Brief historical retrospect}

In recent years, ambitions on quantum computer and other quantum information
devices have driven many people to develop quantum information theory \cite%
{Nielsen1}. Though a practicable quantum computer seems still a dream,
progresses in quantum information theory have developed other related fields
forward. A noticeable one is the relation between quantum entanglement and
quantum phase transitions \cite%
{AOsterloh2002,TJOsbornee,GVidal03,SQSu2006,HDChen10215,SJGuXXZ,SJGuXXZ2,SJGUPRL,Larsson_ent,SSDeng06,YChen07,SJGUCPL,PDSacramento07}%
. Since the entanglement is regarded as a purely quantum correlation and is
absent in classical systems, people think that the entanglement should play
an important role in quantum phase transitions. Though a unified theory on
the role of entanglement in quantum phase transitions is still unavailable,
some definitive conclusions have been commonly accepted \cite{LAmico08}.

Another attractive approach is the quantum fidelity \cite%
{AUhlmann76,PAlberti83,PMAlberti831,PMAlberti832,WKWootters81,RJozsa94,BSchumacher95,CAFuchs,Bures,Rastegin,JLChen02,TGorinPRep,Mendonca,XWang08071781}%
, a concept also emerging in quantum information theory. The fidelity
measures the similarity between two states, while quantum phase transitions
are intuitively accompanied by an abrupt change in the structure of the
ground-state wavefunction, this primary observation motivates people to
explore the role of fidelity in quantum phase transitions \cite%
{HTQuan2006,Zanardi06}. Since the fidelity is purely a quantum information
concept, where no \emph{a priori} knowledge of any order parameter and
changes of symmetry of the system is assumed, it would be a great advantage
if one can use it to characterize the quantum phase transitions. Many works
have been done along this stream \cite%
{HTQuan2006,Zanardi06,HQZhou0701,PZanardi0606130,MCozzini07,MCozzini072,Buonsante1, PZanardi032109,WLYou07,PZanardi0701061,HQZhou07042940,HQZhou07042945,LCVenuti07, SChen07,SJGu072,ATribedi08,PZanardi062318,MFYang07,YCTzeng08,HQZhou08030585,YCTzeng082,JZhangPRL100501, WQNingJPC,NPaunkovic07,NPaunkovic08,HMKwok07,Zhou_GVidal,HQZhou07114651, JOFjerestad,SChen08,DFAbasto081,LCVenuti08012473,XWang08032940,AHamma07, JHZhao0803,SYang08,DFAbasto08,HMKwok08,JMa08,HMKThesis,HTQuan08064633, LCVenuti08070104,XMLu08071370,SJGu08073491,ZMaJLChen08080984,JZhang08081536, XWang08081816,XWang08081817,LGong115114,SJGUDFF,YYZhang08094426,XWang08094898,YCLiPLA,YCLiPRB,WLYouRFS,YZYouCDGong,CYLeungpreprint,DFAbasto08094740,FMCucchiettiPRA032337,YCOuJPA2455,ZGYuanPRA012102, DRossiniPRA032333,HTQuanPRA012104,LCWangPLA362,YCLiPRA032117,XXYiEPJD355,ZGYuanPRA042118, CCormickPRA022317,DRossiniPRA052112,WGWangPRE056218,CYLaiPRB205419}%
.

The motivation of the fidelity approach to quantum phase transitions can be
traced back to the work of Quan \textit{et al} \cite{HTQuan2006} in
determining two ground-state phases of the one-dimensional transverse-field
Ising model by the Loschmidt echo. The Loschmidt echo \cite{APeresb} has
been introduced to describe the hypersensitivity of the time evolution to
perturbations experienced by the environmental system. They found the
quantum critical behavior of the environmental system strongly affects its
capability of enhancing the decay of Loschmidt echo. Since the Loschmidt
echo is defined as the overlap between two time-dependent states
corresponding to two points separated slightly by a target spin with Ising
interaction, its decay around the critical point represents a large distance
between two states. Subsequently, Zanardi and Paunkovi\'{c} \cite{Zanardi06}
proposed out that a static fidelity might be a good indicator for quantum
phase transitions with examples of the one-dimensional transverse-field XY
model and the Dicke model. Similar idea was also proposed by Zhou and
Barjaktarevic \cite{HQZhou0701}. Motivated by these works, the fidelity
approach to quantum phase transitions was quickly applied to free fermionic
systems \cite{PZanardi0606130} and graphs \cite{MCozzini07}, matrix-product
states \cite{MCozzini072}, and the Bose-Hubbard model \cite{Buonsante1}. An
attempt to understand quantum phase transitions from the thermal fidelity
was also made \cite{PZanardi032109}. At that time, the successes of the
fidelity in these studies \cite%
{HTQuan2006,Zanardi06,HQZhou0701,PZanardi0606130,MCozzini07,MCozzini072,Buonsante1,PZanardi032109}
gave peoples a deep impression that the fidelity is able to characterize any
quantum phase transition, including those cannot be described in the
framework of Landau-Ginzburg-Wilson theory, such as the
Beresinskii-Kosterlitz-Thouless transtions and topological transitions. Two
groups addressed the role of the leading term of the fidelity in the quantum
critical phenomena. Zanardi \textit{et al} introduced, based on the
differential-geometry approach, the Riemannian metric tensor \cite%
{PZanardi0701061} inherited from the parameter space to denote the leading
term in the fidelity, and argued that the singularity of this metric
corresponds to quantum phase transitions. While You \textit{et al}
introduced another concept, the so-called fidelity susceptibility (FS) \cite%
{WLYou07}, and established a general relation between the leading term of
the fidelity and the structure factor (correlation functions) of the driving
term in the Hamiltonian. Both of them obtained also that, if one extend the
fidelity to thermal states, the leading term of the fidelity between two
neighboring thermal states is simply the specific heat. In the following, we
will use \textquotedblleft fidelity susceptibility" to name the leading term
of the fidelity because it not only denotes mathematically the fluctuation
of the driving term, such as the specific heat derived from the internal
energy, but also is closer to the picture of condensed matter physics, i.e.
the response of the fidelity to driving parameter. From then on, the field
of the fidelity approach to quantum (or thermal) phase transitions can be
divided roughly into two streams. The first stream still focuses on the
fidelity itself, for which the distance between two points in the parameter
space is still important, while the second stream pays particular attention
to the leading term of the fidelity.

Along the first stream, a connection between the fidelity, scaling and
renormalization was introduced by Zhou \cite{HQZhou07042945,HQZhou07042940},
in which the fidelity between two reduced states of a part of the system
described by a reduced-density matrix was proposed. Zhou \emph{et al} \cite%
{HQZhou07114651} tried to understand the fidelity from a geometric
perspective. In works of Zhou and his colleagues, the fidelity is averaged
over the system size, and is named as fidelity per site. They found that the
fidelity per site is a very useful tool for various interacting systems.
Interestingly, the fidelity per site, as an analog of the free energy per
site, can be computed in the context of tensor network algorithms\cite%
{Zhou_GVidal,GVidalPRL147902,GVidalPRL040502}.

While along the second stream, several questions appeared at that time. 1)
Since the leading term of the fidelity is a combination of correlation
functions, which seems a tool widely used only in the Landau-Ginzburg-Wilson
theory, is the fidelity still able to describe the
Beresinskii-Kosterlitz-Thouless and topological phase transitions? 2) What
is the scaling behavior of the fidelity and its relation to the universality
class? 3) How about the thermal phase transitions and those quantum phase
transitions induced by the continuous ground-state level-crossing where the
perturbation method is not applicable. Most subsequent works are more or
less related to these questions, though some topics are still controversial.

Based on the general relation between the leading term of the fidelity and
correlation functions of the driving term \cite{WLYou07}, Venuti and Zanardi
\cite{LCVenuti07} applied the traditional scaling transformation, and
obtained an interesting scaling relation between the dynamic exponent, the
dimension of the system, and the size exponents of the fidelity. A similar
scaling relation was also obtained numerically by Gu \textit{et al }\cite%
{SJGu072} in their studies on the one-dimensional asymmetric Hubbard model.
Both relations imply that the fidelity susceptibility might not have
singular behavior in some cases, such the Beresinskii-Kosterlitz-Thouless
transition occurring in the asymmetric Hubbard model at half-filling \cite%
{SJGu072}.

On the other hand, Yang \cite{MFYang07} tried to understand the singular
behavior of the fidelity susceptibility from the ground-state energy density
and pointed out that the fidelity susceptibility might not be able to detect
the high-order phase transitions. A little surprising is that their example,
i.e. the effective model of the one-dimensional XXZ chain, which undergoes a
Beresinskii-Kosterlitz-Thouless transition of infinite order at the
isotropic point, shows singular behavior in the fidelity susceptibility.
Similar analysis on the Luttinger Liquid model with a wave functional
approach was also done by Fjrestad \cite{JOFjerestad}. The further
investigations on spin-1 XXZ chain with uniaxial anisotropy by Yang \textit{%
et al} \cite{YCTzeng08,YCTzeng082} supported partially their previous
conclusion and the scaling relation obtained by Venuti and Zanardi \cite%
{LCVenuti07}.

Later, Chen \emph{et al} \cite{SChen08} addressed the feasibility of the
fidelity susceptibility in quantum phase transitions of various order by the
perturbation theory, and concluded that the fidelity susceptibility cannot
describe the phase transition of infinite order. This conclusion conflicts
with both Yang's works on the one-dimensional XXZ model \cite{MFYang07} and
the subsequent studies on the one-dimensional Hubbard model \cite%
{LCVenuti08012473}, but supports previous conclusion obtained by You \emph{%
et al}\cite{WLYou07}. Therefore, the issue on fidelity in describing
high-order phase transitions seems still controversial.

Recently it was realized that the fidelity susceptibility can be either
intensive, extensive, or superextensive, then the critical exponents of the
rescaled fidelity susceptibility at both sides of the critical point can be
different \cite{SJGu08073491}. In addition to the fidelity susceptibility,
the sub-leading term of the fidelity might appear when parameters are
changed along a critical manifold \cite{LCVenuti08070104}.

It became a branch of the story when Hamma \emph{et al} \cite{AHamma07}
firstly touched the feasibility of the fidelity in topological phase
transitions. They found that though the fidelity shows an obvious drop
around the critical point of a topological transition, it cannot tell the
type of transition. Almost one year later, three groups revisited the role
of fidelity in the topological transitions. Zhou \textit{et al} \cite%
{JHZhao0803} studied the fidelity in the Kitaev honeycomb model and found
that fidelity has shows singular behavior at the critical point. Yang
\textit{et al} \cite{SYang08} studied the fidelity susceptibility in the
same model and obtained various critical exponents, they also witnessed a
kind of long-range correlation in the ground state of Kitaev honeycomb
model. While Abasto \textit{et al} \cite{DFAbasto08} studied the fidelity in
the deformed Kitaev toric model and obtained a form of fidelity between
thermal states. The three groups drew a similar conclusion that the fidelity
can describe the topological phase transitions occurring in the both models.

A noteworthy advance in the fidelity approach is the success of using the
state overlap to detect quantum critical point by a
nuclear-magnetic-resonance quantum simulator\cite{JZhangPRL100501}. It was
observed that the different types of quantum phase transitions in the
transverse-field Ising model can be witnessed in experiments. Such an
advance is remarkable. It makes the fidelity approach to quantum phase
transitions no longer purely theoretical.

On the other hand, the global-state fidelity cannot characterize those
quantum phase transitions induced by continuous level-crossing due to its
collapse at each crossing point. Kwok \emph{et al} \cite{HMKwok08} firstly
tackled this type of phase transition with the strategy of the reduced
fidelity, which actually was introduced in previous works \cite%
{HQZhou07042945,NPaunkovic07}. Meanwhile, Ma \emph{et al} \cite{JMa08} also
studied the critical behavior of the reduced fidelity in the
Lipkin-Meshkov-Glick model. The reduced fidelity was latter applied to the
one-dimensional transverse-field Ising \cite{XWang08081816,YCLiPLA} and XY
models\cite{WLYouRFS}, the dimerized Heisenberg chain\cite{XWang08081817},
and the one-dimensional extended Hubbard model \cite{YCLiPLA}.

Despite of the absence of the thermal phase transition in the
one-dimensional XY model, the thermal-state fidelity was firstly used to
study the crossover occurring in the low-temperature critical region \cite%
{PZanardi032109}. Interestingly, the leading term of the thermal-state
fidelity was later found to be just the specific heat \cite%
{PZanardi0701061,WLYou07}. The thermal-state fidelity was also applied to
the BCS superconductivity and the Stoner-Hubbard model with the mean-field
approach \cite{NPaunkovic08}. Moreover, Quan and Cucchietti \cite%
{HTQuan08064633} tried to find the advantages and disvantages of the
fidelity approach to the thermal phase transitions.

Finally, though we focus on the fidelity between the static ground state
only, we would like to mention that the Loschmidt echo has also been widely
applied to study the quantum phase transitions \cite%
{FMCucchiettiPRA032337,YCOuJPA2455,ZGYuanPRA012102,DRossiniPRA032333,HTQuanPRA012104,LCWangPLA362,YCLiPRA032117,XXYiEPJD355,ZGYuanPRA042118, CCormickPRA022317,DRossiniPRA052112,WGWangPRE056218,CYLaiPRB205419}%
. In studies of the Loschmidt echo, one needs to consider the dynamic
behavior of the fidelity of a target object, for instance, a spin coupled
with all other spins in the Ising chain. Then the decoherence property
should be taken into account. These issues are beyond the scope of this
review.

\subsection{About the review}

The main purpose of this review is to gather these distributed works into a
unified paradigm, then provides interested readers, especially beginners, a
systematic framework of the fidelity approach to quantum phase transitions.
Some practical and numerical methods, such as the exact diagonalization and
density matrix renormalization group, are introduced too. We try to keep the
treatment as simple as the subject allows, showing most calculations in
explicit detail. Since the field is still quickly developing, such a review
is far from completeness. We hope that the article can offer some
introductory essays first, then to arouse more wonderful ideas.

The article is organized as follows. In Section \ref{sec:qfidelity}, we give
a brief overview on the fidelity measure and its properties in an adiabatic
evolution exampled by a 1/2 spin subjected to an external field. In Section %
\ref{sec:fs}, we introduce in considerable detail the general relations
between the fidelity and quantum phase transitions, and try to illustrate
the role of fidelity in quantum phase transitions by the one-dimensional
transverse-field Ising model. In Section \ref{sec:scaling}, we focus on the
leading term of the fidelity, i.e. the fidelity susceptibility, and discuss
its general properties around the critical point. We also use the
one-dimensional transverse-field Ising model and the Lipkin-Meshkov-Glick
model as examples for the fidelity susceptibility in describing the
universality class. In Section \ref{sec:otherfidelity}, we review other
types of fidelity in the quantum phase transition, such as the fidelity per
site, partial-state fidelity, thermal-state fidelity, operator fidelity, and
density-functional fidelity. In Section \ref{sec:survey}, we give a survey
on the fidelity approach to quantum phase transitions in various strongly
correlated system. In Section \ref{sec:num}, we show how to calculate the
fidelity and fidelity susceptibility via some numerical methods. An outlook
and a summary will be presented in the concluding section.


\section{Quantum fidelity: a measure of similarity between states}

\label{sec:qfidelity}

In this section, we introduce briefly the concept of quantum fidelity and
discuss its properties in a simple quantum-state adiabatic evolution of a
1/2 spin subjected to an external field.

\subsection{Pure state and mixed state fidelity}

In quantum physics, an overlap between two quantum states usually denotes
the transition amplitude from one state to the another \cite%
{AUhlmann76,PAlberti83,PMAlberti831,PMAlberti832}. While from the point view
of information theory, the overlap can measure the similarity (closeness)
between two states \cite{WKWootters81,RJozsa94,BSchumacher95}. That is the
overlap gives unity if two states are exactly the same, while zero if they
are orthogonal. Such an interpretation has a special meaning in quantum
information theory \cite{Nielsen1} since physicists in the field (for
examples, Ref \cite{SBose207901,ADantan050502,JZhang170501}) hope that a
quantum state can be transferred over a long distance without loss of any
information. The overlap between the input and output states becomes a
useful measure of the loss of information during the transportation. The
overlap is used to define the fidelity in quantum information theory.

To be precise, if we define the overlap between two pure states as
\begin{equation}
f(\Psi ^{\prime },\Psi )=\langle \Psi ^{\prime }|\Psi \rangle ,
\end{equation}%
the fidelity is simply the modulus of the overlap, i.e.%
\begin{equation}
F(\Psi ^{\prime },\Psi )=|\langle \Psi ^{\prime }|\Psi \rangle |
\end{equation}%
where $|\Psi \rangle ,|\Psi ^{\prime }\rangle $ are the input and output
states respectively, and both of them are normalized. The fidelity has a
geometric meaning as well. Since a pure state in quantum mechanics
mathematically is a vector in the Hilbert space, then according to Linear
algebra, an inner product of two vectors $\mathbf{a}$, $\mathbf{b}$ is
\begin{equation}
\mathbf{a\cdot b}=ab\cos (\theta )
\end{equation}%
where $a(b)$ is the magnitude of $\mathbf{a(b)}$, and $\theta $ is the angle
between them. In quantum mechanics, wave functions are usually normalized,
and the fidelity represents the angle distance between two states.

The fidelity has the following expected properties (\emph{axioms}) \cite%
{RJozsa94}
\begin{eqnarray}
0 &\leq &F(\Psi ^{\prime },\Psi )\leq 1,  \label{eq:axioms1} \\
F(\Psi ^{\prime },\Psi ) &=&F(\Psi ,\Psi ^{\prime }), \\
F(U\Psi ^{\prime },U\Psi ) &=&F(\Psi ^{\prime },\Psi ), \\
F(\Psi _{1}\otimes \Psi _{2},\Psi _{1}^{\prime }\otimes \Psi _{2}^{\prime })
&=&F(\Psi _{1}^{\prime },\Psi _{1})F(\Psi _{2}^{\prime },\Psi _{2}),
\label{eq:axioms4}
\end{eqnarray}%
where $U$ denotes a unitary transformation and $\Psi _{1(2)}$ is the state
of one subsystem. For pure states, the global phase difference may affect
the overlap, but not the fidelity.

\vspace{5mm} \fbox{%
\parbox{7cm}{ \textbf{Example:}

The quantum state of a single spin can be expressed in the basis $\{
|\uparrow\rangle, |\downarrow\rangle \}$. For two normalized states of the
spin, say
\begin{eqnarray}
|\Psi (\theta )\rangle &=&\cos \theta |\uparrow \rangle +\sin \theta
|\downarrow \rangle ,  \notag \\
|\Psi (\theta ^{\prime })\rangle &=&\cos \theta ^{\prime }|\uparrow \rangle
+\sin \theta ^{\prime }|\downarrow \rangle ,  \notag
\end{eqnarray}the fidelity between them is
\begin{equation}
F(\Psi (\theta ^{\prime }),\Psi (\theta ))=|\cos (\theta -\theta ^{\prime })|.
\nonumber
\end{equation}
}}

\vspace{5mm}

The quantum fidelity between two mixed states ($\rho ,\rho ^{\prime }$) is
defined as \cite{AUhlmann76}
\begin{equation}
F(\rho ,\rho ^{\prime })=\text{tr}\sqrt{\rho ^{1/2}\rho ^{\prime }\rho ^{1/2}%
}.  \label{eq:fidelity_bure}
\end{equation}%
Here $\rho (\rho ^{\prime })$ is semi-positive defined and normalized, i.e.
tr$\rho =$tr$\rho ^{\prime }=1$. The definition satisfies the expected
properties of the fidelity, i.e. Eqs. (\ref{eq:axioms1}-\ref{eq:axioms4}).

It is not easy to evaluated the fidelity between two arbitrary mixed states.
Nevertheless, there are some special useful cases:

1) If both states are pure $F(\rho ,\rho ^{\prime })=|\langle \Psi ^{\prime
}|\Psi \rangle |$,

2) If one of state is pure, i.e. $\rho =|\Psi \rangle \langle \Psi |$, \
then $F(\rho ,\rho ^{\prime })=\sqrt{\langle \Psi |\rho ^{\prime }|\Psi
\rangle }$, which is simply the square root of the expectation value of $%
\rho ^{\prime }$ \cite{BSchumacher95},

3) If both of states are diagonal in the same basis, such as the thermal
equilibrium state, the fidelity (or classical fidelity) can be calculated as
\begin{equation}
F(\rho ,\rho ^{\prime })=\sum_{j}\sqrt{\rho _{jj}\rho _{jj}^{\prime }}.
\end{equation}

\vspace{5mm} \fbox{%
\parbox{7cm}{ \textbf{Example:}

If a spin is coupled to environment, it can be described by a reduced-density
matrix. For two reduced-density matrices
\begin{equation} \nonumber
\rho =\left(
\begin{array}{cc}
a & 0 \\
0 & b\end{array}\right) ,\;\;\rho ^{\prime }=\left(
\begin{array}{cc}
c & 0 \\
0 & d\end{array}\right)
\end{equation}
the fidelity can be calculated as
\begin{equation} \nonumber
F(\rho ,\rho ^{\prime })=\sqrt{ac}+\sqrt{bd}.
\end{equation}
}}\vspace{5mm}

Though the fidelity itself is not a metric, it can be used to define a
metric on the set of quantum state, i.e.
\begin{eqnarray}
\theta _{B}(\rho ,\rho ^{\prime }) &=&\cos ^{-1}[F(\rho ,\rho ^{\prime })],
\\
D_{B}(\rho ,\rho ^{\prime }) &=&\sqrt{2-2F(\rho ,\rho ^{\prime })}, \\
S_{B}(\rho ,\rho ^{\prime }) &=&\sqrt{1-[F(\rho ,\rho ^{\prime })]^{2}},
\end{eqnarray}%
called commonly as Bures angle, Bures distance \cite{Bures}, and sine
distance \cite{Rastegin}, respectively.

Besides the above well-accepted definitions, there are some alternative
definitions of the fidelity. For example, Chen \emph{et al} \cite{JLChen02}
proposed
\begin{eqnarray}
&&|F(\rho ,\rho ^{\prime })|^{2}=  \label{eq:fidelitydef_chen} \\
&&\frac{1-r}{2}+\frac{1+r}{2}\left[ \text{tr}(\rho \rho ^{\prime })+\sqrt{1-%
\text{tr}(\rho ^{2})}\sqrt{1-\text{tr}(\rho ^{\prime 2})}\right] ,  \notag
\end{eqnarray}%
where $r=1/(d-1)$ with $d$ being the dimension of the system. This
definition has a hyperbolic geometric interpretation, and is reduced to Eq. %
\ref{eq:fidelity_bure} in the special case of $d=2$. The definition [Eq. (%
\ref{eq:fidelitydef_chen})] was recently simplified to \cite{Mendonca}
\begin{equation}
|F(\rho ,\rho ^{\prime })|^{2}=\text{tr}(\rho \rho ^{\prime })+\sqrt{1-\text{%
tr}(\rho ^{2})}\sqrt{1-\text{tr}(\rho ^{\prime 2})}.
\label{eq:fidelitydef_chen2}
\end{equation}%
Obviously, one of advantages of the above definitions is that the fidelity
can be easily evaluated for arbitrary mixed states. Nevertheless, it seems
that for two density matrices of two sets of mutually independent events,
Eq. (\ref{eq:fidelitydef_chen2}) gives a nonzero value. Therefore, another
definition of the fidelity was proposed \cite{XWang08071781}, i.e.%
\begin{equation}
F(\rho ,\rho ^{\prime })=\frac{|\text{tr}(\rho \rho ^{\prime })|}{\sqrt{%
\text{tr}(\rho ^{2})\text{tr}(\rho ^{\prime 2})}}.
\end{equation}

The fidelity has been widely used in many fields. In quantum information
science, the fidelity between quantum states have been proved useful resources
in approaching a number of fundamental problems such as quantifying
entanglement \cite{VVedral97,VVedral8,HorodeckiRev}. There are also many
interesting works on the fidelity in adiabatic processes. For example, the
adiabatic fidelity was used to describe atom-to-molecule conversion
\cite{SYMeng07090359,LHLu053611} in atomic systems and the time evolution in a
Bose-Einstein condensate\cite{JLiu063623,KJHughes035601,GManfredi050405}.
Physicists working on quantum chaos \cite{APeresb,GCasati86,GBenenti03} use
quantum fidelity (Loschmidt echo) to measures the hypersensitivity to small
perturbations of quantum dynamics. In the latter case, the fidelity usually
depends on the time. Interested readers can find more details about the
Loschmidt echo in a recent review article by Gorin \textit{et al} \cite%
{TGorinPRep}. In the fidelity approach to quantum phase transitions, which
will be introduced in this review, the fidelity depends on the adiabatic
parameter (or driving parameter) of the Hamiltonian, and is usually static.

\begin{figure}[tbp]
\includegraphics[bb=0 0 200 220, width=8.5 cm, clip] {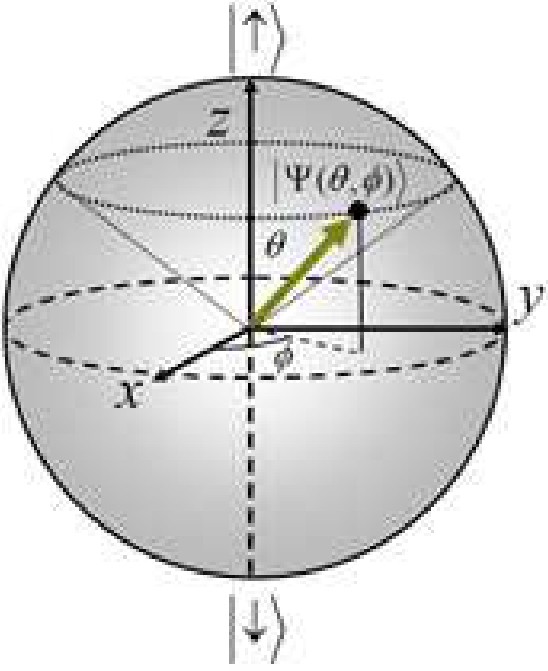}
\caption{ (Color online) A single 1/2 spin state defined on a Bloch sphere.}
\label{figure_bloch}
\end{figure}

\subsection{Quantum state overlap and adiabatic evolution}

To well understand the fidelity in the ground-state state evolution, in this
subsection, we take a 1/2 spin subjected an external magnetic field as a
warm-up example. The Hamiltonian of a free spin under an arbitrary field $%
\mathbf{B}(B\sin \theta \cos \phi ,B\sin \theta \sin \phi ,B\cos \theta )$
with magnitude $B$ is
\begin{equation}
H=-\mathbf{B}\cdot \mathbf{\sigma ,}  \label{eq:Hamiltonian_spin1}
\end{equation}%
where $\sigma =(\sigma ^{x},\sigma ^{y},\sigma ^{z})$ are the Pauli
matrices. In $\sigma ^{z}$ basis $|\uparrow \rangle ,|\downarrow \rangle $,
Pauli matrices take the form%
\begin{equation}
\sigma ^{x}=\left(
\begin{array}{cc}
0 & 1 \\
1 & 0%
\end{array}%
\right) ,\sigma ^{y}=\left(
\begin{array}{cc}
0 & -i \\
i & 0%
\end{array}%
\right) ,\sigma ^{z}=\left(
\begin{array}{cc}
1 & 0 \\
0 & -1%
\end{array}%
\right) .
\end{equation}%
Then the Hamiltonian (\ref{eq:Hamiltonian_spin1}) matrix can be rewritten as
\begin{equation}
H=-B\left(
\begin{array}{cc}
\cos \theta & e^{-i\phi }\sin \theta \\
e^{i\phi }\sin \theta & -\cos \theta%
\end{array}%
\right) .
\end{equation}%
The Hamiltonian can be easily diagonalized and the spin's ground state, with
eigenenergy $E_{0}=-B$, is
\begin{equation}
|\Psi (\theta ,\phi )\rangle =\cos \frac{\theta }{2}|\uparrow \rangle
+e^{i\phi }\sin \frac{\theta }{2}|\downarrow \rangle .
\end{equation}%
Here $\theta $ and $\phi $ define a point on the unit three-dimensional
Bloch sphere to which the spin points to(see Fig. \ref{figure_bloch}). The
state can be multiplied by an arbitrary global phase. Obviously, $\theta $
and $\phi $ can be regarded as adiabatic parameters. \textit{For simplicity
and without loss of generality, we fixed} $\theta $ first$.$ Then the
overlap between two states corresponding to two points on the ring of a
given $\theta $ is
\begin{eqnarray}
f(\theta ,\phi ;\theta ,\phi ^{\prime }) &=&\langle \Psi (\theta ,\phi
)|\Psi (\theta ,\phi ^{\prime })\rangle  \notag \\
&=&\cos ^{2}\frac{\theta }{2}+e^{i(\phi -\phi ^{\prime })}\sin ^{2}\frac{%
\theta }{2}.
\end{eqnarray}%
There are two parts in the overlap. The real part denotes the difference in
the geometrical structure, while the imaginary one corresponds to the
overall phase difference.

Though the overlap shows also the similarity between two states, it does not
show the response of the state at a given point to the adiabatic parameter $%
\phi $. For this purpose, we expand the overlap around a given $\phi $ as
\begin{eqnarray}
&&f(\theta ,\phi ;\theta ,\phi +\delta \phi )  \notag \\
&=&\langle \Psi (\theta ,\phi )|\Psi (\theta ,\phi )\rangle +\delta \phi
\left\langle \Psi (\theta ,\phi )\left\vert \frac{\partial }{\partial \phi }%
\Psi (\theta ,\phi )\right. \right\rangle  \notag \\
&&+\frac{(\delta \phi )^{2}}{2}\left\langle \Psi (\theta ,\phi )\left\vert
\frac{\partial ^{2}}{\partial \phi ^{2}}\Psi (\theta ,\phi )\right.
\right\rangle +\cdots ,
\end{eqnarray}%
where
\begin{eqnarray}
\left\langle \Psi (\theta ,\phi )\left\vert \frac{\partial }{\partial \phi }%
\Psi (\theta ,\phi )\right. \right\rangle &=&i\sin ^{2}\frac{\theta }{2}, \\
\left\langle \Psi (\theta ,\phi )\left\vert \frac{\partial ^{2}}{\partial
\phi ^{2}}\Psi (\theta ,\phi )\right. \right\rangle &=&-\sin ^{2}\frac{%
\theta }{2}.
\end{eqnarray}%
The linear term is the Berry adiabatic connection, which contribute a
Pancharatnam-Berry phase \cite{Pancharatnam56,Berry84} to the spin as the
magnetic field rotates adiabatically around cone direction (the dotted
circle in Fig. \ref{figure_bloch}), i.e.

\begin{eqnarray}
\gamma (\theta ,\phi ) &=&-i\int_{0}^{\phi }\left\langle \Psi (\theta ,\phi
^{^{\prime }})\left\vert \frac{\partial }{\partial \phi ^{^{\prime }}}\Psi
(\theta ,\phi ^{^{\prime }})\right. \right\rangle d\phi ^{^{\prime }},
\notag \\
&=&-\phi \sin ^{2}\frac{\theta }{2}.
\end{eqnarray}%
The phase equals to the solid angle of the cone if the spin rotates one
periodicity. The Berry connection must be a purely imagnary number because
of
\begin{equation}
\left\langle \Psi (\theta ,\phi )\left\vert \frac{\partial }{\partial \phi }%
\Psi (\theta ,\phi )\right. \right\rangle +\left\langle \left. \frac{%
\partial }{\partial \phi }\Psi (\theta ,\phi )\right\vert \Psi (\theta ,\phi
)\right\rangle =0.
\end{equation}

The global phase can be rectified by a gauge transformation $e^{-i\gamma
(\theta ,\phi )}$, which can compensate the geometric phase $\gamma (\theta
,\phi )$ accumulated during the adiabatic evolution. The new state becomes
\begin{equation}
|\Psi (\theta ,\phi )\rangle =e^{-i\gamma (\theta ,\phi )}\left( \cos \frac{%
\theta }{2}|\uparrow \rangle +e^{i\phi }\sin \frac{\theta }{2}|\downarrow
\rangle \right) .
\end{equation}%
Then the overlap between two geometrically similar states becomes

\begin{eqnarray}
f(\theta ,\phi ;\theta ,\phi +\delta \phi ) &=&\exp \left[ i\delta \phi \sin
^{2}\frac{\theta }{2}\right]  \notag \\
&&\times \left( \cos ^{2}\frac{\theta }{2}+e^{-i\delta \phi }\sin ^{2}\frac{%
\theta }{2}\right)  \notag \\
&=&1-\frac{(\delta \phi )^{2}}{2}\sin ^{2}\frac{\theta }{2}\cos ^{2}\frac{%
\theta }{2}+\cdots .
\end{eqnarray}%
The most relevant term is then the second derivative of the overlap.
Moreover, the gauge transformation not only eliminates the Berry adiabatic
connection, but also modifies the second order term. After the phase
rectification, the second-order term is reduced to a minimum. On the other
hand, the phase rectification denotes mathematically a rotation in the
complex plane, which makes the overlap be a purely geometric quantity.

If we take the modulus of the overlap, it becomes the fidelity
\begin{eqnarray}
|f|^{2} &=&\left( 1+i\delta \phi \sin ^{2}\frac{\theta }{2}-\frac{(\delta
\phi )^{2}}{2}\sin ^{2}\frac{\theta }{2}+\cdots \right) ^{2}  \notag \\
&=&1-(\delta \phi )^{2}\sin ^{2}\frac{\theta }{2}\cos ^{2}\frac{\theta }{2}%
+\cdots .
\end{eqnarray}%
Then the fidelity, if we express it in a series form, becomes%
\begin{equation}
F=|f|=1-\frac{(\delta \phi )^{2}}{2}\sin ^{2}\frac{\theta }{2}\cos ^{2}\frac{%
\theta }{2}+\cdots .
\end{equation}%
Therefore, the leading response of the fidelity to the adiabatic parameter
is its second derivative. This is quite natural because the fidelity can not
be large than its upper limit 1, it must be an even function of the
perturbation of the adiabatic parameter. The leading term is called fidelity
susceptibility in some literatures because it is physically a kind of
structure of the driving term,
\begin{eqnarray}
\chi _{F} &=&\sin ^{2}\frac{\theta }{2}\cos ^{2}\frac{\theta }{2}  \notag \\
&=&\frac{1}{4}\sin ^{2}\theta .  \label{eq:spinfdsf}
\end{eqnarray}%
Though the phase rectification can change the Berry adiabatic connection and
the second derivative of the overlap, the fidelity susceptibility does not
change. This phenomenon is due to the simple reason a gauge transformation
cannot affect the modulos of the overlap.

On the other hand, when we study quantum phase transitions occurring in a
quantum-many body system, the ground-state wavefunction is usually defined
in the real space, then the imaginary part of the overlap does not appear.
If the adiabatic parameter is defined on the flat manifold, the linear
correction is zero. The second term is the most important. It denotes the
leading response of the wave function to the adiabatic parameter. Though for
the present case it is simply a constant due to the rotational symmetry of $%
\phi $, it might become singular for a many-body system in the thermodynamic
limit.

Now we consider another case of fixing both $\theta $ and $\phi $, and
changing the magnitude of the external field. If $B>0$, the ground state is%
\begin{equation}
|\Psi (\theta ,\phi )\rangle =\cos \frac{\theta }{2}|\uparrow \rangle
+e^{i\phi }\sin \frac{\theta }{2}|\downarrow \rangle ,
\end{equation}%
with eigenenergy $-B$, while if $B<0$, the ground state becomes%
\begin{equation}
|\Psi (\theta ,\phi )\rangle =e^{i\phi }\sin \frac{\theta }{2}|\uparrow
\rangle -\cos \frac{\theta }{2}|\downarrow \rangle ,
\end{equation}%
with eigenenergy $B$. A ground-state level-crossing occurs at the point $B=0$%
. Then the fidelity shows a very sharp drop at $B=0$ due to the
level-crossing between two orthogonal states. While if we expand the
fidelity in term of $B$, one may find that either the fidelity
susceptibility or the Berry adiabatic connection is zero except for $B=0$.
The point is a singular for both of the fidelity susceptibility and the
Berry adiabatic connection. In many studies on the Pancharatnam-Berry phase,
this level-crossing point is regarded as a monopole in the parameter space.

\begin{figure}[tbp]
\includegraphics[bb=0 -10 330 120, width=8.5 cm, clip] {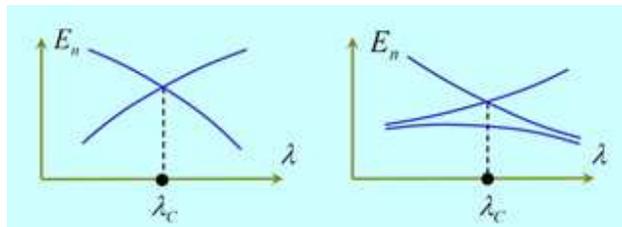}
\caption{ (Color online) A sketch of a ground-state level crossing (LEFT)
and the first-excited state level-crossing(RIGHT) as the system's driving
parameter varies.}
\label{figure_qptlc}
\end{figure}

\begin{figure}[tbp]
\includegraphics[bb=0 -10 330 120, width=8.5 cm, clip] {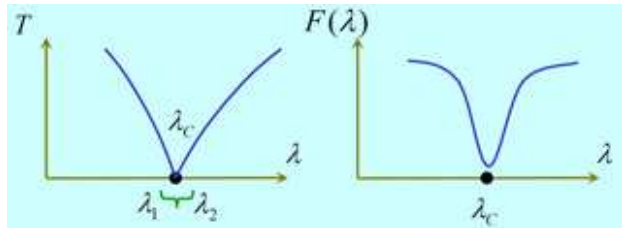}
\caption{ (Color online) A sketch of a quantum phase transition occurred at $%
\protect\lambda_c$ (LEFT) and corresponding expected behavior of the
fidelity $F=\langle\Psi_0(\protect\lambda_1)|\Psi_0(\protect\lambda%
_2)\rangle $ as a function of $\protect\lambda=(\protect\lambda_1+\protect%
\lambda_2)/2$ for a fixed $\protect\delta\protect\lambda=\protect\lambda_2-%
\protect\lambda_1$ (RIGHT).}
\label{figure_qptf}
\end{figure}


\section{Fidelity and quantum phase transitions}

The fidelity and its leading term introduced in the last section is
illustrative. In this section, we try to establish a bridge between quantum
phase transitions and the fidelity in considerable detail through the
one-dimensional transverse-field Ising model.

\label{sec:fs}\bigskip

\subsection{Quantum phase transitions: fidelity perspective}

Without loss of generality, the Hamiltonian of a general quantum many-body
system, which might undergo a quantum phase transition in parameter space,
can be written as
\begin{equation}
H(\lambda )=H_{0}+\lambda H_{I},  \label{eq:generalHamiltonian}
\end{equation}%
where $H_{I}$ is the driving Hamiltonian and $\lambda $ denotes its
strength. According to quantum mechanics, the system satisfies the Schr\"{o}%
dinger equation
\begin{equation}
H(\lambda )|\Psi _{n}(\lambda )\rangle =E_{n}|\Psi _{n}(\lambda )\rangle
,\,n=0,1,\dots ,
\end{equation}%
where $E_{n}$ is the eigenenergy and set to an increasing order $%
E_{0}<E_{1}\leq E_{2}\cdots $, and $|\Psi _{n}(\lambda )\rangle $ defines a
set of orthogonal complete bases in the Hilbert space, i.e.%
\begin{equation}
\sum_{n}|\Psi _{n}(\lambda )\rangle \langle \Psi _{n}(\lambda )|=I.
\end{equation}

As the driving parameter $\lambda $ varies, the energy spectra are changed
correspondingly. The quantum phase transition occurs as the ground-state
energy undergoes a significant change at a certain point. Precisely, its
first- or higher-order derivative with respect to the driving parameter
becomes discontinuous at the transition point. There are two distinct ways.
The first one is the energy level-crossing occurring in the ground state
(left plot of Fig. \ref{figure_qptlc}). The second is that the
level-crossing occurs only in the low-lying excitations \cite{GSTian2003},
and the ground state keeps nondegenerate (right plot of Fig. \ref%
{figure_qptlc}). For both cases, the structures of the ground-state
wavefunction become qualitatively different across the transition point.
That is, if we compare two ground states on both sides of the transition
point, their distance is very large; while if we compare two ground states
in the same phase, their distance is relatively small. Therefore, if we
calculate the fidelity between two ground states, i.e., the fidelity of $%
\Psi _{0}(\lambda _{1})$ and $\Psi _{0}(\lambda _{2})$ at two slightly
separated points $\lambda _{1(2)}$ with fixed $\delta =\lambda _{1}-\lambda
_{2}$, it should manifest a minimum at the transition point, as shown in
Fig. \ref{figure_qptf}. Such a fascinating perspective for quantum phase
transitions was first observed in the one-dimensional transverse-field Ising
model \cite{HTQuan2006,Zanardi06}.

Obviously, the fidelity between two ground states does not bear any apparent
information about the difference in order properties between two phases.
Instead, it is a pure geometric quantity of quantum states. In its approach
to quantum phase transitions, one of obvious advantage is that \emph{no
priori} knowledge of order parameter and symmetry-breaking is required. For
example, if a quantum phase transition is induced by the ground-state level
crossing, then the two crossing states at the transition point are
orthogonal, then the overlap between them is zero; while the fidelity almost
equals to one in other region away from the crossing point. Therefore, it is
believed that the fidelity can describe quantum phase transitions in its own
way.

\subsection{Example: the one-dimensional transverse-field Ising model}

\begin{figure}[tbp]
\includegraphics[bb=0 -10 240 200, width=8.5 cm, clip] {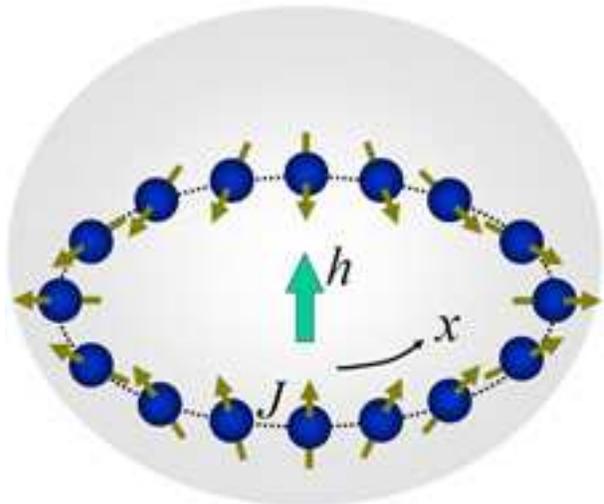}
\caption{ (Color online) A sketch of the one-dimensional transverse-field
Ising model with periodic boundary conditions. Two arbitrary neighboring
spins interact with each other by the Ising interaction $\protect\sigma^x_j
\protect\sigma^x_{j+1}$ (dotted line). All spins are subject to an external
field $h$ along $z$ direction. }
\label{figure_isingmodel}
\end{figure}

\begin{figure}[tbp]
\includegraphics[bb=0 -10 320 200, width=8.5 cm, clip] {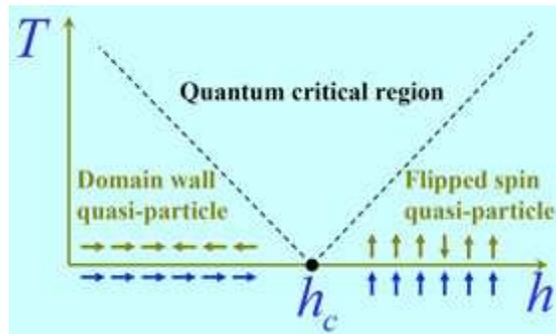}
\caption{ (Color online) A schematic ground-state phase diagram of the
one-dimensional transverse-field Ising model. At the left hand side of $h_c$%
, the ground state is in a ferromagnetic long-range order phase whose
low-lying excitations are flipped spin quasi-particle. While at the right
hand side of $h_c$, the ground state is a fully polarized phase whose
low-lying excitations are domain wall quasi-particles.}
\label{figure_isingphase}
\end{figure}

The one-dimensional transverse-field Ising model \cite%
{PPfeuty70,RJElliott70,RJullien78} is one of the simplest models which can be
solved exactly \cite{EBarouch70,EBarouch71} in the field of condensed matter
physics. Due to its simplicity and clear physical pictures, the model is often
used as a starting model to test new physical ideas and approaches, among which
the fidelity does not make an exception. The following procedure is standard,
and the final expression of fidelity is obtained by Zanardi and Paunkovi\'{c}
\cite{Zanardi06}.

The Hamiltonian of the one-dimensional transverse-field Ising model with
periodic boundary conditions reads%
\begin{eqnarray}
H &=&-\sum_{j=1}^{N}\left( \sigma _{j}^{x}\sigma _{j+1}^{x}+h\sigma
_{j}^{z}\right) ,  \label{eq:Hamiltonian_Ising0} \\
\sigma _{1}^{x} &=&\sigma _{N+1}^{x},
\end{eqnarray}%
where $h$ is the transverse field and $N$ is the number of spins. As
inferred from the model's name, the Hamiltonian describes a chain of spins
with the nearest-neighboring Ising interaction along $x$-direction, and all
spins are subject to a transverse magnetic field $h$ along the $z$-direction
(Fig. \ref{figure_isingmodel}).

The Hamiltonian is invariant under translational operation. Moreover, unlike
usual spin systems,\ the $z$-component of total spins in this model, i.e.
\begin{equation}
\sigma ^{z}=\sum_{j=1}^{N}\sigma _{j}^{z},
\end{equation}%
is not conserved. Instead, if we introduce
\begin{eqnarray}
\sigma ^{+} &=&\frac{1}{2}\left( \sigma ^{x}+i\sigma ^{y}\right) ,\text{ }%
\sigma ^{-}=\frac{1}{2}\left( \sigma ^{x}-i\sigma ^{y}\right) , \\
\sigma ^{+} &=&\left(
\begin{array}{cc}
0 & 1 \\
0 & 0%
\end{array}%
\right) ,\text{ }\sigma ^{-}=\left(
\begin{array}{cc}
0 & 0 \\
1 & 0%
\end{array}%
\right) ,
\end{eqnarray}%
and
\begin{equation}
\sigma ^{-}|\uparrow \rangle =|\downarrow \rangle ,\sigma ^{+}|\downarrow
\rangle =|\uparrow \rangle ,
\end{equation}%
the Hamiltonian (\ref{eq:Hamiltonian_Ising0}) can be transformed into
\begin{eqnarray}
H &=&-\sum_{j=1}^{N}\left[ \left( \sigma _{j}^{+}\sigma _{j+1}^{-}+\sigma
_{j}^{-}\sigma _{j+1}^{+}\right) \right.  \notag \\
&&\left. +\left( \sigma _{j}^{+}\sigma _{j+1}^{+}+\sigma _{j}^{-}\sigma
_{j+1}^{-}\right) +h\sigma _{j}^{z}\right] ,  \label{eq:Hamiltonian_Ising2}
\end{eqnarray}%
then we can see that the off-diagonal terms in the Hamiltonian (\ref%
{eq:Hamiltonian_Ising2}) either exchange the state of a pair of
anti-parallel spins, or flip two upward spins to downward or vice versa. So
they do not change the parity of the system. This property defines a
classification of subspaces based on the parity operators, i.e.%
\begin{equation}
P=\prod_{j=1}^{N}\sigma _{j}^{z},
\end{equation}%
and the Hamiltonian cannot change the parity of the state, i.e.
\begin{equation}
\lbrack H,P]=0.
\end{equation}%
Therefore, we have two subspaces corresponding to parity $P=\pm 1$
respectively.

The ground state of the one-dimensional transverse-field Ising model can be
understood from its two limiting cases. If $h=0$, the Hamiltonian becomes
the classical one-dimensional Ising model. Defining the eigenstates of $%
\sigma ^{x}$ as%
\begin{equation}
|\rightarrow \rangle =\frac{1}{\sqrt{2}}\left( |\uparrow \rangle
+|\downarrow \rangle \right) ,|\leftarrow \rangle =\frac{1}{\sqrt{2}}\left(
|\uparrow \rangle -|\downarrow \rangle \right) ,
\end{equation}%
the doubly degenerate ground states of the Hamiltonian take the form%
\begin{eqnarray}
|\Psi _{1}\rangle &=&|\rightarrow \rightarrow \rightarrow \cdots \rightarrow
\rangle , \\
|\Psi _{2}\rangle &=&|\leftarrow \leftarrow \leftarrow \cdots \leftarrow
\rangle ,
\end{eqnarray}%
which are of ferromagnetic order. The ground-state properties change as the
external field $h$ turns on. Because of
\begin{equation}
\sigma ^{z}|\rightarrow \rangle =|\leftarrow \rangle ,\sigma ^{z}|\leftarrow
\rangle =|\rightarrow \rangle ,
\end{equation}%
the magnetic field mixes $|\Psi _{1}\rangle $ and $|\Psi _{2}\rangle $ and
the ground state becomes non-degenerate for a finite system. Despite of
this, the ground-state property does not change qualitatively. The ground
state still manifests the ferromagnetic long-range order. Precisely, the
correlation function
\begin{equation}
\langle \Psi _{0}|\sigma _{j}^{x}\sigma _{j+r}^{x}|\Psi _{0}\rangle -\langle
\Psi _{0}|\sigma _{j}^{x}|\Psi _{0}\rangle \langle \Psi _{0}|\sigma
_{j+r}^{x}|\Psi _{0}\rangle ,  \label{Eq:Ising_correlation function}
\end{equation}%
does not vanish even if $r\rightarrow \infty $. The correlation function,
therefore, can be used as an order parameter to describe the phase in the
small $h$ region. While if $h\rightarrow \infty $, the Ising interaction is
neglectable, all spins are fully polarized along $z$-direction. The ground
state is non-degenerate and takes the form%
\begin{equation}
|\Psi _{0}\rangle =|\uparrow \uparrow \uparrow \cdots \uparrow \uparrow
\rangle .
\end{equation}%
In this limit, the correlation function Eq. (\ref{Eq:Ising_correlation
function}) does not show long-range behavior. Therefore, a quantum phase
transition between an ordered phase to a disordered phase is expected to
occur as $h$ changes from zero to infinite. A schematic ground-state phase
diagram of the model is shown in Fig. \ref{figure_isingphase}.

In order to discuss the fidelity in the ground state, we now diagonalize the
Hamiltonian in detail. We need three transformations, i.e., the
Jordan-Wigner transformation \cite{PJordan28}, Fourier transformation, and
Bogoliubov transformation.

\textit{The Jordan-Wigner transformation: }The Jordan-Wigner transformation
maps 1/2 spins to spinless fermions, that is%
\begin{eqnarray}
\sigma _{n}^{+} &=&\exp \left[ i\pi \sum_{j=1}^{n-1}c_{j}^{\dagger }c_{j}%
\right] c_{n}=\prod_{j=1}^{n-1}\sigma _{j}^{z}c_{n}, \\
\sigma _{n}^{-} &=&\exp \left[ -i\pi \sum_{j=1}^{n-1}c_{j}^{\dagger }c_{j}%
\right] c_{n}^{\dagger }=\prod_{j=1}^{n-1}\sigma _{j}^{z}c_{n}^{\dagger }, \\
\sigma _{n}^{z} &=&1-2c_{n}^{\dagger }c_{n},
\end{eqnarray}%
where $c_{n}^{\dagger }\ $and $c_{n}$ are fermionic operators and satisfy
the anticommutation relations
\begin{eqnarray}
\{c_{n}^{\dagger },c_{m}\} &=&\delta _{nm},\text{ \ } \\
\{c_{n},c_{m}\} &=&\{c_{n}^{\dagger },c_{m}^{\dagger }\}=0.
\end{eqnarray}%
After the Jordan-Wigner transformation, the Hamiltonian becomes%
\begin{eqnarray}
H &=&-\sum_{j=1}^{N-1}\left[ \left( c_{j}^{\dagger }c_{j+1}+c_{j+1}^{\dagger
}c_{j}\right) +\left( c_{j}^{\dagger }c_{j+1}^{\dagger }+c_{j+1}c_{j}\right) %
\right]  \notag \\
&&+\left( c_{1}^{\dagger }c_{N}+c_{N}^{\dagger }c_{1}\right) \exp \left[
i\pi \sum_{j=1}^{N}c_{j}^{\dagger }c_{j}\right]  \notag \\
&&+\left( c_{N}^{\dagger }c_{1}^{\dagger }+c_{1}c_{N}\right) \exp \left[
i\pi \sum_{j=1}^{N}c_{j}^{\dagger }c_{j}\right]  \notag \\
&&-\sum_{j=1}^{N}h\left( 1-2c_{j}^{\dagger }c_{j}\right) .
\end{eqnarray}%
The exponential factor
\begin{equation*}
P=\exp \left[ i\pi \sum_{j=1}^{N}c_{j}^{\dagger }c_{j}\right] ,
\end{equation*}%
is nothing but the parity of the system which is a constant, i.e. for
periodic boundary conditions $P=-1$ and antiperiodic boundary conditions $%
P=1 $. The Hamiltonian can be simplified as%
\begin{eqnarray}
H &=&-\sum_{j=1}^{N}\left[ \left( c_{j}^{\dagger }c_{j+1}+c_{j+1}^{\dagger
}c_{j}\right) +\left( c_{j}^{\dagger }c_{j+1}^{\dagger }+c_{j+1}c_{j}\right) %
\right]  \notag \\
&&-\sum_{j=1}^{N}h\left( 1-2c_{j}^{\dagger }c_{j}\right) .
\end{eqnarray}

\textit{Fourier transformation: }Since the Hamiltonian is invariant under
translational operation, we can perform standard Fourier transformation. For
the present case, the transformations are
\begin{eqnarray}
c_{j} &=&\frac{1}{\sqrt{N}}\sum_{k}e^{-ikj}c_{k},\text{ \ \ }  \notag \\
c_{j}^{\dagger } &=&\frac{1}{\sqrt{N}}\sum_{k}e^{ikj}c_{k}^{\dagger },
\end{eqnarray}%
where the momentum $k$s are chosen under conditions:%
\begin{equation}
k=\left\{
\begin{array}{cc}
\frac{(2n+1)\pi }{N} & P=1 \\
\frac{2n\pi }{N} & P=-1%
\end{array}%
\right. ,
\end{equation}%
with $n=0,1,2,\cdots N-1$. Then the Hamiltonian can be transformed into $k$%
-space form,
\begin{eqnarray}
H &=&-\sum_{k}\left[ (2\cos k-2h)c_{k}^{\dagger }c_{k}+i\sin k\left(
c_{-k}^{\dagger }c_{k}^{\dagger }+c_{-k}c_{k}\right) \right]  \notag \\
&&-Nh.  \label{eq:Hamiltonian_Isingk}
\end{eqnarray}

\textit{Bogoliubov transformation:} Obviously, the quadratic Hamiltonian can
be further diagonalized under the famous Bogoliubov transformation:

\begin{eqnarray}
c_{k} &=&u_{k}b_{k}+iv_{k}b_{-k}^{\dagger },  \notag \\
c_{k}^{\dagger } &=&u_{k}b_{k}^{\dagger }-iv_{k}b_{-k},  \notag \\
c_{-k} &=&u_{k}b_{-k}-iv_{k}b_{k}^{\dagger },  \notag \\
c_{-k}^{\dagger } &=&u_{k}b_{-k}^{\dagger }+iv_{k}b_{k},  \label{eq:Bogtran}
\end{eqnarray}%
where $b_{k}$ and $b_{k}^{\dagger }$ are also fermionic operator and satisfy
the same anticommutation relation as $c_{k}$ and $c_{k}^{\dagger }$. Because
of this, one can find the coefficients in the transformation (\ref%
{eq:Bogtran}) should satisfy the following condition

\begin{equation}
v_{k}=-v_{-k},\text{ \ \ }u_{k}^{2}+v_{k}^{2}=1.
\end{equation}%
So we can introduce trigonal relation
\begin{equation}
v_{k}=\sin \theta _{k},\text{\ }u_{k}=\cos \theta _{k}.
\end{equation}%
Inserting the Bogoliubov transformation into Eq. (\ref{eq:Hamiltonian_Isingk}%
), the coefficients are determined by\
\begin{eqnarray}
\cos 2\theta _{k} &=&\frac{\cos (k)-h}{\sqrt{1-2h\cos (k)+h^{2}}},  \notag \\
\sin 2\theta _{k} &=&\frac{\sin (k)}{\sqrt{1-2h\cos (k)+h^{2}}},
\end{eqnarray}%
such that the Hamiltonian becomes a quasi-free fermion system,%
\begin{equation}
H=\sum_{k}\epsilon (k)\left( 2b_{k}^{\dagger }b_{k}-1\right) ,
\end{equation}%
where%
\begin{equation}
\epsilon (k)=\sqrt{1-2h\cos (k)+h^{2}}
\end{equation}%
is the dispersion relation of the quasi particles. The dispersion relation
shows that the thermodynamic system is gapless only at $h=1$, and gapped in
\ both phases of $0<h<1$ and $h>1$. Therefore, the quantum phase transition
occurs at the point $h=1$.

\textit{The ground state:} The ground state of the model is defined as the
vacuum state of $b_{k}|\Psi _{0}\rangle =0$ where%
\begin{eqnarray}
b_{k} &=&\cos \theta _{k}c_{k}-i\sin \theta _{k}c_{-k}^{\dagger },  \notag \\
b_{-k} &=&\cos \theta _{k}c_{-k}+i\sin \theta _{k}c_{k}^{\dagger }.
\end{eqnarray}%
Since the condition%
\begin{equation}
b_{k}\left(
\begin{array}{c}
a|0\rangle _{k}|0\rangle _{-k}+b|1\rangle _{k}|0\rangle _{-k} \\
+c|0\rangle _{k}|1\rangle _{-k}+d|1\rangle _{k}|1\rangle _{-k}%
\end{array}%
\right) =0
\end{equation}%
gives
\begin{equation}
a=\cos \theta _{k},b=0,c=0,d=i\sin \theta _{k},
\end{equation}%
the ground state takes the form%
\begin{equation}
|\Psi _{0}(h)\rangle =\prod_{k>0}\left( \cos \theta _{k}|0\rangle
_{k}|0\rangle _{-k}+i\sin \theta _{k}|1\rangle _{k}|1\rangle _{-k}\right) .
\label{eq:gswavefunctionIsing}
\end{equation}%
The low-lying excitation can be obtained by applying $b_{k}^{\dagger }$ to
the ground state. In the ferromagnetic phase, the excitation is visualized
as a quasi-particle of a flipped spin; and as a domain wall quasi-particle
in the fully polarized phase (Fig. \ref{figure_isingphase}).
\begin{figure}[tbp]
\includegraphics[width=8cm]{Ising_fidelity.eps}
\caption{ (Color online) The fidelity of the one-dimensional
transverse-field Ising model for various system sizes. Here $\protect\delta %
h=0.01$.}
\label{figure_Ising_fidelity}
\end{figure}

\begin{figure}[tbp]
\includegraphics[width=8cm]{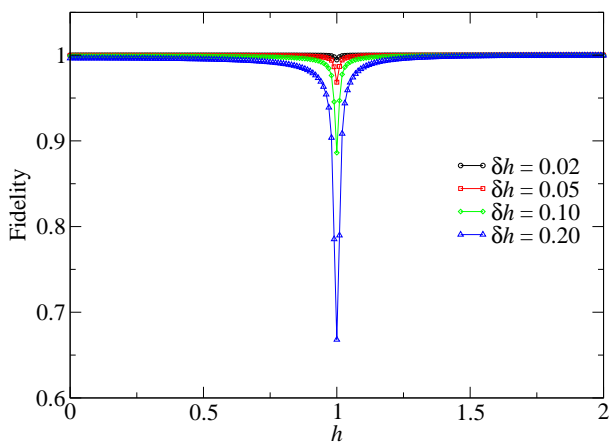}
\caption{ (Color online) The fidelity of the one-dimensional
transverse-field Ising model for various $\protect\delta h$. Here $N=290$.}
\label{figure_Ising_fidelity2}
\end{figure}

\textit{Fidelity:} Once the ground state is obtained explicitly, the
fidelity between $h$ and $h^{\prime }$ can be calculated as \cite{PJordan28}
\begin{equation}
F(h,h^{\prime })=|\langle \Psi _{0}(h^{\prime })|\Psi _{0}(h)\rangle
|=\prod_{k>0}\cos (\theta _{k}-\theta _{k}^{\prime }).
\label{eq:Ising_fidelityexpression}
\end{equation}%
As is always emphasized, the fidelity is purely a geometrical quantity since
it is an inner product between two vectors. Eq. (\ref%
{eq:Ising_fidelityexpression}) refresh our mind on this point because the
expression is just an angle between two vectors. It is also consistent with
the fourth fidelity axiom of Eq. (\ref{eq:axioms4}) because the ground-state
wavefunction (\ref{eq:gswavefunctionIsing}) is already a product state.

Fig. \ref{figure_Ising_fidelity} shows the ground-state fidelity of the
transverse-field Ising model as a function of $h$ with parameter difference $%
\delta h=0.01$. The numerical results of a smaller sample, say 20 sites,
have also been compared with exact numerical computations and the agreement
is essentially perfect (see Table II of section \ref{sec:num}). As expected,
the quantum critical region is clearly marked by a sudden drop of the value
of fidelity. The behavior can be ascribed to a dramatic change in the
structure of the ground state of the system during the quantum phase
transition. The drop becomes sharper and sharper as the system size
increases. Meanwhile the fidelity in the non-critical region is also reduced
though the reduced magnitude is smaller than that at the critical point.
This property can be interpreted due to the increasing of number of degree
of freedom. Actually, in the thermodynamic limit, the fidelity between two
different ground states might be zero, no matter how small the difference in
parameter $\delta h$ is. That is the two ground states are orthogonal to
each other. This phenomena has been studied in quantum many-body systems,
and is known as \textit{the Anderson} \textit{orthogonality catastrophe}\cite%
{PWAnderson67}. Fig. \ref{figure_Ising_fidelity2} shows the fidelity for a
given size system but various $\delta h$. The figure is easy to be
understood. The larger the distance between two points in the parameter
space, the larger the distance between the two corresponding ground states.

\section{Fidelity susceptibility, scaling and universality class}

\label{sec:scaling}

\subsection{The leading term of the fidelity and dynamic structure factor}

\textit{The differential form: }A sudden drop of the fidelity caused by the
ground-state level-crossing is too obvious to be interesting enough. People
are interested in those ground-state wavefunctions which are differentiable
in parameter space. Therefore, the overlap between two ground states at $%
\lambda $ and $\lambda +\delta \lambda $ can be defined as%
\begin{equation}
f(\lambda ,\lambda +\delta \lambda )=\langle \Psi _{0}(\lambda )|\Psi
_{0}(\lambda +\delta \lambda )\rangle .
\end{equation}%
Performing series expansion, the overlap becomes
\begin{eqnarray}
f(\lambda ,\lambda +\delta \lambda ) &=&1+\delta \lambda \left\langle \Psi
_{0}(\lambda )\left\vert \frac{\partial }{\partial \lambda }\Psi
_{0}(\lambda )\right. \right\rangle \\
&&+\frac{(\delta \lambda )^{2}}{2}\left\langle \Psi _{0}(\lambda )\left\vert
\frac{\partial ^{2}}{\partial \lambda ^{2}}\Psi _{0}(\lambda )\right.
\right\rangle +\cdots .
\end{eqnarray}%
The fidelity, as the absolute value of the overlap, then becomes%
\begin{eqnarray}
&&|f(\lambda ,\lambda +\delta \lambda )|^{2}  \notag \\
&=&1+\delta \lambda \left( \left\langle \Psi _{0}\left\vert \frac{\partial }{%
\partial \lambda }\Psi _{0}\right. \right\rangle +\left\langle \left. \frac{%
\partial }{\partial \lambda }\Psi _{0}\right\vert \Psi _{0}\right\rangle
\right)  \notag \\
&&+(\delta \lambda )^{2}\left( \left\langle \left. \frac{\partial }{\partial
\lambda }\Psi _{0}\right\vert \Psi _{0}\right\rangle \left\langle \Psi
_{0}\left\vert \frac{\partial }{\partial \lambda }\Psi _{0}\right.
\right\rangle \right.  \notag \\
&&+\frac{1}{2}\left\langle \Psi _{0}\left\vert \frac{\partial ^{2}}{\partial
\lambda ^{2}}\Psi _{0}\right. \right\rangle +\frac{1}{2}\left. \left\langle
\left. \frac{\partial ^{2}}{\partial \lambda ^{2}}\Psi _{0}\right\vert \Psi
_{0}\right\rangle \right) +\cdots .
\end{eqnarray}%
The linear correction must be zero. There are two reasons. The first is due
to the normalization condition, i.e.
\begin{equation}
\left\langle \Psi _{0}\left\vert \frac{\partial }{\partial \lambda }\Psi
_{0}\right. \right\rangle +\left\langle \left. \frac{\partial }{\partial
\lambda }\Psi _{0}\right\vert \Psi _{0}\right\rangle =\frac{\partial }{%
\partial \lambda }\langle \Psi _{0}|\Psi _{0}\rangle =0.
\end{equation}%
The second is that the $|f|$ must be small than 1, then the leading term
must be an even function of $\delta \lambda $. Therefore,
\begin{equation}
F(\lambda ,\lambda +\delta \lambda )=1-\frac{(\delta \lambda )^{2}}{2}\chi
_{F}+\cdots ,
\end{equation}%
where $\chi _{F}$ denotes the fidelity susceptibility of the ground state,
\begin{eqnarray}
\chi _{F} &=&\left\langle \frac{\partial }{\partial \lambda }\Psi
_{0}\left\vert \frac{\partial }{\partial \lambda }\Psi _{0}\right.
\right\rangle  \notag \\
&&-\left\langle \left. \frac{\partial }{\partial \lambda }\Psi
_{0}\right\vert \Psi _{0}\right\rangle \left\langle \Psi _{0}\left\vert
\frac{\partial }{\partial \lambda }\Psi _{0}\right. \right\rangle .
\end{eqnarray}%
This is the differential form the fidelity susceptibility.

On the other hand, if the ground-state wavefunction is defined in the
multi-dimensional parameter space, say $\lambda =\{\lambda
_{a}\},a=1,2,\cdots \Lambda $, the overlap between two states at $\lambda $
and $\lambda ^{\prime }=\lambda +\delta \lambda $ is
\begin{eqnarray}
f(\lambda ,\lambda +\delta \lambda ) &=&1+\sum_{a}\delta \lambda
_{a}\left\langle \Psi _{0}\left\vert \frac{\partial }{\partial \lambda _{a}}%
\Psi _{0}\right. \right\rangle   \notag \\
&&+\sum_{ab}\frac{\delta \lambda _{a}\delta \lambda _{b}}{2}\left\langle
\Psi _{0}\left\vert \frac{\partial }{\partial \lambda _{a}}\frac{\partial }{%
\partial \lambda _{b}}\Psi _{0}\right. \right\rangle +\cdots .  \notag \\
&&  \label{eq:fidelity_multidim}
\end{eqnarray}%
The fidelity susceptibility becomes%
\begin{eqnarray}
\chi _{F} &=&\sum_{ab}\frac{\partial \lambda _{a}}{\partial \lambda }\frac{%
\partial \lambda _{b}}{\partial \lambda }\left( \frac{1}{2}\left\langle
\frac{\partial }{\partial \lambda _{a}}\Psi _{0}\left\vert \frac{\partial }{%
\partial \lambda _{b}}\Psi _{0}\right. \right\rangle \right.   \notag \\
&&+\frac{1}{2}\left\langle \frac{\partial }{\partial \lambda _{b}}\Psi
_{0}\left\vert \frac{\partial }{\partial \lambda _{a}}\Psi _{0}\right.
\right\rangle  \\
&&\left. -\left\langle \left. \frac{\partial }{\partial \lambda _{a}}\Psi
_{0}\right\vert \Psi _{0}\right\rangle \left\langle \Psi _{0}\left\vert
\frac{\partial }{\partial \lambda _{b}}\Psi _{0}\right. \right\rangle
\right) ,  \label{eq:fidelitys_multidim}
\end{eqnarray}%
where the vector $\partial \lambda _{a}/\partial \lambda $ denotes the
direction of the short displacement $\delta \lambda $ in parameter space.
The term in the parenthesis of Eq. (\ref{eq:fidelitys_multidim}) is called
quantum metric tensor\cite{MVBerry89b,JPProvost} or the Riemann metric
tensor \cite{PZanardi0701061}%
\begin{eqnarray}
g_{ab} &=&\frac{1}{2}\left\langle \frac{\partial }{\partial \lambda _{a}}%
\Psi _{0}\left\vert \frac{\partial }{\partial \lambda _{b}}\Psi _{0}\right.
\right\rangle +\frac{1}{2}\left\langle \frac{\partial }{\partial \lambda _{b}%
}\Psi _{0}\left\vert \frac{\partial }{\partial \lambda _{a}}\Psi _{0}\right.
\right\rangle   \notag \\
&&-\left\langle \left. \frac{\partial }{\partial \lambda _{a}}\Psi
_{0}\right\vert \Psi _{0}\right\rangle \left\langle \Psi _{0}\left\vert
\frac{\partial }{\partial \lambda _{b}}\Psi _{0}\right. \right\rangle .
\end{eqnarray}%
The quantum metric tensor is symmetric under exchange of the index $a$ and $b
$. It is the real part of a more generalized quantum geometric tensor \cite%
{MVBerry89b} of the ground state. Precisely, if we defined the projection
operator%
\begin{equation}
P\equiv I-\left\vert \Psi _{0}\rangle \langle \Psi _{0}\right\vert ,
\label{eq:gsprojectorout}
\end{equation}%
which projects out the ground state, the quantum geometric tensor then is
defined as
\begin{equation}
T_{ab}=\left\langle \left. \frac{\partial }{\partial \lambda _{a}}\Psi
_{0}\right\vert P\left\vert \frac{\partial }{\partial \lambda _{b}}\Psi
_{0}\right. \right\rangle .
\end{equation}%
Therefore,
\begin{equation}
g_{ab}=\text{Re}T_{ab},
\end{equation}%
and the imagnary part of $T_{ab}$ defines a 2-form phase,
\begin{equation}
V_{ab}=2\text{Im}T_{ab}.
\end{equation}%
The 2-form phase $V_{ab}$ plays a very important role in geometric phase.
Its flux gives the Berry phase. While the quantum geometric tensor provides
a natural means of measuring distance along the evolution path in parameter
space. The distance between two ground states can be expressed in the
differential-geometrical form, i.e%
\begin{equation}
ds^{2}=\sum_{ab}g_{ab}\delta \lambda _{a}\delta \lambda _{b}.
\end{equation}%
In addition, if the ground state of the system evolves adiabatically from $%
\lambda $ to $\lambda ^{\prime }$ at a given path $S$, the quantum distance $%
R_{q}$ in the parameter space is
\begin{equation}
R_{q}=\oint_{S}\sqrt{\sum_{ab}g_{ab}d\lambda _{a}d\lambda _{b}}.
\end{equation}%
Therefore, if we do geodesics,
\begin{equation}
\delta R_{q}=0.
\end{equation}%
we can in principle find the shortest path connecting the two ground states
at $\lambda $ and $\lambda ^{\prime }$.

\vspace{5mm} \fbox{%
\parbox{7cm}{ \textbf{Example:}

Take the spin in an external field as an example, its ground state is
\begin{equation}
|\Psi (\theta ,\phi )\rangle =\cos \frac{\theta }{2}|\uparrow \rangle +e^{i\phi
}\sin \frac{\theta }{2}|\downarrow \rangle . \nonumber
\end{equation}The quantum metric tensor takes the form
\begin{equation}
g=\left(
\begin{array}{cc}
1 &  \\
& \sin \theta\end{array}\right) . \nonumber
\end{equation}so\begin{equation}
ds^{2}=d\theta ^{2}+\sin \theta d\phi ^{2} \nonumber
\end{equation}which is the metric on the sphere of parameters. }}\vspace{5mm}

\textit{The perturbation form: }We concern mainly on the fidelity in
continuous phase transitions. That is, the ground state of the Hamiltonian
is nondegenerate for a finite system. Therefore, as the point $\lambda
+\delta \lambda $ closing to $\lambda $, the ground-state wavefunction can
be obtained, to the first order, as
\begin{equation}
|\Psi _{0}(\lambda +\delta \lambda )\rangle =|\Psi _{0}(\lambda )\rangle
+\delta \lambda \sum_{n\neq 0}\frac{H_{I}^{n0}(\lambda )|\Psi _{n}(\lambda
)\rangle }{E_{0}(\lambda )-E_{n}(\lambda )},
\end{equation}%
where
\begin{equation}
H_{I}^{n0}=\langle \Psi _{n}(\lambda )|H_{I}|\Psi _{0}(\lambda )\rangle
\end{equation}%
is the hoping matrix of the driving Hamiltonian $H_{I}$. Therefore, if we
normalized the wavefunction $|\Psi _{0}(\lambda +\delta \lambda )\rangle $,
the fidelity becomes, to the leading order,
\begin{equation}
{F^{2}}=1-\delta \lambda ^{2}\sum_{n\neq 0}\frac{|\langle \Psi _{n}(\lambda
)|H_{I}|\Psi _{0}(\lambda )\rangle |^{2}}{[E_{n}(\lambda )-E_{0}(\lambda
)]^{2}}+\cdots .  \label{eq:fidelityexp}
\end{equation}%
Obviously, the fidelity depends both on $\lambda $ and $\delta \lambda $.
The most relevant term is the leading term in Eq. (\ref{eq:fidelityexp}),
i.e. the second order derivative of the fidelity with respect to $\delta
\lambda $. The term actually defines the response of the fidelity to a small
change in $\lambda $. The fidelity susceptibility can be obtained as
\begin{eqnarray}
\chi _{F}(\lambda ) &\equiv &\lim_{\delta \lambda \rightarrow 0}\frac{-2\ln F%
}{\delta \lambda ^{2}} \\
&=&-\frac{\partial ^{2}F}{\partial (\delta \lambda )^{2}}.
\end{eqnarray}%
With Eq. (\ref{eq:fidelityexp}), it can be rewritten as \cite%
{WLYou07,PZanardi0701061}
\begin{equation}
\chi _{F}(\lambda )=\sum_{n\neq 0}\frac{|\langle \Psi _{n}(\lambda
)|H_{I}|\Psi _{0}(\lambda )\rangle |^{2}}{[E_{n}(\lambda )-E_{0}(\lambda
)]^{2}}.  \label{eq:FSperturbation}
\end{equation}%
This is the summation form of the fidelity susceptibility. The form
establishes a relation between the structure difference of two wavefunctions
and low-lying energy spectra.

\vspace{5mm} \fbox{%
\parbox{7cm}{ \textbf{Example:}

To understand Eq. (\ref{eq:FSperturbation}), we still take the spin in an
external field [Eq. (\ref{eq:Hamiltonian_spin1})] as an example. The driving
term in the Hamiltonian at a fix point can be obtained as
\begin{equation}
\frac{\partial H}{\partial \phi }=iB\left(
\begin{array}{cc}
0 & e^{-i\phi }\sin \theta \\
-e^{i\phi }\sin \theta & 0\end{array} \nonumber
\right) .
\end{equation}The excited state is\begin{equation}
|\Psi _{1}(\theta ,\phi )\rangle =e^{i\phi }\sin \frac{\theta }{2}|\uparrow
\rangle -\cos \frac{\theta }{2}|\downarrow \rangle \nonumber
\end{equation}Then the hoping matrix between the ground state and excited state takes the
form\begin{equation}
\left\langle \Psi _{1}(\theta ,\phi )\left\vert \frac{\partial H}{\partial \phi
}\right\vert \Psi _{0}(\theta ,\phi )\right\rangle =iBe^{i\phi }\sin \theta
\nonumber
\end{equation}The energy different between two state is $2B$, then the fidelity
susceptibility becomes\begin{equation}
\chi _{F}=\frac{1}{4}\sin ^{2}\theta \nonumber
\end{equation}which is the same as Eq. (\ref{eq:spinfdsf}).

}}\vspace{5mm}

On the other hand, according the perturbation theory, the second order
perturbation to the ground-state energy takes the form
\begin{equation}
E_{0}^{(2)}=\sum_{n\neq 0}\frac{|\langle \Psi _{n}(\lambda )|H_{I}|\Psi
_{0}(\lambda )\rangle |^{2}}{E_{0}(\lambda )-E_{n}(\lambda )}.
\label{eq:E2perturbation}
\end{equation}%
Obviously, Eq. (\ref{eq:FSperturbation}) and Eq. (\ref{eq:E2perturbation})
are very similar in their form except for different exponents in both
denominators \cite{SChen08}. Therefore, one might expect that the origin of
the singularity of the fidelity susceptibility and $E_{0}^{(2)}$ are both
due to the vanishing of the energy gap though the fidelity susceptibility
shows a sharper peak than $E_{0}^{(2)}$. For the finite-order phase
transition, however, $E_{0}^{(2)}$ can be still a continuous function of the
driving parameter, then there is no reason to require that the fidelity
susceptibility shows singular behavior in high-order($>2$) quantum phase
transitions. It was also pointed out later that the fidelity susceptibility
might be related to the third energy perturbation \cite{LCVenuti08012473},
\begin{equation}
\chi _{F}=\frac{1}{H_{I}^{00}}\sum_{i,j>0}\frac{%
H_{I}^{0i}H_{I}^{ij}H_{I}^{j0}}{(E_{i}-E_{0})(E_{j}-E_{0})}-\frac{E_{0}^{(3)}%
}{E_{0}^{(1)}}.
\end{equation}%
Therefore, the fidelity susceptibility might not be able to witness those
phase transitions of infinite order \cite{WLYou07,SChen08}, such as the
Beresinskii-Kosterlitz-Thouless transition.

\textit{The fidelity susceptibility as a kind of fluctuation}: The hoping
matrix $\langle \Psi _{n}|H_{I}|\Psi _{0}\rangle $ implies dynamics
behaviors of the fidelity susceptibility. Similar to the linear response
theory, one can define the \textit{dynamic fidelity susceptibility} as

\begin{equation}
\chi _{F}(\omega )=\sum_{n\neq 0}\frac{|\langle \Psi _{n}|H_{I}|\Psi
_{0}\rangle |^{2}}{[E_{n}-E_{0}]^{2}+\omega ^{2}}.
\end{equation}%
Performing a Fourier transformation, the dynamic fidelity susceptibility
becomes
\begin{equation}
\chi _{F}(\tau )=\sum_{n\neq 0}\frac{\pi |\langle \Psi _{n}|H_{I}|\Psi
_{0}\rangle |^{2}}{E_{n}-E_{0}}e^{-(E_{n}-E_{0})|\tau |}.
\end{equation}
The energy difference in the denominators can be canceled if one take a
derivative with respect to $\tau $, the dynamic fidelity susceptibility then
is

\begin{equation}
\frac{\partial \chi _{F}(\tau )}{\partial \tau }=-\pi G_{I}(\tau )\theta
(\tau )+\pi G_{I}(-\tau )\theta (-\tau ).  \notag  \label{eq:fed_fluctuation}
\end{equation}%
Here $\theta (\tau )$ is the step function
\begin{equation}
\theta (\tau )=\left\{
\begin{array}{cc}
1 & \tau >0 \\
1/2 & \tau =0 \\
0 & \tau <0%
\end{array}%
\right.
\end{equation}%
and%
\begin{eqnarray}
G_{I}(\tau ) &=&\langle \Psi _{0}|H_{I}(\tau )H_{I}(0)|\Psi _{0}\rangle
-\langle \Psi _{0}|H_{I}|\Psi _{0}\rangle ^{2} \\
H_{I}(\tau ) &=&e^{H(\lambda )\tau }H_{I}e^{-H(\lambda )\tau },
\end{eqnarray}%
with $\tau $ being the imaginary time. Performing an inverse Fourier
transformation, we can obtain%
\begin{equation}
\chi _{F}(\omega )=\frac{1}{\omega }\int_{0}^{\infty }\sin (\omega \tau
)G_{I}(\tau )d\tau .
\end{equation}%
The fidelity susceptibility then becomes%
\begin{equation}
\chi _{F}=\lim_{\omega \longrightarrow 0}\frac{1}{\omega }\int_{0}^{\infty
}\sin (\omega \tau )G_{I}(\tau )d\tau .
\end{equation}%
For any finite system, the correlation function $G_{I}(\tau )$ decays in the
large $\tau $ limit, the above limit satisfies the Lebesgue's convergent
theorem, the fidelity susceptibility, finally, has the form \cite{WLYou07}

\begin{equation}
\chi _{F}=\int_{0}^{\infty }\tau G_{I}(\tau )d\tau .  \label{eq:fsstrcture}
\end{equation}%
Therefore, the fidelity susceptibility is nothing but a kind of dynamics
structure factor of the driving Hamiltonian.

The Eq. (\ref{eq:fsstrcture}) is remarkable because it connects the fidelity
to dynamical response of the system by the driving Hamiltonian $H_{I}$. In
this way, the adiabatic evolution of the ground state and the fidelity
susceptibility are expressed in terms of standard quantities in linear
response theory and their physics content is clarified. Traditionally,
quantum phase transitions are said to be driven by quantum fluctuations,
which originate from the Heisenberg uncertainty relation. In the Hamiltonian
(\ref{eq:generalHamiltonian}),
\begin{equation}
\lbrack H_{0},H_{I}]\neq 0.\;
\end{equation}%
The the second order perturbation to the ground-state energy
\begin{eqnarray}
E_{0}^{(2)} &=&\int \left[ \langle \Psi _{0}|H_{I}(\tau )H_{I}(0)|\Psi
_{0}\rangle -\langle \Psi _{0}|H_{I}|\Psi _{0}\rangle ^{2}\right] d\tau
\notag \\
&&
\end{eqnarray}%
is also a kind of fluctuation. Clearly, both\ $E_{0}^{(2)}$ and $\chi _{F}$
become zero if $[H_{0},H_{I}]=0$. Therefore, the expression (\ref%
{eq:fsstrcture}) provides a new angle to understand the role of quantum
fluctuation in quantum phase transitions.

\textit{The quantum metric tensor}: In case that the Hamiltonian is defined
in a high-dimensional parameter space, the fidelity susceptibility becomes a
metric tensor. The Hamiltonian in $\Lambda $-dimensional parameter space
reads,
\begin{equation}
H=\sum_{a}\lambda _{a}H_{a},\;\;a=1,\dots ,\Lambda ,
\end{equation}%
where $\lambda _{a}$s are coupling parameters. Clearly
\begin{equation}
H_{a}=\frac{\partial H}{\partial \lambda _{a}}.\;
\end{equation}%
In the parameter space, we can always let the ground state of the system
evolves along a certain path, i.e
\begin{equation}
\lambda _{a}=\lambda _{a}(\lambda ),
\end{equation}%
where $\lambda $ plays a kind of driving parameter along the evolution line.
Therefore, the driving term in the Hamiltonian at a given point $\lambda $
is
\begin{equation}
H_{I}=\frac{\partial H}{\partial \lambda }=\sum_{a}\frac{\partial \lambda
_{a}}{\partial \lambda }H_{a}.\;\;
\end{equation}%
Then the fidelity susceptibility along this line can be calculated as%
\begin{eqnarray}
\chi _{F}(\lambda ) &=&\sum_{n\neq 0}\frac{|\langle \Psi _{n}(\lambda
)|H_{I}|\Psi _{0}(\lambda )\rangle |^{2}}{[E_{n}(\lambda )-E_{0}(\lambda
)]^{2}}  \notag \\
&=&\sum_{ab}\frac{\partial \lambda _{a}}{\partial \lambda }\frac{\partial
\lambda _{b}}{\partial \lambda }\sum_{n\neq 0}\frac{H_{a}^{0n}H_{b}^{n0}}{%
[E_{n}(\lambda )-E_{0}(\lambda )]^{2}},
\end{eqnarray}%
where $H_{a(b)}^{mn}=\langle \Psi _{m}(\lambda )|H_{a(b)}|\Psi _{n}(\lambda
)\rangle $. So the quantum metric tensor takes the form%
\begin{equation}
g_{ab}=\sum_{n}\frac{H_{a}^{0n}H_{b}^{n0}}{[E_{n}(\lambda )-E_{0}(\lambda
)]^{2}}.
\end{equation}%
This is the perturbative form for the quantum metric tensor. \
\begin{equation}
g_{ab}=\int \tau G_{ab}(\tau )d\tau ,  \label{eq:gabcorrelation}
\end{equation}%
where $G_{ab}(\tau )$ is a time-dependent correlation function,
\begin{eqnarray}
G_{ab}(\tau ) &=&\theta (\tau )(\langle \Psi _{0}|H_{a}(\tau )H_{b}(0)|\Psi
_{0}\rangle  \notag \\
&&-\langle \Psi _{0}|H_{a}(0)|\Psi _{0}\rangle \langle \Psi
_{0}|H_{b}(0)|\Psi _{0}\rangle ).
\end{eqnarray}%
Clearly, Eq. (\ref{eq:gabcorrelation}) is an extension of Eq. (\ref%
{eq:fsstrcture}). Both of them clarify the physics content of the fidelity
approach and the role of quantum fluctuation in the adiabatic evolution.

\subsection{Scaling analysis and universality class}

In physics, if a physical quantity depends linearly on the system size, the
quantity is extensive (additive), such as energy; and if it is independent
of the system size, it is intensive, such as energy density. If we define $%
\Delta $ as the energy gap between the ground state and the lowest
excitation with nonzero $\langle \Psi _{n}(\lambda )|H_{I}|\Psi _{0}(\lambda
)\rangle $, the fidelity susceptibility satisfies the following inequalities
\begin{eqnarray}
\chi _{F} &\leq &\frac{1}{\Delta }\sum_{n\neq 0}\frac{|\langle \Psi
_{n}(\lambda )|H_{I}|\Psi _{0}(\lambda )\rangle |^{2}}{[E_{n}(\lambda
)-E_{0}(\lambda )]}=-\frac{1}{2\Delta }\frac{\partial ^{2}E(\lambda )}{%
\partial \lambda ^{2}},  \notag \\
&\leq &\frac{1}{\Delta ^{2}}\sum_{n\neq 0}|\langle \Psi _{n}(\lambda
)|H_{I}|\Psi _{0}(\lambda )\rangle |^{2}, \\
&=&\frac{1}{\Delta ^{2}}[\langle \Psi _{0}(\lambda )|H_{I}^{2}|\Psi
_{0}(\lambda )\rangle -\langle \Psi _{0}(\lambda )|H_{I}|\Psi _{0}(\lambda
)\rangle ^{2}].  \notag
\end{eqnarray}
The inequalities tell us some useful information about the properties of the
fidelity susceptibility away from the critical point:

\begin{enumerate}
\item {If the system is gapped and }$H_{I}$ behaves like a single particle{,
$\chi _{F}$ is an intensive quantity.}

\item {If the system is gapped, the fidelity susceptibility share the same
(or smaller) dependence on the system size as the second order derivative of
the ground state energy.}

\item {If $\chi _{F}$ is a superextensive quantity, the ground state of the
system should be gapless.}
\end{enumerate}

Therefore, unlike conventional physical quantities, the fidelity
susceptibility can either be an extensive quantity, like the energy, or has
other type of dependence on the system size. To have an explicit view of the
dependence of the fidelity susceptibility on the system size, it is useful
to perform scaling transformation, which is firstly carried out by Venuti
and Zanardi \cite{LCVenuti07}. Without loss of generality, the interaction
Hamiltonian can be written as a summation of local terms, i.e.
\begin{equation}
H_{I}=\sum_{r}V(r),
\end{equation}%
and number of site
\begin{equation}
N=L^{d},
\end{equation}%
where $d$ is the real dimension of the system. Then the fidelity
susceptibility becomes \cite{WLYou07}
\begin{equation}
\frac{\chi _{F}}{L^{d}}=\sum_{r}\int \tau C(r,\tau )d\tau ,
\end{equation}%
where%
\begin{gather}
C(r,\tau )=\langle \Psi _{0}(\lambda )|V(r,\tau )V(0,0)|\Psi _{0}(\lambda
)\rangle  \notag \\
-\langle \Psi _{0}(\lambda )|V(r,0)|\Psi _{0}(\lambda )\rangle \langle \Psi
_{0}(\lambda )|V(0,0)|\Psi _{0}(\lambda )\rangle .
\end{gather}%
In the vicinity of the critical point, the scaling transformation goes \cite%
{MNBarberb,MAContinentinob}
\begin{equation}
r^{\prime }=s\,r,\;\;\tau ^{\prime }=s^{\zeta }\tau ,\;\;V(r^{\prime
})=s^{-\Delta _{V}}V(r),
\end{equation}%
where $s>1$, $\zeta $ is the dynamic exponent, and $\Delta _{V}$ is the
scaling dimension of $V(r)$, one can find that
\begin{equation}
\frac{\chi _{F}^{\prime }}{(L^{\prime }/s)^{d}}=\frac{1}{s^{2\zeta -2\Delta
_{V}}}\sum_{r^{\prime }}\int \tau ^{\prime }C(r^{\prime },\tau )d\tau
^{\prime }.
\end{equation}%
Therefore
\begin{equation}
\frac{\chi _{F}^{\prime }}{(L^{\prime })^{d}}=\frac{1}{s^{d+2\zeta -2\Delta
_{V}}}\frac{\chi _{F}}{L^{d}},
\end{equation}%
which defines the scaling transformation of the fidelity susceptibility.
Around the critical point, the correlation length is divergent and the only
length scale is the system size itself, the fidelity susceptibility scales
like
\begin{equation}
\frac{\chi _{F}}{L^{d}}\sim L^{d+2\zeta -2\Delta _{V}}.
\label{eq:scalingeqfdds0}
\end{equation}%
The expression is interesting. It establishes a relation between the size
dependence of the fidelity susceptibility, the dynamic exponent, and the
scaling dimension of the driving term.

In most cases, we can regard $\chi _{F}/L^{d}$ as an intensive quantity in
general quantum phases because the system is usually gapped. Therefore, the
singular behavior of the fidelity susceptibility is due to
\begin{equation}
d+2\zeta >2\Delta _{V}  \label{eq:scalingeqfdds}
\end{equation}%
at the critical point. If the fidelity susceptibility diverges as%
\begin{equation}
\frac{\chi _{F}}{L^{d}}\sim \frac{1}{|\lambda -\lambda _{c}|^{\alpha }}.
\end{equation}%
the scaling exponents satisfy%
\begin{equation}
\alpha =\frac{d+2\zeta -2\Delta _{V}}{\nu },  \label{eq:scalingeqfdds3}
\end{equation}%
where $\nu $ is the critical exponent of the correlation length. A similar
relation was also obtained by Gu \textit{et al} \cite{SJGu072} in their
studies on the one-dimensional asymmetric Hubbard model.

However, Eqs. (\ref{eq:scalingeqfdds0}-\ref{eq:scalingeqfdds3}) are not
universally true in all quantum phase transitions. Considering a general
correlation function
\begin{equation}
C(r,\tau )=\frac{1}{r^{2\Delta _{V}}}f(r\tau ^{1/\zeta })
\end{equation}%
the fidelity susceptibility, in the thermodynamic limit, becomes%
\begin{eqnarray}
\frac{\chi _{F}}{L^{d}} &=&\sum_{r}\int \frac{\tau }{r^{2\Delta _{V}}}%
f(r\tau ^{1/\zeta })d\tau , \\
&\sim &\sum_{r}\frac{1}{r^{2\Delta _{V}-2\zeta }} \\
&\propto &\left\{
\begin{array}{cc}
L^{d+2\zeta -2\Delta _{V}}, & 2\Delta _{V}-2\zeta \neq d \\
\text{ln}L, & 2\Delta _{V}-2\zeta =d%
\end{array}%
\right. .\text{ }
\end{eqnarray}%
Such an explict expression implies that the fidelity susceptibility is not
alway extensive. Even in a gapped phase, as we can see from the
Lipkin-Meshkov-Glick model, it can be intensive. Therefore, if we define the
size exponent above (below) the critical point as $d^{+}(d^{-})$, that is
the fidelity susceptibility%
\begin{equation}
\chi (\lambda )\propto L^{d^{\pm }}.
\end{equation}%
Hence $\chi _{F}/L^{d^{\pm }}$ is an intensive quantity in corresponding
phases. Physically, since the fidelity susceptibility denotes the response
of the ground state to the adiabatic parameter and\ $d_{a}^{\pm }$ is the
dimensional dependence of the adiabatic evolution, $d_{a}^{\pm }$ has been
called \textit{adiabatic dimension}.

Therefore, the singular part of the fidelity susceptibility as an intensive
quantity, around the critical point behaves \cite{SJGu072}
\begin{equation}
\frac{\chi _{F}}{L^{d_{a}^{\pm }}}\sim \frac{1}{|\lambda -\lambda
_{c}|^{\alpha ^{\pm }}},  \label{eq:criticalexponentsfds}
\end{equation}%
where $\alpha ^{\pm }$ denote the critical exponents of the fidelity
susceptibility above or below the critical point. On the other hand, if the
fidelity susceptibility around the critical point shows a peak for a finite
system, its maximum point at $\lambda _{\mathrm{max}}$ scales like
\begin{equation}
\chi (\lambda =\lambda _{\mathrm{max}})\propto L^{d_a^c },
\end{equation}%
where $d_a^c$ denotes the critical adiabatic dimension, and can be either
analyzed from Eq. (\ref{eq:scalingeqfdds0}) or obtained from numerical
scaling analysis. The following function can include the above two
asymptotic behaviors,
\begin{equation}
\frac{\chi (\lambda ,L)}{L^{d_{a}^{\pm }}}=\frac{A}{L^{-d_a^c +d_{a}^{\pm
}}+B(\lambda -\lambda _{\mathrm{max}})^{\alpha ^{\pm }}},
\label{eq:scalefun}
\end{equation}%
where $A$ is a constant, $B$ is a nonzero function of $\lambda $ around the
critical point, and both of them are independent of the system size.
According to Eq. (\ref{eq:scalefun}), the rescaled FS is a universal
function of the rescaled driving parameter $L^{\nu }(\lambda -\lambda _{%
\mathrm{max}})$, i.e.,
\begin{equation}
\frac{\chi _{F(\lambda )}(\lambda =\lambda _{\mathrm{max}},L)-\chi
_{F(\lambda )}(\lambda ,L)}{\chi _{F(\lambda )}(\lambda ,L)}=f[L^{\nu
}(\lambda -\lambda _{\mathrm{max}})],
\end{equation}%
where $\nu $ is the critical exponent of the correlation length. Then we
have \cite{SJGu072,SJGu08073491}
\begin{equation}
\alpha ^{\pm }=\frac{d_a^c -d_{a}^{\pm }}{\nu }.  \label{eq:crticalrelations}
\end{equation}%
Therefore, unlike the second derivative of the ground-state energy, the
fidelity susceptibility might have different critical exponents at both
sides of the critical point. The above procedure is useful to determine the
critical exponent from numerical computations, and has been used in some
models. Nevertheless, it is still not complete. In some cases, the fidelity
susceptibility shows logarithmic divergence around the critical point. Then
we have $\alpha =0$ , and Eqs. (\ref{eq:criticalexponentsfds}-\ref%
{eq:crticalrelations}) should be changed correspondingly. For this case, at
one side of the critical point, if $\chi _{F}/L^{d_{a}}$ is intensive, and
\begin{equation}
\frac{\chi (\lambda =\lambda _{\mathrm{max}})}{L^{d_{a}}}\propto \ln L,
\end{equation}%
around the critical point. The logarithemic divergence implies that
\begin{eqnarray}
&&1-\exp [\chi _{F(\lambda )}(\lambda ,L)-\chi _{F(\lambda )}\left( \lambda
=\lambda _{\mathrm{max}},L\right) ]  \notag \\
&=&f[L^{\nu }(\lambda -\lambda _{\mathrm{max}})],
\end{eqnarray}%
\ should be a universal function.

Clearly, unlike conventional physical quantity that is either intensive or
extensive, the fidelity susceptibility, as analyzed above, manifests
distinct scaling behavior. This property might be due to both the relevance
of the driving Hamiltonian under the renormalization group transformation
and distinct dynamic exponent in different phases.

\subsection{Example A: the one-dimensional transverse-field Ising model}

\begin{figure}[tbp]
\includegraphics[width=8cm]{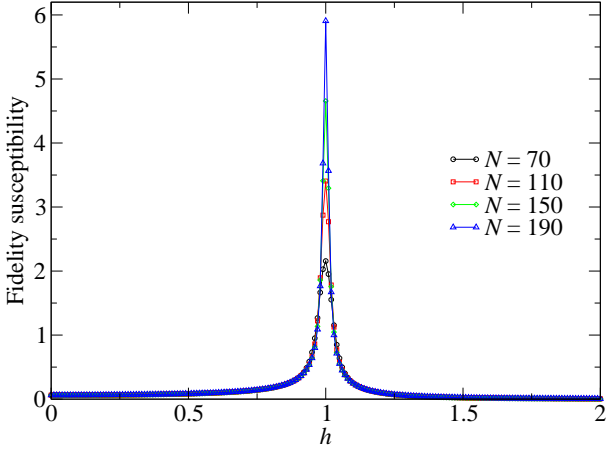}
\caption{ (Color online) The fidelity susceptibility $\protect\chi_F/N$ of
the one-dimensional transverse-field Ising model for various system sizes.}
\label{figure_Ising_fs}
\end{figure}

The one-dimensional transverse-field Ising model gives us a simple example.
For the Ising model, the fidelity is
\begin{equation}
F(h,h^{\prime })=|\langle \Psi _{0}(h^{\prime })|\Psi _{0}(h)\rangle
|=\prod_{k>0}\cos (\theta _{k}-\theta _{k}^{\prime }).
\end{equation}%
At the point $h$, the fidelity susceptibility can be calculated as%
\begin{equation}
\chi _{F}=\sum_{k>0}\left( \frac{d\theta _{k}}{d\lambda }\right) ^{2},
\label{eq:fsisingana}
\end{equation}%
where%
\begin{equation}
\frac{d\theta _{k}}{d\lambda }=\frac{1}{2}\frac{\sin k}{1+h^{2}-2h\cos k}.
\end{equation}%
To find the scaling behavior of the fidelity susceptibility, let us first
consider the case of $h=1$, then
\begin{eqnarray}
\chi _{F} &=&\frac{1}{16}\sum_{k}\frac{\sin ^{2}k}{(1-\cos k)^{2}}, \\
&\simeq &\frac{N}{2}\int_{\pi /N}^{\pi (N-1)/N}\frac{1}{k^{2}}dk, \\
&\varpropto &N^{2}.
\end{eqnarray}%
Therefore, we have
\begin{equation}
d_a^c =2,
\end{equation}%
for the one-dimensional transverse-field Ising model (see Fig. \ref%
{figure_Ising_fs}). On the other hand, if $h\neq 1$, the fidelity
susceptibility becomes%
\begin{equation*}
\chi _{F}=\frac{N}{2}\int_{\pi /N}^{\pi (N-1)/N}\left( \frac{\sin k}{%
1+h^{2}-2h\cos k}\right) ^{2}dk.
\end{equation*}%
Obviously, there is no pole in the denominator of integrand. $\chi _{F}$
then is an extensive quantity (see Fig. \ref{figure_Ising_fs}),
\begin{equation}
\frac{\chi _{F}}{N}=\frac{1}{2}\int_{0}^{1}\left[ \frac{\sin (2\pi x)}{%
1+h^{2}-2h\cos (2\pi x)}\right] ^{2}dx.  \label{eq:fidelitycird}
\end{equation}%
So the adiabatic dimenion for the Ising model is $d_{a}^{\pm }=1$.
Explicitly, the integration can be evaluated by the residue theorem \cite%
{HMKThesis}. Let
\begin{eqnarray*}
\sin (2\pi x) &=&\frac{1}{2i}\left( z-\frac{1}{z}\right) , \\
\cos (2\pi x) &=&\frac{1}{2}\left( z+\frac{1}{z}\right) ,
\end{eqnarray*}%
Eq. (\ref{eq:fidelitycird}) becomes a contour integration along a unit
circle. Then, we can obtain%
\begin{equation*}
\frac{\chi _{F}}{N}=\frac{1}{16(1-h^{2})},
\end{equation*}%
for $h<1$, and%
\begin{equation*}
\frac{\chi _{F}}{N}=\frac{1}{16h^{2}(h^{2}-1)}.
\end{equation*}%
for $h>1$. At both sides of $h_{c}=1$, we have
\begin{equation}
\alpha ^{\pm }=1,
\end{equation}%
as the critical exponents. One can observe that in the both phases of the
Ising model, the fidelity susceptibility is an extensive quantity. In
addition, taking into account that the exponents correlation length $\nu =1$%
. The scaling relation $\alpha ^{\pm }=(d_a^c-d_{a}^{\pm })/\nu $ is
satisfied.

\begin{figure}[tbp]
\includegraphics[bb=0 -10 200 160, width=8.5 cm, clip] {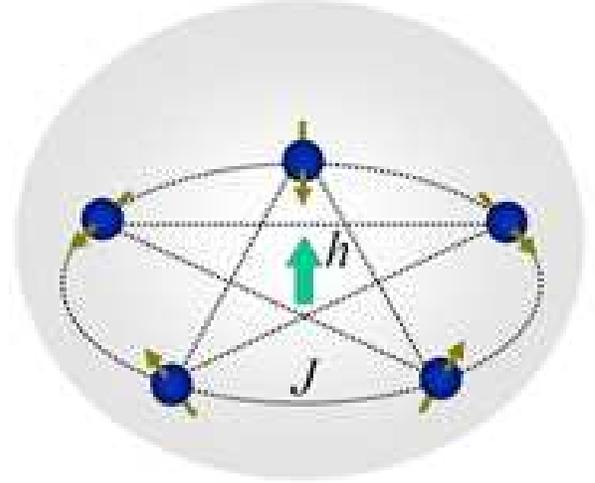}
\caption{ (Color online) A sketch of the Lipkin-Meshkov-Glick model of 5
spins in which all 5 spins are mutually interact with each other (dotted
lines) and subject to an external field $h$ along $z$ direction. }
\label{figure_lmgmodel}
\end{figure}

\begin{figure}[tbp]
\includegraphics[width=8.5cm]{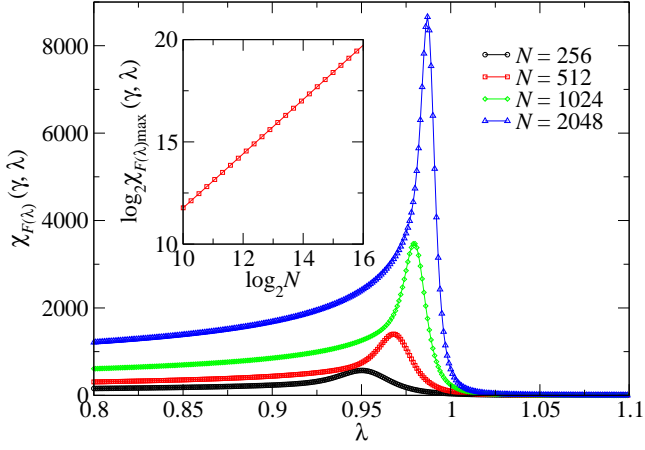}
\caption{(Color online) The fidelity susceptibility in the ground state of
the Lipkin-Meshkov-Glick model as a function of $\protect\lambda$ at $%
\protect\gamma = 0.5$. The inset denotes the scaling behavior of the maximum
of the fidelity susceptibility, in which the slope of the line represents
the size exponent of the fidelity susceptibility (From Ref. \protect\cite%
{HMKwok07}).}
\label{figure_fs_h05}
\end{figure}

\begin{figure}[tbp]
\includegraphics[width=8cm]{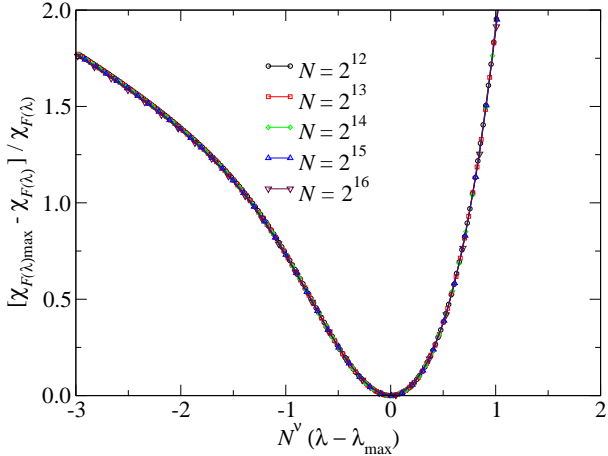}
\caption{(Color online) The finite size scaling analysis is performed for
the case of power-law divergence at $\protect\gamma= 0.5$ for system sizes $%
N=2^{n} (n=12, 13, 14, 15, 16)$. The fidelity susceptibility is considered
as a function of system size and driving parameter is a function of $N^%
\protect\nu (\protect\lambda-\protect\lambda_{\mathrm{max}})$ only, with the
critical exponent $\protect\nu\simeq 0.665$ (From Ref. \protect\cite%
{HMKwok07}).}
\label{figure_3}
\end{figure}

\subsection{Example B: the Lipkin-Meshkov-Glick model}

The one-dimensional transverse-field Ising model is not clear enough to
explain the role played by the adiabatic dimension. For this purpose, we now
take the Lipkin-Meshkov-Glick model \cite%
{Lipkin1965,Meshkov1965,Meshkov19652} as an example. The
Lipkin-Meshkov-Glick model was originally introduced by Lipkin, Meshkov, and
Glick \cite{Lipkin1965,Meshkov1965,Meshkov19652} to describe a collective
motion in nuclei. The model consists of a cluster of mutually interacting
spins in a transverse magnetic field $\lambda $. Its Hamiltonian reads
\begin{eqnarray}
H &=&-\frac{1}{N}\sum\limits_{i<j}{\left( {\sigma _{i}^{x}\sigma
_{j}^{x}+\gamma \sigma _{i}^{y}\sigma _{j}^{y}}\right) }-\lambda
\sum\limits_{i}{\sigma _{i}^{z},} \\
&=&-\frac{{2}}{N}\left( {S_{x}^{2}+\gamma S_{y}^{2}}\right) -2\lambda S_{z}+%
\frac{1}{2}\left( {1+\gamma }\right) ,  \label{eq:HamiltonianLMGmodel}
\end{eqnarray}%
where $S_{\kappa }=\sum_{i}\sigma _{i}^{\kappa }/2(\kappa =x,y,z)$ are spin
1/2 operators, $S_{\pm }=S_{x}\pm iS_{y}$, and $N$ the number of spins. The
prefactor $1/N$ is to ensure finite energy per spin in the thermodynamic
limit $N\rightarrow \infty $. In the anisotropic case: $\gamma \neq 1$, the
Hamiltonian commutes with both the total spin $S^{2}$ and the parity ${%
P=\prod\limits_{i}{\sigma _{i}^{z}}}$, i.e.
\begin{equation}
\left[ {H,S^{2}}\right] =\left[ {H,P}\right] =0.
\end{equation}%
The symmetries significantly reduce the dimension of Hilbert space. For the
present case, the ground state is a ferromagnetic when $\lambda >0$. Then
the dimension of the subspace in which the ground state locates is $N/2$.
This reduces the complexity of the problem, and one can study numerically a
sample up to $2^{16}$ spins using the standard diagonalization method.

In the thermodynamic limit, the ground state of the system undergoes a
second order quantum phase transition at $\lambda _{c}=1$. If $\lambda
>\lambda _{c}$, the system is fully magnetized, while $0<\lambda <\lambda
_{c}$ it is a symmetry-broken state. This conclusion was early drawn by the
mean-field approaches \cite{Botet82,Botet83}. The finite-size scaling of
this model was studied by the $1/N$ expansion in the Holstein-Primakoff
single boson representation \cite{Holstein1940} and by the continuous
unitary transformations \cite{CUT1,CUT2}. Recent studies also reveals a rich
structure of the spectrum, four regions are distinghished in the parameter
space \cite{JVidalSpectra}. Besides, peoples also found that entanglement
properties \cite{QPT1st,QPT2nd,EntE1,EntE2} and fidelity \cite{HMKwok07} in
the ground state of model can provide us a deep understanding on the quantum
phase transition.

The following results based on the fidelity approach done by Kwok \textit{et
al }\cite{HMKwok07}. Fig. \ref{figure_fs_h05} show the dependence of the
fidelity susceptibility on the driving parameter for various system sizes.
As expected, the peak around the critical point becomes sharper and sharper
as the system size increases. Numerical analysis show that $\chi
_{F}(\lambda =\lambda _{\mathrm{max}})\sim N^{d_a^c}$ with $d_a^c \simeq
1.33 $. The second observation is that the fidelity susceptibility shows
different dependence on the system size in both phases. In the
symmetry-breaking phase, $\chi _{F}(\lambda )\sim N$, while in the classical
phase, $\chi _{F}(\lambda )$ in an intensive quantity. Then adiabatic
dimension of the fidelity susceptibility takes $d^{-}=1$ and $d^{+}=0$
respectively. On the other hand, according to the scaling analysis discussed
above, the rescaled fidelity susceptibility%
\begin{equation*}
\frac{\chi _{F}(\lambda =\lambda _{\mathrm{max}},N)-\chi _{F}(\lambda ,N)}{%
\chi _{F}(\lambda ,N)},
\end{equation*}%
should be a function of $N^{\nu }(\lambda -\lambda _{\mathrm{max}})$. Fig. %
\ref{figure_3} shows this function for various system size. All lines fall
onto a single line for $\nu \simeq 0.665$, Therefore, the exponents for $%
\chi _{F}(\lambda ,N)/N^{d^{\pm }}$ around the critical point, as an
intensive quantity,%
\begin{equation}
\alpha ^{\pm }=\left\{
\begin{array}{cc}
{\ 1/2} & \lambda <1 \\
2 & \lambda >1%
\end{array}%
\right. ..
\end{equation}%
Therefore, the Lipkin-Meshkov-Glick model shows distinct critical exponents
around the critical point. Fortunately, the Lipkin-Meshkov-Glick model is
also an exactly solvable model. The exact results then help us the check the
numerical results obtained by scaling analysis.

\begin{table*}[tbp]
\caption{Critical exponents $d_a^c, \protect\nu, d^\pm, \protect\alpha^\pm$
in various quantum phase transitions. The scaling relation $\protect\alpha%
^\pm = (d_a^c-d^\pm)/\protect\nu$ so far can be tested explicitly in the
first five models. For the Ising model, the Kitaev toric model, and Harper
model, we have $d^+_a=d^-_a$, hence $\protect\alpha^+=\protect\alpha^-$;
while for both the Lipkin-Meshkov-Glick model and the Kitaev honeycomb
model, $d_a^\pm$ and $\protect\alpha^\pm$ are different. }
\label{tab:critcalexp}
\begin{center}
\begin{ruledtabular}
\begin{tabular}{ccccccc}
Model & $d_a^c$ & $\nu$ & $d^+_a$ & $\alpha^+$ & $d^-_a$ & $\alpha^-$ \\
\midrule[0.4pt] \hline 1D Ising model($h_c=1$)\cite{Zanardi06} & 2 & 1 & 1 & 1
& 1 & 1
\\  Deformed Kitaev toric model [$\lambda_{c}
=\frac{1}{2}{\rm ln}(\sqrt{2}+1)$]\cite{DFAbasto08} & ln & 1 & 1 & ln & 1 & ln
\\  Extended Harper model [$\lambda=2\mu$]\cite{LGong115114} & 5(or 2) & 2.5(or 1) & 0 & 2 & 0 &
2
\\ \hline Lipkin-Meshkov-Glick model($h_c=1$)\cite{HMKwok07} & 4/3 & 2/3 & 0 &
2
& 1 & 1/2 \\
 Kitaev honeycomb model($J_{z,c} = 1/2$)\cite{SYang08,SJGu08073491} &
2.50 & 1 & 2 & 1
& 2+ln & 1/2-ln \\
\hline 1D AHM ($t_c = 0.456$ for $n=2/3$ )\cite{SJGu072}  &5.3 & 2.65 & - &-
& 1 & 1.6 \\
 Luttinger model($\lambda_c= 1$ of XXZ model)\cite{MFYang07}  &-  & - & -
&-
& - & 1 \\
 Luttinger model($\lambda_c= -1$ of XXZ model)\cite{MFYang07} & - & - & -
& 1
& - & - \\
\end{tabular}
\end{ruledtabular}
\end{center}
\end{table*}

In the region $\lambda >1$, the Lipkin-Meshkov-Glick model can be
diagonalized using the Holstein-Primakoff transformation \cite{Holstein1940}
\begin{eqnarray}
S_{z} &=&S-a^{\dag }a, \\
S_{+} &=&\left( 2S-a^{\dag }a\right) ^{1/2}a,
\end{eqnarray}%
where $a(a^{\dag })$ is bosonic annihilation(creation) operator satisfying $%
[a,a^{\dag }]=1$. In the large $S(N)$ limit, the Hamiltonian can be
transformed into,%
\begin{equation}
H=-\lambda N+\left[ 2\lambda -\gamma -1\right] a^{\dag }a-\frac{1-\gamma }{2}%
\left( a^{\dag 2}+a^{2}\right) .
\end{equation}%
Obviously, the Hamiltonian is quadratic, and can be diagonalized via the
standard Bogoliubov transformation, i.e.%
\begin{eqnarray}
a^{\dagger } &=&\cosh (\theta )b^{\dagger }+\sinh (\theta )b,  \notag \\
a &=&\sinh (\theta )b^{\dagger }+\cosh (\theta )b.
\end{eqnarray}%
Here $b(b^{\dagger })$ is bosonic annihilation (creation) operator, and the
hyperbolic functions are set to ensure the bosonic commutation relation $%
[b,b^{\dag }]=1$. At the condition%
\begin{equation}
\tanh [2\theta (h\geq 1)]=\frac{1-\gamma }{2\lambda -\gamma -1},
\end{equation}%
the Hamiltonian becomes a diagonal form%
\begin{equation}
H=-\lambda (N+1)+2\sqrt{(\lambda -1)(\lambda -\gamma )}\left( b^{\dagger }b+%
\frac{1}{2}\right) .
\end{equation}%
Therefore, the set of eigenstate of the Hamiltonian is simply denoted as $%
\{\left\vert n\right\rangle \}$ with
\begin{equation}
E_{n}=-\lambda (N+1)+2\sqrt{(\lambda -1)(\lambda -\gamma )}(n+1/2),
\end{equation}%
where $n$ is the number of quasiparticles. The driving term in the
Hamiltonian becomes%
\begin{eqnarray}
-\sum\limits_{i}{\sigma _{z}^{i}} &=&-2S_{z}=2a^{\dag }a-2S,  \notag \\
&=&-2S+2[\cosh (\theta )b^{\dagger }+\sinh (\theta )b]  \notag \\
&&\times \lbrack \sinh (\theta )b^{\dagger }+\cosh (\theta )b],
\end{eqnarray}%
in which only $\sinh (2\theta )b^{\dagger 2}$ is the relevant term acting on
the ground state and\ on projecting excited state. The fidelity
susceptibility can be calculated, to the leading order, as
\begin{equation}
\chi _{{F}}\left( {\gamma ,}\lambda \right) =\frac{(1-\gamma )^{2}}{%
32(1-\lambda )^{2}(\lambda -\gamma )^{2}}.
\end{equation}%
So the critical $\alpha ^{+}=2$. In the region $0<\lambda <1$, the
Lipkin-Meshkov-Glick model can be diagonalized similarly. The fidelity
susceptibility becomes, to the leading order,
\begin{equation}
\chi _{{F}}\left( {\gamma ,}\lambda \right) ={\frac{N}{4\sqrt{(1-\lambda
^{2})(1-\gamma )}}}.
\end{equation}%
The critical exponent $\alpha ^{-}=1/2$. Therefore, the critical exponents
are different on both sides of the critical point. The exact results are the
same as those obtained from the numerical data.

As a brief summary of the scaling and universality of the fidelity
susceptibility, we collect the critical exponents of the fidelity
susceptibility in various quantum phase transitions, and show them in Table. %
\ref{tab:critcalexp}. From the table, we can see that quantum phase
transtions can be classified into two distinct types, depending on whether
the adiabatic dimension changes or not during the transition. For the former
case, the change in the adiabatic dimension plays naturally a role of order
parameter.

\subsection{Higher order of the fidelity}

The fidelity susceptibility denotes only the leading term of the fidelity.
In case that the fidelity susceptibility is invalid, it might be useful to
look into the higher oder term in the fidelity. In this subsection, we
present some basic formulism of the fidely expansion.

The overlap between two wavefunction $|\Psi _{0}(\lambda )\rangle $ and $%
|\Psi _{0}(\lambda +\delta \lambda )\rangle $ can be expanded to an
arbitrary order, i.e.
\begin{equation*}
f(\lambda ,\lambda +\delta \lambda )=1+\sum_{n=1}^{\infty }\frac{(\delta
\lambda )^{n}}{n!}\left\langle \Psi _{0}(\lambda )\left\vert \frac{\partial
^{n}}{\partial \lambda ^{n}}\Psi _{0}(\lambda )\right. \right\rangle .
\end{equation*}%
Therefore, the fidelity becomes
\begin{eqnarray}
&&F^{2}=1+\sum_{n=1}^{\infty }\frac{(\delta \lambda )^{n}}{n!}\left\langle
\Psi _{0}\left\vert \frac{\partial ^{n}}{\partial \lambda ^{n}}\Psi
_{0}\right. \right\rangle   \notag \\
&&+\sum_{n=1}^{\infty }\frac{(\delta \lambda )^{n}}{n!}\left\langle \left.
\frac{\partial ^{n}}{\partial \lambda ^{n}}\Psi _{0}\right\vert \Psi
_{0}\right\rangle  \\
&&+\sum_{m,n=1}^{\infty }\frac{(\delta \lambda )^{m+n}}{m!n!}\left\langle
\Psi _{0}\left\vert \frac{\partial ^{n}}{\partial \lambda ^{n}}\Psi
_{0}\right. \right\rangle \left\langle \left. \frac{\partial ^{m}}{\partial
\lambda ^{m}}\Psi _{0}\right\vert \Psi _{0}\right\rangle   \notag
\end{eqnarray}%
Using the relation for a given $n$%
\begin{equation}
\sum_{m=0}^{n}\frac{n!}{m!(n-m)!}\left\langle \left. \frac{\partial ^{m}}{%
\partial \lambda ^{m}}\Psi _{0}\right\vert \frac{\partial ^{n-m}}{\partial
\lambda ^{n-m}}\Psi _{0}\right\rangle =0
\end{equation}%
we can find that%
\begin{equation}
F^{2}=1-\sum_{l=1}^{\infty }(\delta \lambda )^{l}\chi _{F}^{(l)}
\end{equation}%
where%
\begin{equation}
\chi _{F}^{(l)}=\sum_{l=m+n}\frac{1}{m!n!}\left\langle \left. \frac{\partial
^{m}}{\partial \lambda ^{m}}\Psi _{0}\right\vert P\left\vert \frac{\partial
^{n}}{\partial \lambda ^{n}}\Psi _{0}\right. \right\rangle ,
\label{eq:higherorderdiff}
\end{equation}%
where $P$ is defined in Eq. (\ref{eq:gsprojectorout}). It is easy to check
that $\chi _{F}^{(1)}$ is zero and $\chi _{F}^{(2)}$ the fidelity
susceptibility we discussed in previous subsections.

On the other hand, according the perturbation theory, the ground-state
wavefunction, up to the second order, is%
\begin{eqnarray}
|\Psi _{0}(\delta \lambda )\rangle &=&|\Psi _{0}\rangle +\delta \lambda
\sum_{n\neq 0}\frac{H_{I}^{n0}|\Psi _{n}\rangle }{E_{0}-E_{n}}  \notag \\
&&+\left( \delta \lambda \right) ^{2}\sum_{m,n\neq 0}\frac{%
H_{I}^{nm}H_{I}^{m0}|\Psi _{n}\rangle }{(E_{0}-E_{m})(E_{0}-E_{n})}  \notag
\\
&&-\left( \delta \lambda \right) ^{2}\sum_{n\neq 0}\frac{H^{00}H_{I}^{n0}|%
\Psi _{n}\rangle }{(E_{0}-E_{n})^{2}}  \notag \\
&&-\frac{\left( \delta \lambda \right) ^{2}}{2}\sum_{n\neq 0}\frac{%
H^{0n}H_{I}^{n0}|\Psi _{n}\rangle }{(E_{0}-E_{n})^{2}}.
\end{eqnarray}%
Therefore,

\begin{eqnarray}
\chi _{F}^{(3)} &=&\sum_{n\neq 0}\frac{(H_{I}^{00}-\frac{1}{2}%
H_{I}^{0n})H_{I}^{0n}H_{I}^{n0}}{(E_{0}-E_{n})^{3}}  \notag \\
&&-\sum_{n\neq 0}\sum_{m\neq 0}\frac{H_{I}^{0n}H_{I}^{nm}H_{I}^{m0}}{%
(E_{0}-E_{m})(E_{0}-E_{n})^{2}}.  \label{eq:higheroderperturb}
\end{eqnarray}

Eqs. (\ref{eq:higherorderdiff}) and (\ref{eq:higheroderperturb}) conclude
the main formulism of the higher order expansion of the fidelity. Up to now,
the physical meaning of the high order term in the fidelity is still not
clear.

\section{Fidelity per site, mixed state fidelity, and related}

\label{sec:otherfidelity}

Mathematically, the fidelity can be roughly classfied into two types, the
pure-state fidelity and the mixed-state fidelity. In this section, we
introduce various fidelity measure under different physical conditions
rather than mathematical view. We will see below, various fidelity
expressions can be traced back to the original fidelity definition by
Uhlmann.

\subsection{Fidelity per site}

In quantum many-body systems, the fidelity between two ground states
increases as the system size increases. It usually scales like $\mathcal{F}%
^{L^{d}}$ in the large $L$ limit, where $N=L^{d}$ is the system size and $%
\mathcal{F}$ is called the scaling parameter. Therefore, Zhou, Zhao, and Li
\cite{HQZhou07042940} proposed the scaling parameter (also called the
fidelity per site) might be a good quantity to describe quantum phase
transitions. Precisely, the fidelity per site is defined as%
\begin{equation}
\mathcal{F}(\lambda ,\lambda ^{\prime })=\lim_{N\longrightarrow \infty
}F^{1/N}(\lambda ,\lambda ^{\prime }).
\end{equation}%
The expression can also be written in terms of logarithmic fidelity%
\begin{equation}
\ln \mathcal{F}(\lambda ,\lambda ^{\prime })=\lim_{N\longrightarrow \infty }%
\frac{1}{N}\ln F(\lambda ,\lambda ^{\prime }).  \label{eq:deffidelitypersite}
\end{equation}

Clearly, the fidelity per site has following properties:

1) It is symmetric under interchange $\lambda \longleftrightarrow \lambda
^{\prime }$,

2) $\mathcal{F}(\lambda ,\lambda )=1$,

3) $0\leq \mathcal{F}(\lambda ,\lambda ^{\prime })\leq 1$.

Taking the one-dimensional transverse-field Ising model as an example, the
fidelity between two ground states at $h$ and $h^{\prime }$ is%
\begin{equation}
F(h,h^{\prime })=|\langle \Psi _{0}(h^{\prime })|\Psi _{0}(h)\rangle
|=\prod_{k>0}\cos (\theta _{k}-\theta _{k}^{\prime }),
\end{equation}%
then the logarithmic fidelity becomes%
\begin{eqnarray}
\ln \mathcal{F}(h,h^{\prime }) &=&\lim_{N\longrightarrow \infty }\frac{1}{N}%
\sum_{k>0}\cos (\theta _{k}-\theta _{k}^{\prime }), \\
&=&\frac{1}{2\pi }\int_{0}^{\pi }\cos (\theta _{k}-\theta _{k}^{\prime })dk.
\end{eqnarray}%
Therefore, one can easily discuss the critical behavior of the fidelity per
site or the logarithmic fidelity in quantum phase transitions.

The fidelity per site can be mapped onto the partition function of a
classical statistical vertex lattice model with the same lattice geometry
and dimension \cite{Zhou_GVidal}. The mapping is due to the recent
remarkable finding that any state of a quantum lattice system may be
represented in terms of a tensor network \cite%
{GVidalPRL147902,GVidalPRL040502,VMurgPRA033605}, such as a matrix product
state for one-dimensional systems or a projected entangled-pair state for
systems in $D\geq 2$ dimensions. Then the fidelity between any two groud
states can be calculated in the context of the tensor network algorithms
\cite{GVidalPRL070201}. Clearly, such an approach can be generalized to the
fidelity susceptibility.

On the other hand, in case that the adiabatic dimension and system's
dimension are not the same $d_{a}\neq d$ in a certain phase, the logarithmic
fidelity should be redefined \cite{CYLeungpreprint}, in principle,%
\begin{equation}
\ln \mathcal{F}(\lambda ,\lambda ^{\prime })=\lim_{L\longrightarrow \infty }%
\frac{1}{L^{d_{a}}}\ln F(\lambda ,\lambda ^{\prime })
\end{equation}%
which keeps intensive inside the phase. As a simple example, in the
Lipkin-Meshkov-Glick model, if $\lambda ,\lambda ^{\prime }<1$, then $\ln
F(\lambda ,\lambda ^{\prime })\propto N$; while if $\lambda ,\lambda
^{\prime }>1$, $\ln F(\lambda ,\lambda ^{\prime })$ itself is intensive \cite%
{CYLeungpreprint}.

\subsection{Reduced fidelity}

The reduced fidelity is defined as the fidelity between two reduced-density
matraces corresponding to the states of a part of the system separated by a
small distance in the parameter space. Precisely, if we divide the system
into two parts: A and B, the reduced state of part A can be obtained by
tracing out the degree of freedom of part B, i.e.%
\begin{equation}
\rho _{A}(h)=\text{tr}_{B}\left( |\Psi _{0}(\lambda )\rangle \langle \Psi
_{0}(\lambda )|\right) .
\end{equation}%
Then the fidelity between two reduced states $\rho _{A}(\lambda )$ \ and $%
\rho _{A}(\lambda ^{\prime })$ at $\lambda $ and $\lambda ^{\prime }$ is
simply the mixed-state fidelity
\begin{equation}
F(\lambda ,\lambda ^{\prime })=\text{tr}\sqrt{[\rho _{A}(\lambda
)]^{1/2}\rho _{A}(\lambda ^{\prime })[\rho _{A}(\lambda )]^{1/2}}.
\end{equation}%
In case that the ground state is $N$-fold degenerate, i.e. $|\Psi
_{0}^{i}(h)\rangle ,i=1,..N$, one can either define the thermal ground state
as%
\begin{equation}
\rho (h)=\frac{1}{N}\sum_{i}|\Psi _{0}^{i}(h)\rangle \langle \Psi
_{0}^{i}(h)|,
\end{equation}%
or choose either of them as a physical ground state. The reduced-density
matrix can be obtained in a similar way.

The reduced fidelity approach to quantum phase transitions was firstly
discussed by Zhou \cite{HQZhou07042945} in order to study the relation
between the fidelity per site and the renormalization group flows. The
reduced fidelity was used later to study the quantum phase transitions in a
superconducting lattice with a magnetic impurity inserted at its center \cite%
{NPaunkovic07}. Nevertheless, for a pedagogical purpose, here we introduce
its application to another class of quantum phase transitions induced by a
sequence of ground-state level-crossing, such as the magnetization process.
In this case, the fidelity between two \textquotedblleft
global\textquotedblright\ ground states is not suitable because it drops to
zero at each level-crossing point. For these systems, the Hamiltonian take
typically the form%
\begin{equation}
H(h)=H_{0}-hM,  \label{eq:HamiltonianMag}
\end{equation}%
where $h$ is the external field and $M$ is the magnetization of the ground
state. Unlike the Hamiltonian of Eq. (\ref{eq:generalHamiltonian}) \ where $%
H_{0}$ and $H_{I}$ usually do not commute with each other, $M$ in Eq. (\ref%
{eq:HamiltonianMag}) commutes with $H_{0}$. Then $H_{0}$ and $M$ have the
same eigenvectors. Suppose the magnetization of the system is zero in the
absence of the external field, the system will be magnetized step by step as
the external field increases until it is fully magnetized at a certain
transition point $h_{c}$. Typically, the ground state has two phases, i.e.,
a partially magnetized phase and a fully magnetized phase. For the latter,
the ground state does not change as the external field changes. Then the
fidelity is always one and the fidelity susceptibility is zero. While for
the former, the ground state undergoes infinite level-crossings and the
fidelity drops to zero at each crossing point. Therefore, it is not
convenient to characterize this type of phase transition in terms of global
state fidelity. The difficulty can be overcome if one consider only the
reduced state of a local part of the system, which is described by a
reduced-density matrix. Mathematically, the \textquotedblleft
orthogonality\textquotedblright\ of two pure states is very sensitive to the
distinguishability between the states, such as good quantum numbers or
which-way flag \cite{Whichwayexp}. However, if one consider only the local
part of the system, the rate of \textquotedblleft
orthogonality\textquotedblright\ is reduced because the reduced state is a
mixed state that usually is free of the conserved quantity.

The Lipkin-Meshkov-Glick model still provides a very good example \cite%
{HMKwok08,JMa08}. If $\gamma =1$, the Hamiltonian of Eq. (\ref%
{eq:HamiltonianLMGmodel}) commutes with the $z$-component of the total
spins. Then $S^{z}$ is a conserved quantity. The ground state can be written
in $\left\{ {\left\vert {S,S}^{z}\right\rangle }\right\} $. For $S=N/2$, the
eigenenergies are
\begin{equation}
E\left( {M,h}\right) =\frac{2}{N}\left( {M-\frac{hN}{2}}\right) ^{2}-\frac{N%
}{2}\left( {1+h^{2}}\right) ,
\end{equation}%
and the ground state (with quantum number $M_{0}$) is then obtained
\begin{equation}
M_{0}=\left\{
\begin{array}{l}
\frac{N}{2},h\geq 1 \\
I\left( \frac{hN}{2}\right) ,0\leq h<1%
\end{array}%
\right. ,
\end{equation}%
where $I(x)$ gives the integer part of $x$. We can define the reduced state
of a single spin as%
\begin{equation}
\rho =\frac{1}{2}\left(
\begin{array}{cc}
1+\langle \sigma ^{z}\rangle & 0 \\
0 & 1-\langle \sigma ^{z}\rangle%
\end{array}%
\right) .
\end{equation}%
The fidelity between two reduced states around the each crossing point is
\begin{eqnarray}
F &=&\frac{1}{2}\sqrt{\left( 1+\langle \sigma _{z}\rangle ^{j}\right) \left(
1+\langle \sigma _{z}\rangle ^{j+1}\right) }  \notag \\
&&+\frac{1}{2}\sqrt{\left( 1-\langle \sigma _{z}\rangle ^{j}\right) \left(
1-\langle \sigma _{z}\rangle ^{j+1}\right) },
\end{eqnarray}%
where $\langle \sigma _{z}\rangle ^{j}$ denotes the expectation value of $%
\sigma _{z}$ of $j$th state during the level-crsossing process. The fidelity
susceptibility can be defined in the similar way, in the thermodynamic
limit,
\begin{equation}
\chi _{_{F}}=\lim_{\delta h\rightarrow 0}\frac{-2\text{ln}F}{(\delta h)^{2}},
\end{equation}%
where $\delta h=h_{j+1}-h_{j}$ is the difference of $h$ between two
level-crossing points.

\subsection{Fidelity between thermal states}

At finite temperatures, an equilibrium state of a thermal system is
described by a mixed state \cite{KHuangbook},
\begin{equation}
\rho (\beta )=\frac{1}{Z}\sum_{n}e^{-\beta E_{n}}|\Psi _{n}\rangle \langle
\,\Psi _{n}|,
\end{equation}%
where $\beta =1/T$ is the inverse temperature with Boltzmann constant $%
k_{B}=1$, $Z$ is the partition function defined as
\begin{equation}
Z=\sum_{n}e^{-\beta E_{n}},
\end{equation}%
and $|\Psi _{n}\rangle $ is the eigenstate of the system's Hamiltonian with
eigenvalue $E_{n}$. Therefore, if we choose the temperature as a driving
parameter, the fidelity between two thermal states at $\beta -\delta \beta
/2 $ and $\beta +\delta \beta /2$ is \cite{PZanardi032109}
\begin{eqnarray}
&&F(\beta -\delta \beta /2,\beta +\delta \beta /2)  \notag \\
&=&\text{tr}\sqrt{\rho (\beta -\delta \beta /2)\rho (\beta +\delta \beta /2)}
\\
&=&\frac{Z(\beta )}{\sqrt{Z(\beta -\delta \beta /2)Z(\beta +\delta \beta /2)}%
}.  \label{eq:thermalstatefidelitypf}
\end{eqnarray}%
This result is very interesting. It establishes a relation between the state
overlap and well known thermal quantities. Therefore, one can understand the
state evolution at finite temperature from the knowledge of thermodynamics.
Since the Helmholtz free energy is
\begin{equation}
G=\langle E\rangle -TS=-\frac{1}{\beta }\ln Z,
\end{equation}%
the thermal fidelity susceptibility then becomes \cite%
{WLYou07,PZanardi0701061}
\begin{eqnarray}
\chi _{F} &=&\left. \frac{-2\ln F}{(\delta \beta )^{2}}\right\vert _{\delta
\beta \rightarrow 0}, \\
&=&\beta \left. \frac{\left[ 2G(\beta )-G(\beta +\delta \beta /2)-G(\beta
-\delta \beta /2)\right] }{(\delta \beta )^{2}}\right\vert _{\delta \beta
\rightarrow 0}, \\
&=&\frac{C_{v}}{4\beta ^{2}}.
\end{eqnarray}%
To obtain the above result, we have used the following standard relations in
the thermodynamics \cite{KHuangbook}%
\begin{equation}
S=-\frac{\partial G}{\partial T},C_{v}=-T\frac{\partial S}{\partial T}.
\end{equation}%
Similarly, if the driving term in the Hamiltonian is a Zeemann-like term,
say $-hM$, which is crucial in Landau-Ginzburg-Wilson symmetry-breaking
theory, then the fidelity susceptibility is simply the magnetic
susceptibility $\chi $\cite{WLYou07},
\begin{equation}
\chi _{F}=\left. \frac{-2\ln F_{i}}{\delta h^{2}}\right\vert _{\delta
h\rightarrow 0}=\frac{\beta \chi }{4}.
\end{equation}%
The thermal fidelity susceptibility is similar to the ground-state fidelity
susceptibility. Both of them are a kind of structure factor because of
\begin{eqnarray}
C_{v} &=&\beta ^{2}(\langle E^{2}\rangle -\langle E\rangle ^{2}), \\
\chi &=&\beta (\langle M^{2}\rangle -\langle M\rangle ^{2})
\end{eqnarray}%
in thermodynamics. The main difference is that the ground-state fidelity
susceptibility involves the dynamic behavior of the system.

\begin{figure*}[tbp]
\includegraphics[width=16cm]{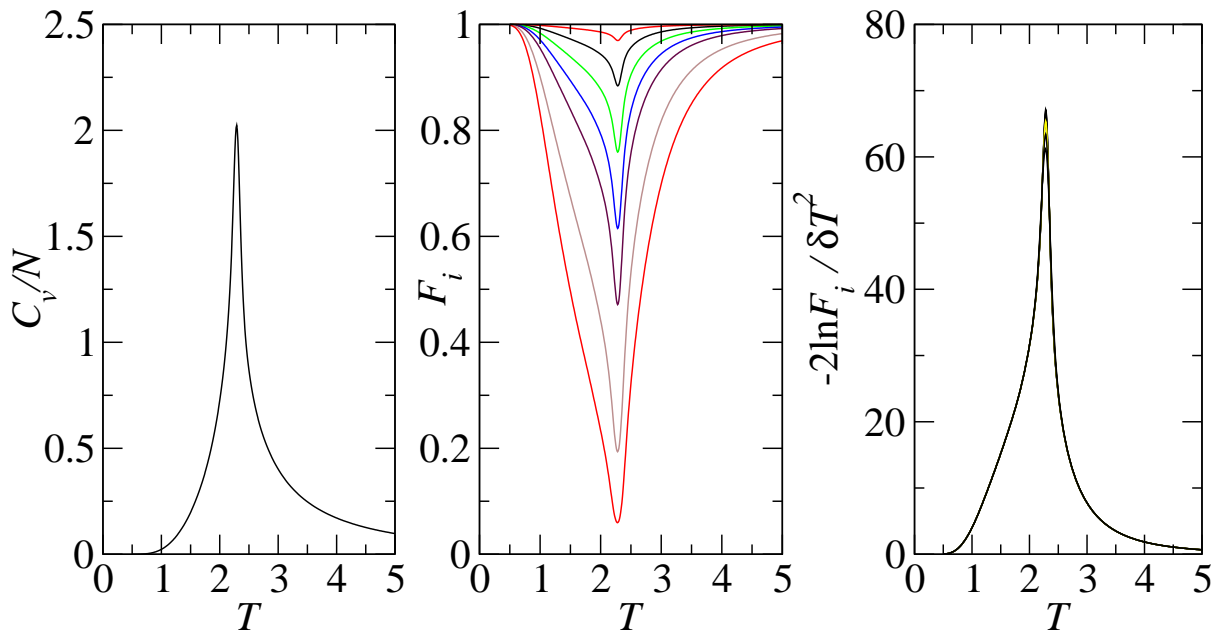}
\caption{(Color online) LEFT: The specific heat as a function of the
temperature $T$ for $40\times 40$ Ising model. MIDDLE: The fidelity between
two states separated by different temperature interval $\protect\delta %
T=0.02, 0.04, 0.06. 0.08, 0.1, 0.15, 0.20$ for lines from up to bottom.
RIGHT: the fidelity susceptibility $\protect\chi_F$ as a function of $T$,
obtained from the data of the middle picture. All lines in the middle
picture collapse onto a single line (From Ref. \protect\cite{WLYou07}). }
\label{fig:fsus_ising}
\end{figure*}

As a simple application, we take the two-dimensional Ising model on a square
lattice as an example. The Hamiltonian reads,
\begin{equation}
H=-\sum_{\mathbf{\langle ij\rangle }}\sigma _{\mathbf{i}}^{z}\sigma _{%
\mathbf{j}}^{z},
\end{equation}%
where the sum is over all pairs of nearest-neighbor sites $\mathbf{i}$ and $%
\mathbf{j}$, and the coupling is set to unit for simplicity. The model is
certainly the most thoroughly researched model in statistical physics \cite%
{LOnsager44,BMMccoyb}. The results for a $40\times 40$-site system are shown
in Fig. \ref{fig:fsus_ising}. Clearly, there is a maximum point in the line
of the specific heat, whose scaling behavior to an infinite system defines
the critical point. Meanwhile, the middle picture in Fig. \ref%
{fig:fsus_ising} shows various fidelity calculated from different
temperature interval. This obvious difference in the fidelity disappears if
we distill the fidelity susceptibility from them, as shown in the right
picture of Fig. \ref{fig:fsus_ising}.

\subsection{Operator fidelity}

The operator fidelity was proposed by Wang, Sun and Wang \cite{XWang08032940}%
. For two arbitrary linear operators $A$, $B$ defined in a $d$-dimensional
Hilbert space, the expectation value of their product are tr$(AB)$, which is
a kind of inner product in the $d^{2}$-dimensional (or $d^{2}-1$ for
Hermitian operators) space. On the other hand, in a $d$ dimensional Hilbert
space, any state acted after a linear operator becomes another state in the
Hilbert space. Then the fidelity between states can be generalized to the
operator level. Specifically, for two unitary evolution operators $U_{0}$
and $U_{1}$, the fidelity between them can be calculated as
\begin{equation}
F^{2}=\frac{1}{d^{2}}|\text{tr}(U_{0}^{\dagger }U_{1})|.
\end{equation}%
Precisely, for a general Hamiltonian, the evolution operator can be
expressed as
\begin{eqnarray}
U(t) &=&1-i\delta \lambda \int_{0}^{t}dt_{1}V(t_{1})  \notag \\
&&-(\delta \lambda
)^{2}\int_{0}^{t}dt_{1}\int_{0}^{t_{1}}dt_{2}V(t_{1})V(t_{2})+\cdots ,
\end{eqnarray}%
where $V(t)=\exp (iH_{0}t)V\exp (-iH_{0}t)$ denotes the perturbation term in
the interaction picture. The trace of the evolution operator is,
\begin{eqnarray}
\overline{\text{tr}}[U(t)] &=&1-i\delta \lambda \overline{\text{tr}}\left[
W(t)\right] -  \label{eq:operatorfided} \\
&&-\frac{(\delta \lambda )^{2}}{2}\overline{\text{tr}}\left[ W(t)^{2}\right]
+\cdots ,
\end{eqnarray}%
where%
\begin{equation}
\overline{\text{tr}}(A)=\frac{1}{d}\text{tr}(A),
\end{equation}%
and%
\begin{equation}
W(t)=\int_{0}^{t}dt_{1}V(t_{1}).
\end{equation}%
The left hand side of Eq. (\ref{eq:operatorfided}) denotes the inner product
of two states at different time. Then the fidelity becomes
\begin{equation}
F^{2}=1-(\delta \lambda )^{2}\left[ \overline{\text{tr}}\left[ W(t)^{2}%
\right] -(\overline{\text{tr}}W(t))^{2}\right] +\cdots .
\end{equation}%
Similarly, the operator fidelity susceptibility can be extracted
\begin{equation}
\chi _{F}=\overline{\text{tr}}\left[ W(t)^{2}\right] -\left[ \overline{\text{%
tr}}W(t)\right] ^{2}.
\end{equation}

The operator fidelity is remarkable. In its approach to quantum phase
transitions, it is a good indicator regardless of the ground-state
degeneracy. Moreover, it also reveals that in the state evolution, the
driving mechanism is due to the fluctuation of $W(t)$.

\subsection{Density-functional fidelity}

The density functional theory (DFT) \cite{Hohenberg,KohnSham} is based on
the Hohenberg-Kohn theorem \cite{Hohenberg}, which asserts that the
ground-state properties are uniquely determined by the density distribution $%
n_{x}$ that minimizes the functional for the ground-state energy $%
E_{0}[n_{x}]$. To date, the DFT becomes the most successful method for
first-principle calculations of the electronic properties of solids. Since
the normalized density distribution captures the most relevant information
about the ground state, Gu \cite{SJGUDFF} tried to link the fidelity and
quantum phase transitions via the DFT.

For a general Hamiltonian system
\begin{equation}
H(\lambda )=H_{0}+\lambda H_{I}+\sum_{x}\mu _{x}n_{x},
\label{eq:Hamiltonian_DFT}
\end{equation}%
where $H_{I}$ is the interaction term and $\lambda $ denotes strength, and $%
\mu _{x}$ is the local (pseudo)potential associated with density
distribution $\{n_{x}\}$. The index $x$ can be discrete or continuous
depending on the system under study. Though in the
local-density-approximation (LDA) calculation, $n_{x}$ usually refers to the
density of electrons in real space, it can also be generalized to\ the
population in configuration space of a reduced-density matrix or the density
of state in energy(momentum) space. The density distribution can be obtained
by the Hellmann-Feymann theorem%
\begin{equation}
n_{x}=\left\langle \Psi _{0}(\lambda )\left\vert \frac{\partial H}{\partial
\mu _{x}}\right\vert \Psi _{0}(\lambda )\right\rangle ,
\end{equation}%
where $|\Psi _{0}(\lambda )\rangle $ is the ground state, and $%
\sum_{x}n_{x}=1$. The density-functional fidelity is defined as the distance
between two density distributions at $\lambda $ and $\lambda ^{\prime }$ in
the parameter space
\begin{equation}
F(\lambda ,\lambda \prime )=\text{tr}\sqrt{n(\lambda )n(\lambda ^{\prime })}.
\end{equation}%
Since the density distribution is experimentally measureable, the
density-functional fidelity then provides a practicable approach for
experimentalist to study quantum phase transitions in perspective of
information theory.

Expanding the density-functional fidelity to the leading order, one can find
the density-functional fideltiy susceptibility has the form%
\begin{equation}
\chi _{F}=\sum_{x}\frac{1}{4n_{x}}\left( \frac{\partial n_{x}}{\partial
\lambda }\right) ^{2}.  \label{eq:dfffs}
\end{equation}%
Therefore, if one regards $\partial n_{x}/\partial \lambda $ as an
independent function besides the density distribution $n_{x}$, \ the
density-functional fidelity susceptibility is a functional of $n_{x}$ and $%
\partial n_{x}/\partial \lambda $, both of which, in principle, maximize the
density-functional fidelity susceptibility at the critical point.

Mathematically, the density-functional fidelity and thermal-state fidelity
is related to the Fisher-Rao distance. For the probability $\{n_{x}\}$
defined in parameter space $\{\lambda _{a}\}$, the Fisher-Rao distance is
defined as
\begin{equation}
ds^{2}=\frac{1}{4}\sum_{x}\frac{dn_{x}dn_{x}}{n_{x}}.
\end{equation}%
The Fisher-Rao metric then can be written as%
\begin{equation}
g_{ab}=\frac{1}{4}\sum_{x}\frac{1}{n_{x}}\frac{dn_{x}}{d\lambda _{a}}\frac{%
dn_{x}}{d\lambda _{b}}.
\end{equation}


\section{Fidelity in strongly correlation systems}

\label{sec:survey}

The fidelity approach to quantum phase transitions has been applied to many
strongly correlated systems besides the one-dimensional transverse-field
Ising model and the Lipkin-Meshkov-Glick model. This section is devoted to a
survey of fidelity in these systems. We mainly focus on those models which
are well studied and whose phase diagrams are known.

\subsection{Pure-state fidelity}

\subsubsection{One-dimensional spin systems}

\begin{figure}[tbp]
\includegraphics[bb=0 0 320 200, width=8.5 cm, clip]{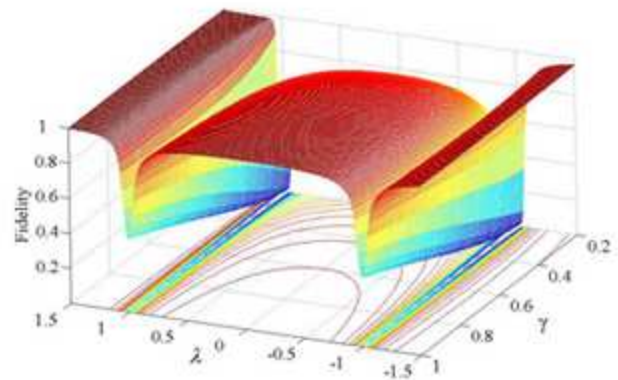}
\caption{(Color online) The fidelity of the one-dimensional transverse-field
XY model as a function of $h$ and $\protect\gamma$, for a system of $N=100$
and $\protect\delta\protect\lambda=\protect\delta\protect\gamma=0.1$. The
colored curves on the $F=0$ plane constitutes a contour map. The deep
grooves in the curved surface of the fidelity separate naturally the whole
region into three phases: one order phase in the middle and two polarized
phases on both sides (Reproduced from Ref. \protect\cite{Zanardi06})}
\label{xy_fidelity.eps}
\end{figure}

\textit{The one-dimensional transverse-field XY model:} The model is an
extended version of the one-dimensional transverse-field Ising model. The XY
model can be exactly solved too \cite{EBarouch71,PJordan28}. Its Hamiltonian
reads
\begin{eqnarray}
H(\gamma ,\lambda ) &=&-\sum_{j=1}^{L}\left( \frac{1+\gamma }{2}\sigma
_{j}^{x}\sigma _{j+1}^{x}+\frac{1-\gamma }{2}\sigma _{j}^{y}\sigma
_{j+1}^{y}+\lambda \sigma _{j}^{z}\right) ,  \notag \\
&&
\end{eqnarray}%
where $\gamma $ defines the anisotropy and $\lambda $ represents the
external magnetic field. Obviously, if $\gamma =1$, the XY model becomes the
transverse-field Ising model which is presented as an example in section \ref%
{sec:fs}. The role of quantum entanglement in the quantum phase transitions
occurred in the XY model has been widely studied \cite%
{AOsterloh2002,TJOsbornee,GVidal03,SQSu2006,HDChen10215}. It was shown that
either the pairwise entanglement \cite{AOsterloh2002,TJOsbornee}, as
measured by the concurrence \cite{SHill97,WKWootters98}, or the two-site
local entanglement \cite{SQSu2006,HDChen10215} shows interesting singular
and scaling behavior around the critical point. The fidelity approach to the
one-dimensional transverse-field XY model was firstly done by Zanardi and
Paunkovi\'{c} \cite{Zanardi06}. They used the XY model as an example to
present a new characterization of quantum phase transitions in terms of the
fidelity. Their physical intuition behind the fidelity approach is in order:
since the fidelity measures the similarity between two states, then the
fidelity between two ground states obtained for two different values of
external parameters should has a minimum around the critical point(See Fig. %
\ref{xy_fidelity.eps}). The work is one of original works in this promising
field.

Meanwhile, Zhou, Zhao, and Li \cite{HQZhou07042940} studied the critical
phenomena in the XY model in terms of the fidelity per site. They found that
the logarithmic function of fidelity per site with respect to the transverse
field is logarithmically divergent at the critical point. The scaling and
universality hypothesis based on the fidelity per site was also confirmed in
the vicinity of the transition point. This work is the first one that
proposed the idea of fidelity per site.

\begin{figure}[tbp]
\includegraphics[bb=0 0 970 330, width=8.5 cm, clip]{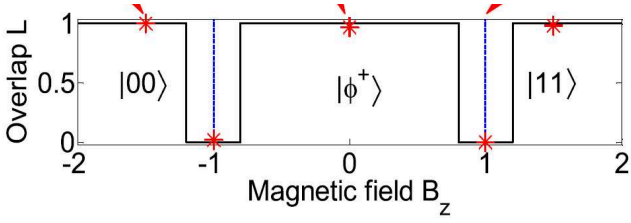}
\caption{(Color online) The theoretical (line) and experimental (asterisks)
fidelity in the quantum phase transition of the Ising dimer induced by the
ground-state level-crossing. (From Ref. \protect\cite{JZhangPRL100501})}
\label{fig:expising1.eps}
\end{figure}

\begin{figure}[tbp]
\includegraphics[bb=0 0 600 360, width=8.5 cm, clip]{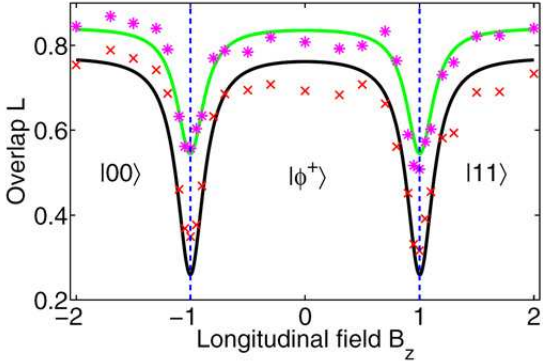}
\caption{(Color online) The experimental fidelity of the Ising dimer
measured by the nuclear-magnetic-resonance experiments for $\protect\delta %
h_x=0.2$ (*) and $\protect\delta h_x=0.3$ ($\times$). (From Ref.
\protect\cite{JZhangPRL100501})}
\label{fig:expising2.eps}
\end{figure}

A noticeable achievement is the experimental detection of the quantum phase
transitions in terms of the fidelity \cite{JZhangPRL100501}.\ Using the
nuclear-magnetic-resonance technique, Zhang \textit{et al} measured the
sensibility of the ground state of the Ising model to perturbations when it
comes to the critical point. Their system can be described by the
Hamiltonian
\begin{equation}
H^{s}=\sigma _{1}^{z}\sigma _{2}^{z}+B_{x}(\sigma _{1}^{x}+\sigma
_{2}^{x})+B_{z}(\sigma _{1}^{z}+\sigma _{2}^{z}).
\end{equation}%
If $B_{x}=0$, the system undergoes a first-order phase transition induced by
the ground-state level-crossing. They found that the ground-state overlap
shows a drop at the crossing point (\ref{fig:expising1.eps}). While if $%
B_{z}=0$, the model becomes the transverse-field Ising model. Though the
second-order quantum phase transition occurs only in the thermodynamic
limit, the fidelity in a small sample can still tell us the significant
change in the structure of the ground state around the critical point (See
Fig. \ref{fig:expising2.eps}).

\textit{The one-dimensional XXZ model:} The Hamiltonian of the
one-dimensional XXZ model reads%
\begin{equation}
H(\lambda )=\sum_{j=1}^{L}\left( \sigma _{j}^{x}\sigma _{j+1}^{x}+\sigma
_{j}^{y}\sigma _{j+1}^{y}+\lambda \sigma _{j}^{z}\sigma _{j+1}^{z}\right) .
\end{equation}%
The XXZ model can be solved by the Bethe-ansatz method \cite%
{HABethe31,CNYang66,MTakahashib}, through which the energy spectra can be
fully determined. The ground-state of the one-dimensional XXZ model consists
three phases. If $\lambda <-1$, all spins are align to the same direction,
the ground-state is in a fully polarized state; when $-1<\lambda <1$, the
quantum fluctuation term dominates, then the ground state is a quantum
fluctuation phase; while if $\lambda >1$, the antiferromagnetic coupling
dominates, the ground state is in an antiferromagnetic state. So there are
two critical points $\lambda =\pm 1$. There are various tools to witness the
quantum phase transition occurred at $\lambda =1$. They include the
vanishing energy gap and divergent correlation length \cite{RJBaxterb} as $%
\lambda \rightarrow 1^{+}$, abrupt change in the spin stiffness \cite%
{BSShastry90,SJGu2002}, and the maximum value of quantum entanglement \cite%
{SJGuXXZ}.

The fidelity approach to the quantum phase transition of the one-dimensional
XXZ model is not easy because the ground-state wave function is not known.
Yang \cite{MFYang07} first used the Luttinger Liquid model to describe the
one-dimensional XXZ model in the quantum fluctuation region \cite%
{TGiamarchi_book}, i.e.
\begin{equation}
H(\lambda )=\frac{u}{2}\int dx\left( K\Pi (x)^{2}+\frac{1}{K}(\partial
_{x}\Phi )^{2}\right) .
\end{equation}%
Here
\begin{eqnarray}
K &=&\frac{\pi }{2}[\pi -\arccos (\lambda )], \\
u &=&\frac{\pi \sqrt{1-\lambda ^{2}}}{2\arccos \lambda },
\end{eqnarray}%
and $\Pi ,\Phi $ are bosonic phase field operators. They obtained the
fidelity as%
\begin{equation}
F(K,K^{\prime })=\prod_{k\neq 0}\frac{2}{\sqrt{K/K^{^{\prime }}}+\sqrt{%
K^{\prime }/K}},
\end{equation}%
and the fidelity susceptibility
\begin{equation}
\frac{\chi _{F}(\lambda )}{L}=\frac{1}{4[\pi -\arccos (\lambda )]^{2}}\frac{1%
}{1-\lambda ^{2}}.
\end{equation}%
Therefore the critical exponent $\alpha =1$. So they conclude that the
fidelity might be able to signal the Beresinskii-Kosterlitz-Thouless phase
transition occurring in the XXZ model. This conclusion is a little
surprising. Because in previous studies, some groups claimed that the
Beresinskii-Kosterlitz-Thouless transitions cannot be signalled from the
singularity of the fidelity susceptibility \cite{WLYou07,SChen08}.

\begin{figure}[tbp]
\includegraphics[bb=0 -10 250 300, width=8.5 cm, clip]{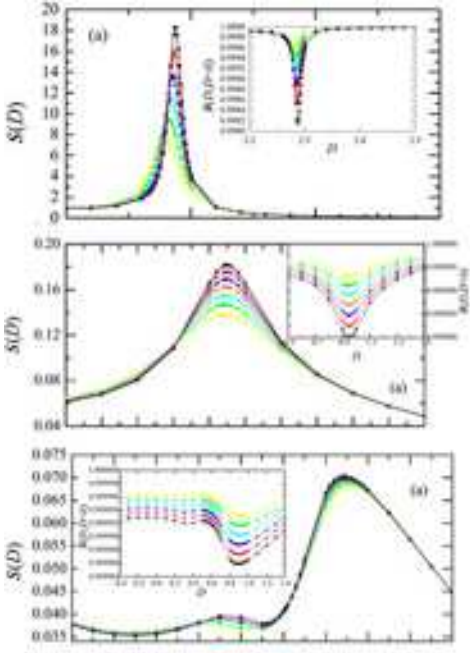}
\caption{(Color online) The fidelity and fidelity susceptibility in the
spin-one anisotropic model (From Ref. \protect\cite{YCTzeng082}).}
\label{fig:spin1_fidelity.eps}
\end{figure}

\textit{The spin-one anisotropic model:}The Hamiltonian of the spin-one
anisotropic model reads%
\begin{equation}
H(\lambda )=\sum_{j=1}^{L}\left[ S_{j}^{x}S_{j+1}^{x}+S_{j}^{y}S_{j+1}^{y}+%
\lambda S_{j}^{z}S_{j+1}^{z}+D(S_{j}^{z})^{2}\right] ,
\end{equation}%
where $S_{j}^{\kappa }(\kappa =x,y,z)$ stands for spin-1 operators, and $%
\lambda $ and D parameterize the Ising-like and the uniaxial interaction.
The ground-state phase diagram of the model consists of six phases\cite%
{HJSchulz86,MdenNijs89}. Specifically, for three cases of $\lambda =2.59,1$,
and 0.5, the system undergoes second-order, third-order, and fifth-order
quantum phase transitions, respectively. For the case of $\lambda =2.59$,
the authors found the fidelity susceptibility as a function of D around $%
D=2.30$. For the second case of $\lambda =1$, though the second-order
derivative of the ground-state energy does not show singular behavior around
$D=0.95$, the fidelity susceptibility is still able to signal the
transition. For the third case of $D=0.5$, however, both the fidelity
susceptibility and the second derivative of the ground-state energy is
continuous in the critical region. The authors then conclude that the
fidelity susceptibility might not be able to characterize the Gaussian
transition. (See Fig. \ref{fig:spin1_fidelity.eps}).

\begin{figure}[tbp]
\includegraphics[bb=0 1 110 90, width=8.5 cm, clip]{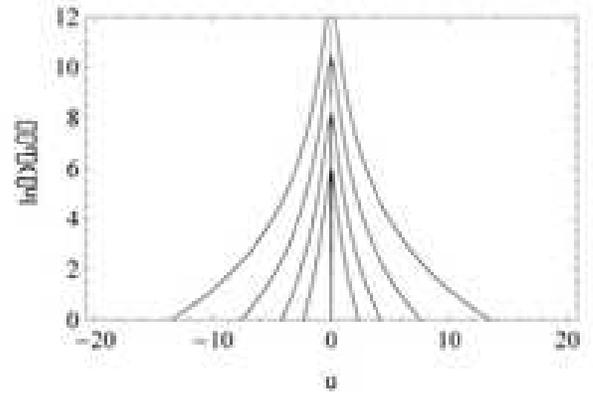}
\caption{(Color online) The fidelity (TOP) as a function of $u$ and $N$ for $%
\protect\delta=0.001$, and the fidelity as a function of $u$ for various
system size ($N=10^3, 10^4, 10^5$, and $10^6$) (From Ref. \protect\cite%
{ATribedi08}).}
\label{fig:spinladder_fidelity.eps}
\end{figure}

\emph{The spin-ladder model:} The Hamiltonian of a general spin ladder model
reads \cite{Kolezhuk98}
\begin{eqnarray}
H &=&\sum_{j=1}[J(S_{1,j}S_{1,j+1}+S_{2,j}S_{2,j+1})+J_{r}S_{1,j}S_{2,j}
\notag \\
&&+V(S_{1,j}S_{1,j+1})(S_{2,j}S_{2,j+1})  \notag \\
&&+J_{d}(S_{1,j}S_{2,j+1}+S_{2,j}S_{1,j+1})  \notag \\
&&+K[(S_{1,j}S_{2,j+1})(S_{2,j}S_{1,j+1})  \notag \\
&&-(S_{1,j}S_{2,j})(S_{1,j+1}S_{2,j+1})],
\end{eqnarray}%
where the indices $1$ and $2$ distinguish the lower and upper legs of the
ladder and $i$ labels the rungs. The ground state of the spin ladder is very
complicated. However, if we redefine the parameters
\begin{eqnarray}
u &=&-u,K=J_{r}=\frac{(u^{2}-1)(u^{2}+3)}{2},J_{d}=0, \\
V &=&\frac{(5u^{4}+2u^{2}+9)}{4},J=\frac{3(u^{4}+10u^{2}+5)}{16}, \\
\epsilon _{1} &=&\frac{(3u^{4}+14u^{2}+15)}{8},\epsilon _{2}=\frac{%
(5u^{4}+18u^{2}+9)}{8},
\end{eqnarray}%
the Hamiltonian becomes a one-parameter model. The ground state of the
system can be explicitly written as a matrix product state%
\begin{equation*}
|\Psi _{0}(u,u^{\prime })\rangle =\frac{1}{\sqrt{N_{c}}}\text{tr}%
[g_{1}(u)g_{2}(u^{\prime })\cdots g_{2N-1}(u)g_{2N}(u^{\prime })],
\end{equation*}%
where $N_{c}$ is the normalization constant. Such a ground state undergoes
two second-order quantum phase transitions at $u=0$ and $u=\infty $. At $u=0$%
, the ground state changes from the dimerized phase to the Haldane phase.
The latter can be described by an effective Hamiltonian of the $S=1$
Affleck-Kennedy-Lieb-Tasaki chain \cite{AKLT}. Fig. \ref%
{fig:spinladder_fidelity.eps} shows the fidelity approach to the phase
transition occurring at $u=0$. Clearly the fidelity shows a minimum and the
fidelity susceptibility shows a sharp peak at $u=0$, corresponding to an
abrupt change in the ground-state wavefunction.

\begin{figure}[tbp]
\includegraphics[width=8cm]{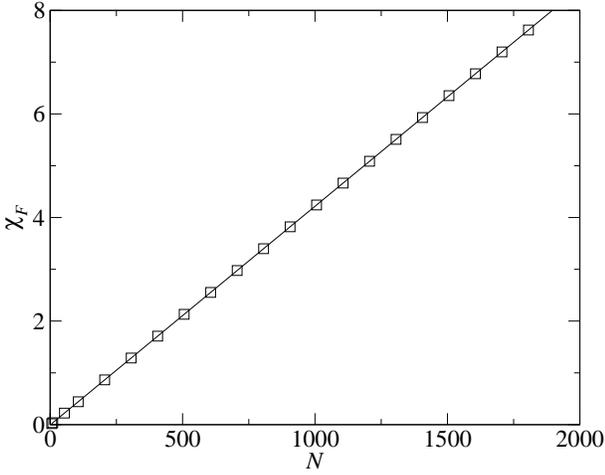}
\caption{(Color online) The dependence of the fidelity susceptibility of the
Hubbard model on the system size at $U=0$ (From Ref. \protect\cite{WLYou07}%
). }
\label{fig:chi_hub.eps}
\end{figure}

\subsubsection{One-dimensional fermionic systems}

\emph{The one-dimensional Hubbard model:} The Hamiltonian of the
one-dimensional Hubbard model \cite{Hubbard} reads
\begin{equation}
H_{\mathrm{HM}}=-\sum_{j=1}^{L}\sum_{\delta =\pm 1}\sum_{\sigma
}tc_{j,\sigma }^{\dagger }c_{j+\delta ,\sigma }+U\sum_{j=1}^{L}n_{j,\uparrow
}n_{j,\downarrow },
\end{equation}%
where $c_{j,\sigma }^{\dagger }$ and $c_{j,\sigma },\sigma =\uparrow
,\downarrow $ are creation and annihilation operators for fermionic atoms
with spin $\sigma $ at site $j$ respectively, $n_{\sigma }=c_{\sigma
}^{\dagger }c_{\sigma }$, and $U$ denotes the strength of on-site
interaction. Besides the obvious SU(2) symmetry in the spin sector, the
model has charge SU(2) symmetry \cite{hubCNYang89}. The global symmetry \cite%
{hubCNYang90} of the model is SO(4) since half of the irreducible
representations of SU(2)$\otimes $SU(2) are excluded. The Hubbard model was
solved by the Bethe-ansatz method \cite{EHLieb68,TDeguchi00,FHLEsslerb}. For
the half-filled case, the system undergoes a Mott-insulator transition at
the critical point $U_{c}=0$. According to the exact solution, the
ground-state energy can be differentiated to an arbitrary order, therefore
the transition is of infinite order \cite{TDeguchi00,FHLEsslerb}%
(Beresinskii-Kosterlitz-Thouless like). The role of quantum entanglement in
the phase transitions occurred in the Hubbard model was addressed by Gu
\emph{et al} \cite{SJGUPRL}, Deng \emph{et al}, \cite{SSDeng06} and Larsson
and Johannesson\cite{Larsson_ent}. The fidelity approach to the
one-dimensional Hubbard was firstly studied by You \emph{et al} \cite%
{WLYou07}. They found that the fidelity susceptibility is not singular at
the transition point and concluded that it might not be able to signal the
transition. Especially at the critical point $U_{c}=0$, the Hamiltonian is
diagonal, and the fidelity susceptibility per system length, as an intensive
quantity, is a constant (See Fig. \ref{fig:chi_hub.eps}). The fidelity in
the Hubbard model was later revisited by Venuti \emph{et al} \cite%
{LCVenuti08012473}, with the strategy of bosonization and Bethe-ansatz
techniques. They showed that the metal-insulator phase transition can be
insightfully analyzed in term of the fidelity. The fidelity susceptibility
shows divergences depending on the path approaching to the critical point.
Bosonization results shows that the fidelity susceptibility may diverge as
\begin{equation*}
\chi _{F}\propto U^{-4},
\end{equation*}%
if the doping rate is proportional to $\sqrt{U}\exp [-2\pi /U]$. The authors
also performed exact diagonalization for a system up to 14 sites. Based on
the scaling analysis, their results indicate that the fidelity might be
super-extensive, hence divergent in the thermodynamic limit. However, they
also stated that the available sizes of the exact diagonalization are too
small to provide a conclusive answer. Therefore, whether the fidelity
susceptibility shows singular behavior in the one-dimensional Hubbard model
is still not conclusively answered.

\begin{figure}[tbp]
\includegraphics[bb=0 1 330 100, width=8.5 cm, clip] {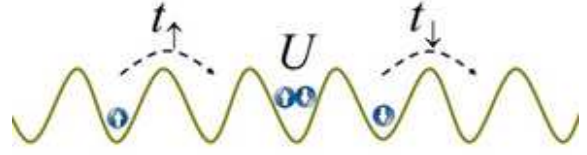}
\caption{ (Color online) A sketch of the asymmetric Hubbard model which can
be realized in an optical lattice. The solid sinusoid denotes the periodic
potential formed by two interfering standing laser waves. Two species of
fermionic atoms are supposed to be trapped in the potential. In case of $%
t_\uparrow=t_\downarrow$, the asymmetric Hubbard model becomes the Hubbard
model.}
\label{figure_ahm}
\end{figure}

\begin{figure}[tbp]
\includegraphics[width=8cm]{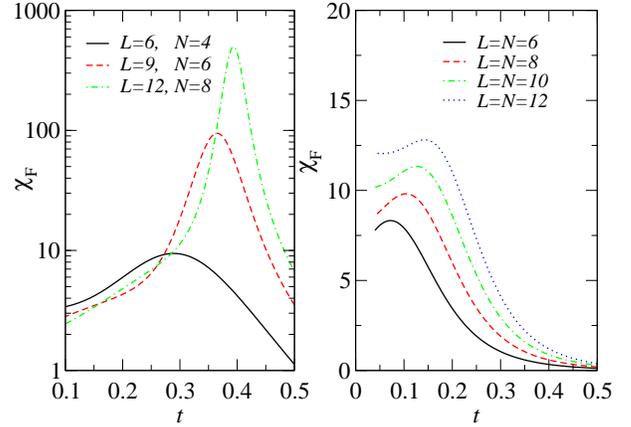}
\caption{(Color online) The scaling behavior of the FS as a function of $t$
for the cases of $n=2/3$ (LEFT) and $n=1$ (RIGHT). Here $U=30$. (From Ref.
\protect\cite{SJGu072})}
\label{fig:fidelity_ahm.eps}
\end{figure}

\emph{The one-dimensional asymmetric Hubbard model:} The Hamiltonian of the
asymmetric Hubbard model \cite{GFath95,GFath96} reads
\begin{equation}
H_{\mathrm{AHM}}=-\sum_{j=1}^{L}\sum_{\delta =\pm 1}\sum_{\sigma }t_{\sigma
}c_{j,\sigma }^{\dagger }c_{j+\delta ,\sigma }+U\sum_{j=1}^{L}n_{j,\uparrow
}n_{j,\downarrow },
\end{equation}%
where $t_{\sigma }$ is $\sigma $-dependent hoping integral. Not much
attention was paid to the model in the last century because we did not have
a realistic system that the model can be applied. However, in describing a
mixture of two species of fermionic atoms in optical lattices which has been
realized by recent experiments on the cold atoms \cite{CChin04}, the model
itself becomes a current research interest \cite%
{MACazalilla05,SJGu05,ZGWang07}. The ground-state phase diagram of the
asymmetric Hubbard model can be understood from its two limiting cases, i.e.
the Falicov-Kimball (FK) model region \cite{LMFalicov69,TKennedy86} ($%
t_{\downarrow }=0$) and the Hubbard model region (HM) \cite{Hubbard} ($%
t_{\uparrow }=t_{\downarrow }$). The schematic phase diagram of the model
was obtained from the renormalization group analysis \cite{MACazalilla05}.
Subsequently, a quantitative phase diagram was also obtained by analyzing
structure factor with the density-matrix renormalization group \cite{SJGu05}
and bosonization techniques \cite{ZGWang07}. The fidelity approach to the
model was firstly done by Gu \emph{et al }\cite{SJGu072}. The authors found
that the fidelity susceptibility can be used to identify the universality
class of the quantum phase transitions in this model. That is the quantum
phase transitions occurred at different band filling can be characterized by
the critical exponents of the fidelity susceptibility. Fig. \ref%
{fig:fidelity_ahm.eps} shows the fidelity susceptibility as a function of $%
t_{\downarrow }/t_{\uparrow }$ for the cases of $n=2/3$ and $n=1$, and
various system sizes. Obviously, the fidelity susceptibility diverges
quickly as the system is away from the half-filling, and relatively slow at
half-filling. The scaling analysis reveals that the maximum value of the
fidelity susceptibility scales like
\begin{equation*}
\chi _{F}\propto L^{5.3}
\end{equation*}%
at $n=2/3$ case, while
\begin{equation*}
\chi _{F}\propto L
\end{equation*}%
at half-filling. Then the intensive fidelity susceptibility shows
singularity away from half-filling, while no singularity at half-filling.
These observations support their conclusions on the role of fidelity in
describing the universality class.

\begin{figure}[tbp]
\includegraphics[bb=0 1 170 100, width=8.5 cm, clip]{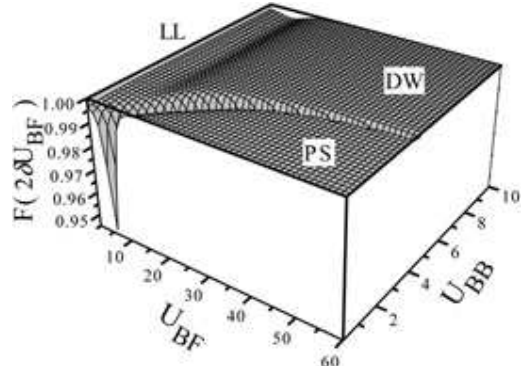}
\caption{(Color online) The ground-state phase diagram defined on the $%
U_{BB} $-$U_{BF}$ plane of the one-dimensional Bose-Fermi Hubbard model in
terms of the fidelity (Reproduced from Ref. \protect\cite{WQNingJPC}).}
\label{fig:bfm_fig1.eps}
\end{figure}

\emph{The one-dimensional Bose-Fermi mixture:} The simplest one-dimensional
Bose-Fermi Hubbard model can be modeled by
\begin{eqnarray}
H_{\text{BF}} &=&-\sum\limits_{i=1}^{N}(t_{\mathrm{F}}c_{i}^{\dag
}c_{i+1}+t_{\mathrm{B}}b_{i}^{\dag }b_{i+1}+\mathrm{{H.c.})}  \notag \\
&&+U_{\mathrm{BF}}\sum\limits_{i=1}^{N}c_{i}^{\dag }c_{i}b_{i}^{\dag
}b_{i}+U_{\mathrm{BB}}\sum\limits_{i=1}^{N}b_{i}^{\dag }b_{i}(b_{i}^{\dag
}b_{i}-1).  \label{eq:HamiltonianBFMix}
\end{eqnarray}%
Here $b_{i}$ ($b_{i}^{\dag }$) and $c_{i}$ ($c_{i}^{\dag }$) are the bosonic
and fermionic annihilation (creation) operators at site $i$, respectively. $%
t_{\mathrm{F}}$($t_{\mathrm{B}}$) is the hoping integral of fermions
(bosons). $U_{\mathrm{BF}}$ denotes the on-site interaction between fermion
and bosons, and $U_{\mathrm{BB}}$ the interaction between bosons. The ground
state of the model has been studied by the quantum Monte-Carlo method \cite%
{LPolletPRL190402}. Several phases, including Luttinger liquid phase,
density wave phase, phase separation state, and Ising phase, are predicted.
The fidelity approach to the quantum phase transitions occurring in the
ground state was done by Ning \textit{et al }\cite{WQNingJPC}. As shown in
Fig. \ref{fig:bfm_fig1.eps}, Luttinger liquid phase, density wave phase,
phase separation state can be observed from the behaviors of the fidelity.
However, it is difficult to find the phase boundary between the Ising phase
and density wave phase. The authors interpret that the phase transition
between the Ising phase and density wave phase is of the
Beresinskii-Kosterlitz-Thouless type since in the large $U_{\mathrm{BB}}$
and $U_{\mathrm{BF}}$ limit, the effective Hamiltonian of Eq. (\ref%
{eq:HamiltonianBFMix}) becomes the one-dimensional XXZ model. Instead, they
found that the concurrence \cite{SHill97,WKWootters98}, as a measure of
entanglement between two 1/2 spins, can locate the transition point because
the concurrence shows a maximum at the isotropic point of the XXZ model \cite%
{SJGuXXZ}. Nevertheless, as we mentioned before, the role of fidelity in the
Beresinskii-Kosterlitz-Thouless phase transitions is still controversial.

\begin{figure}[tbp]
\includegraphics[bb=0 0 200 400, width=8.5 cm, clip] {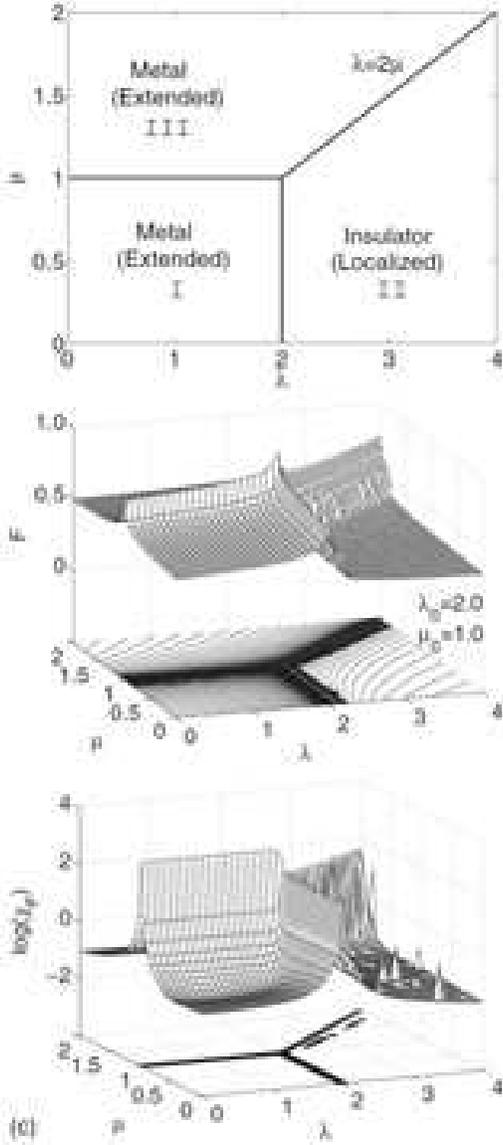}
\caption{(Color online) (a) The ground-state phase diagram in the $\protect%
\lambda-\protect\mu$ plane. (b)The fidelity $F(\protect\lambda, \protect\mu;
\protect\lambda_0, \protect\mu_0)$ for $(\protect\lambda_0, \protect\mu%
_0)=(2.0, 1.0)$. (c) The fidelity susceptibility and its contour map as a
function of $\protect\lambda$ and $\protect\mu$ along the direction $(1/%
\protect\sqrt{5}, -2/\protect\sqrt{5})$. (From Ref. \protect\cite%
{LGong115114}).}
\label{fig:harper.eps}
\end{figure}

\textit{The extended Harper model}: The Harper model \cite%
{JHHan94,YTakada04,KIno06} was proposed to describe electons in a
two-dimensional periodic potential under a magnetic field, for understanding
Hofstadter-butterfly energy spectrum \cite{DRHofstadter76}. For a system of
electrons moving on a triangle lattice in a magnetic field, the Hamiltonian
can be written as\cite{KIno06}%
\begin{eqnarray}
H_{\text{Harp}} &=&-\sum_{n}\left[ t_{a}+t_{c}e^{-2\pi i\phi (n-1/2)+ik_{y}}%
\right] c_{n}^{\dagger }c_{n-1}  \notag \\
&&-\sum_{n}\left[ t_{a}+t_{c}e^{2\pi i\phi (n+1/2)-ik_{y}}\right]
c_{n}^{\dagger }c_{n+1}  \notag \\
&&-\sum_{n}t_{b}\cos (2\pi \phi n+k_{y})c_{n}^{\dagger }c_{n},
\end{eqnarray}%
where $t_{a},t_{b},t_{c}$ are hoping amplitude for each bond on the
triangular lattice, $n$ lattice index, $\phi /2$ magnetic flux piercing each
triangle, and $k_{y}$ the momentum in $y$ direction. Since there is no
interaction between electrons, the model can be solved exactly. The ground
state of the model consists of three different phase, two conducting phases
and one insulating phase. Defining $\lambda =2t_{b}/t_{a},\mu =t_{c}/t_{a}$,
the ground-state phase diagram is shown in Fig. \ref{fig:harper.eps}(a) \cite%
{KIno06}. Therefore, in addition to metal-insulator transitions, there is
also an interesting metal-metal transition.

The fidelity approach to quantum phase transitions between the three phases
of the Harper model was done by Gong and Tong \cite{LGong115114}. They
studied the fidelity between the ground state at $(\lambda ,\mu )$ and
various reference states. For example, Fig. \ref{fig:harper.eps}(b) shows
the fidelity of a reference state at $(\lambda _{0},\mu _{0})=(2.0,1.0)$
which is a tricritical point in the phase diagram. At that point, the
fidelity is 1, while away from the tricritical point, the ground state
changes quickly, then the fidelity show distinct behavior in the parameter
space. One can find that the ground-state phase diagram can be sketched out
by the contour map of the fidelity. Such a property also is consistent with
the primary motivation of the fidelity approach. The authors studied also
the fidelity susceptibility along various evolution directions. Fig. \ref%
{fig:harper.eps}(c) shows the fidelity susceptibility in the $\lambda -\mu $
plane along $(1/\sqrt{5},-2/\sqrt{5})$ direction. In this case, the fidelity
susceptibility are a combination of two elements in the quantum geometric
tensor. Since the direction is not parallel to any transition line, the
fidelity susceptibility reaches maximum at the boundary between three
phases. Interestingly, the authors found the critical exponents of the
fidelity susceptibility depend on the different choice of system size. For
example, if the system size equal to Fibonacci number $%
F_{m}(F_{m}=F_{m-1}+F_{m-2})$ with $m=3l+1$, the adiabatic dimension at the
critical point $d_{a}^{c}=4.9371$ and $\nu =2.4718$; while form $m\neq 3l+1$%
, $d_{a}^{c}=2$ and $\nu =1$. Therefore, the critical exponent $\alpha
=d_{a}^{c}/\nu =2$ in both cases.

\subsubsection{The fidelity in topological quantum phase transitions}

Topological phase transitions are very special in quantum critical
phenomena. They do not rely on any local order parameters nor on a symmetry
breaking mechanism, hence cannot be described by Landau-Ginzburg-Wilson
paradigm. A typical example of these novel phases is the fractional quantum
Hall state, in which electrons in two dimension are strongly correlated and
their fluctuations are entirely quantum in nature, therefore, the
Landau-Ginzburg-Wilson theory, which is based on a classical local order,
might fail. The fidelity approach provides an alternative method to study
these fascinating phase transitions.

\begin{figure}[tbp]
\includegraphics[bb=0 -10 350 190, width=8.5 cm, clip] {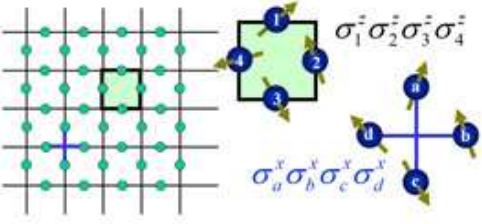}
\caption{ (Color online) A sketch of the Kitaev toric model: The square
lattice (LEFT) is defined on a torus. All 1/2 spins locate at the middle
point (solid dot) of bonds. Four neighboring spins interact with other by
plaquette or star operators (RIGHT) depending on their locations. }
\label{figure_kitaevtoricmodel}
\end{figure}

\textit{The deformed Kitaev toric model:} The deformed Kitaev toric model
\cite{Kitaev} is defined on a square lattice with spin-1/2 degrees of
freedom residing on the bonds (Fig. \ref{figure_kitaevtoricmodel}, left).
The Hamiltonian reads%
\begin{eqnarray}
H_{\text{DKT}} &=&-\lambda _{0}\sum_{p}B_{p}-\lambda _{1}\sum_{s}A_{s}
\notag \\
&&+\lambda _{1}\sum_{s}\exp \left( -\beta \sum_{i\in s}\sigma
_{i}^{z}\right) , \\
&=&H_{\text{Kitaev}}+\lambda _{1}\sum_{s}\exp \left( -\beta \sum_{i\in
s}\sigma _{i}^{z}\right) .
\end{eqnarray}%
Here $A_{s}=\prod_{i\in s}\sigma _{i}^{x}$ and $B_{p}=\prod_{i\in p}\sigma
_{i}^{z}$ are the star and plaquette operators (Fig. \ref%
{figure_kitaevtoricmodel}, right) of the Kitaev model, $\lambda _{0,1}>0$
and $\beta $ is a parameter tuning the system across a topological phase
transition. The ground state of the deformed Kitaev toric model has been
obtained exactly \cite{CCastelnovo2008}%
\begin{eqnarray}
\left\vert \Psi _{0}(\beta )\right\rangle &=&\sum_{g\in G}\frac{\exp \left[
-\beta \sum_{i}\sigma _{i}^{z}(g)/2\right] }{\sqrt{Z(\beta )}}g\left\vert
0\right\rangle , \\
Z(\beta ) &=&\sum_{g\in G}\exp \left[ -\beta \sum_{i}\sigma _{i}^{z}(g)/2%
\right] ,
\end{eqnarray}%
where G is the Abelian group of all spin-flip operators obtained as products
of star-type operators. The fidelity approach to the topological phase
transition occurring in this model has been done \cite{AHamma07,DFAbasto08}.

The ground state of the model consists of two distinct phases\cite%
{CCastelnovo2008}. If $\beta =0$, the model is pure Kitaev toric model and
its ground state is a closed string condensed phase, in which each $x(z)$
string is a collection of spins that are flipped in the $\sigma ^{z}(\sigma
^{x})$ basis. While if $\beta $ is very large, the ground state favors a
fully polarized phase. A quantum phase transition is found to be occurred at
the critical point $\beta _{c}=(1/2)\ln (\sqrt{2}+1)$. The system shows a
topological order if $\beta <\beta _{c}$. The fidelity between two states at
the points $\beta \pm \delta \beta $ was obtained by Abasto, Hamma, and
Zanardi \cite{DFAbasto08}%
\begin{eqnarray}
&&F(\beta -\delta \beta /2,\beta +\delta \beta /2)  \notag \\
&=&\langle \Psi _{0}(\beta -\delta \beta /2)\left\vert \Psi _{0}(\beta
+\delta \beta /2)\right\rangle , \\
&=&\sum_{g\in G}\frac{\exp \left[ -\beta \sum_{i}\sigma _{i}^{z}(g)\right] }{%
\sqrt{Z(\beta -\delta \beta /2)Z(\beta +\delta \beta /2)}},
\end{eqnarray}%
and the fidelity susceptibility is
\begin{eqnarray}
\chi _{F} &=&\left. \frac{-2\ln F}{(\delta \beta )^{2}}\right\vert _{\delta
\beta \rightarrow 0}, \\
&=&\frac{C_{v}}{4\beta ^{2}}.
\end{eqnarray}%
where $C_{v}$ denotes the specific heat of the two-dimensional Ising model.
Then the fidelity susceptibility shows a logarithmic divergence at the
critical point $\beta _{c}$. These findings are very interesting. The
corresponding to the thermal phase transition occurring in the
two-dimensional Ising model reveals that the topological phase transition in
the deformed Kitaev toric model could be detected by the local magnetization%
\cite{CCastelnovo2008}.

\begin{figure}[tbp]
\includegraphics[bb=0 -10 340 180, width=8.5 cm, clip] {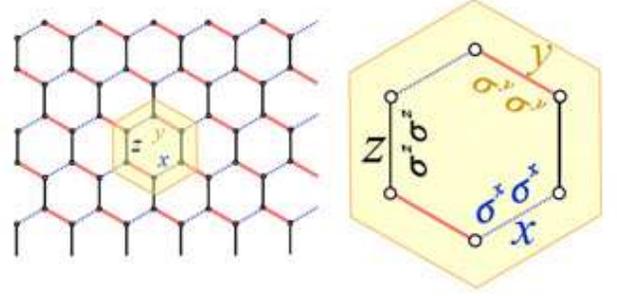}
\caption{ (Color online) A sketch of the Kitaev honeycomb model: Spins
locate at the vertices of a honeycomb lattice (LEFT). Each spin interacts
with three neighboring spins through three types of bonds, i.e. ``$x(y,z)$"
bonds depending on their direction (RIGHT).}
\label{figure_kitaevhombmodel}
\end{figure}

\begin{figure}[tbp]
\includegraphics[bb=30 370 550 765, width=8.5 cm, clip] {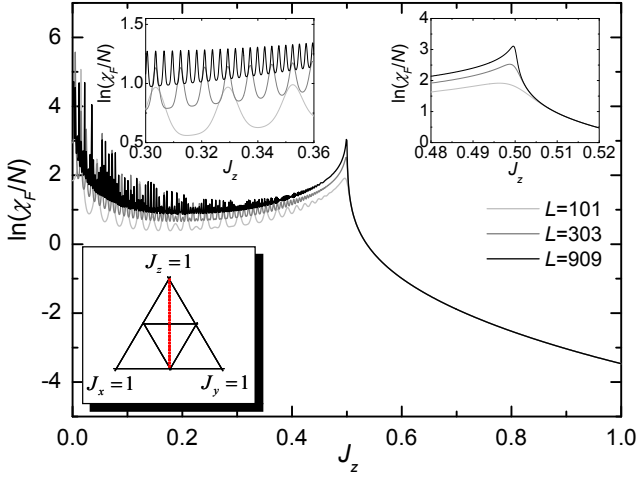}
\caption{(Color online) The fidelity susceptibility as a function of $J_z$
along the dashed line shown in the triangle for various system sizes $L=101,
303, 909$. Both up insets correspond to enlarged picture of two small
portions (From Ref. \protect\cite{SYang08}).}
\label{Kitaev_figure_fs}
\end{figure}

\textit{The Kitaev honeycomb model:} The Kitaev honeycomb model was also
introduced by Kitaev in search of topological order and anyonic statistics.
The model is associated with a system of 1/2 spins which are located at the
vertices of a honeycomb lattice (Fig. \ref{figure_kitaevhombmodel}: left).
Each spin interacts with three nearest neighbored spins through three types
of bonds, called \textquotedblleft $x$($y,z$)-bonds" depending on their
direction (Fig. \ref{figure_kitaevhombmodel}: right). The model Hamiltonian
\cite{Kitaev} is as follows:%
\begin{eqnarray}
H_{\text{KH}} &=&-J_{x}\sum_{x\text{-bonds}}\sigma _{j}^{x}\sigma
_{k}^{x}-J_{y}\sum_{y\text{-bonds}}\sigma _{j}^{y}\sigma _{k}^{y}  \notag \\
&&-J_{z}\sum_{z\text{-bonds}}\sigma _{j}^{z}\sigma _{k}^{z}, \\
&=&-J_{x}H_{x}-J_{y}H_{y}-J_{z}H_{z},  \label{eq:Hamiltonian_Kitaev}
\end{eqnarray}%
where $j,k$ denote two ends of the corresponding bond, and $J_{a}(a=x,y,z)$
are coupling constants. The ground state of the Kitaev honeycomb model
consists of two phases, i.e., a gapped A phase with Abelian anyon
excitations and a gapless B phase with non-Abelian anyon excitations. A
quantum phase transition occurred between these two phase is believed to be
a topological phase transition because no local operators can be used to
describe such a transition.

The Kitaev honeycomb model can be diagonalized exactly in the vortex-free
subspace. The ground state can be written as%
\begin{equation}
\left\vert \Psi _{0}\right\rangle =\prod_{\mathbf{q}}\frac{1}{\sqrt{2}}%
\left( \frac{\sqrt{\epsilon _{\mathbf{q}}^{2}+\Delta _{\mathbf{q}}^{2}}}{%
\Delta _{\mathbf{q}}+\text{i}\epsilon _{\mathbf{q}}}a_{-\mathbf{q},1}+a_{-%
\mathbf{q},2}\right) \left\vert 0\right\rangle ,
\end{equation}%
where
\begin{eqnarray}
\epsilon _{\mathbf{q}} &=&J_{x}\cos q_{x}+J_{y}\cos q_{y}+J_{z}, \\
\Delta _{\mathbf{q}} &=&J_{x}\sin q_{x}+J_{y}\sin q_{y}. \\
q_{x\left( y\right) } &=&\frac{2n\pi }{L},n=-\frac{L-1}{2},\cdots ,\frac{L-1%
}{2},
\end{eqnarray}%
and $a_{-\mathbf{q},1(2)}$ are Majorana operators for two sites of a single
bound. Then the fidelity between two states is \cite{JHZhao0803,SYang08}
\begin{equation}
F^{2}=\prod_{\mathbf{q}}\cos ^{2}\left( \theta _{\mathbf{q}}-\theta _{%
\mathbf{q}}^{\prime }\right) .
\end{equation}%
with
\begin{eqnarray}
\cos \left( 2\theta _{\mathbf{q}}\right) &=&\frac{\epsilon _{\mathbf{q}}}{E_{%
\mathbf{q}}},\sin \left( 2\theta _{\mathbf{q}}\right) =\frac{\Delta _{%
\mathbf{q}}}{E_{\mathbf{q}}},  \notag \\
\cos \left( 2\theta _{\mathbf{q}}^{\prime }\right) &=&\frac{\epsilon _{%
\mathbf{q}}^{\prime }}{E_{\mathbf{q}}^{\prime }},\sin \left( 2\theta _{%
\mathbf{q}}^{\prime }\right) =\frac{\Delta _{\mathbf{q}}^{\prime }}{E_{%
\mathbf{q}}^{\prime }}.  \label{eq:kitaevthetaqxy}
\end{eqnarray}%
The fidelity depends on the positions of two states in the parameter space.
Therefore, in order to extract the fidelity susceptibility, we must know the
direction of line connecting the two points. If we define the ground-state
phase diagram on the plane $J_{x}+J_{y}+J_{z}=1$, and consider a certain
line $J_{x}=J_{y}$ along which the ground state of the system evolves at
zero temperature. The fidelity susceptibility becomes \cite{SYang08}
\begin{equation}
\chi _{F}=\frac{1}{16}\sum_{\mathbf{q}}\left[ \frac{\sin q_{x}+\sin q_{y}}{%
\epsilon _{\mathbf{q}}^{2}+\Delta _{\mathbf{q}}^{2}}\right] ^{2}.
\end{equation}%
Fig. \ref{Kitaev_figure_fs} shows the fidelity susceptibility's dependence
on the driving parameter (red line in the figure) for various system size.
The authors also performed the scaling analysis, and found that the fidelity
susceptibility scaling like
\begin{equation}
\frac{\chi _{F}}{L^{2}}\sim \frac{1}{(J_{z}-1/2)^{1/2}},
\end{equation}%
around the critical point $J_{z}=0.5^{+}$. While in the gapless phase, Gu
and Lin \cite{SJGu08073491} found that the fidelity susceptibility is
superextensive, i.e.
\begin{equation}
\chi _{F}\sim L^{2}\ln L.
\end{equation}%
Then the critical exponent is quite different at the other side of the
critical point, i.e.
\begin{equation}
\frac{\chi _{F}|J_{z}-1/2|^{1/2}}{L^{2}\ln L}\sim \ln |J_{z}-1/2|,
\end{equation}%
\bigskip

Meanwhile, the topological phase transition was studied in terms of the
fidelity per site by Zhao and Zhou \cite{JHZhao0803}. According to the
definition of Eq. (\ref{eq:deffidelitypersite}), the fidelity per site in
the ground state of the Kitaev honeycomb model has the form%
\begin{eqnarray}
\ln d(J,J^{\prime }) &=&\frac{1}{L^{2}}\sum_{\mathbf{q}}\ln \left[ \cos
(\theta _{\mathbf{q}}-\theta _{\mathbf{q}}^{\prime })\right] , \\
&=&\frac{1}{(2\pi )^{2}}\int_{0}^{\pi }dq_{x}\int_{0}^{\pi }dq_{y}\ln \left[
\cos (\theta _{\mathbf{q}}-\theta _{\mathbf{q}}^{\prime })\right] ,  \notag
\\
&&
\end{eqnarray}%
where $\theta _{\mathbf{q}}$ is determined by Eq. (\ref{eq:kitaevthetaqxy}).
The above expression makes it possible to study the scaling and critical
behavior of the fidelity per site easily. More precisely, if $J^{\prime }$
is fixed, they found that the fidelity per site is logarithmically divergent
when $J$ is varied such that a critical point is crossed. For example, if
only $J_{x}$ of $J$ is considered as a driving parameter, they found the
fidelity per site%
\begin{equation}
\left. \frac{d^{2}\ln d(J,J^{\prime })}{dJ_{x}^{2}}\right\vert
_{J_{x}=J_{xm}}\sim \ln L,
\end{equation}%
where $J_{xm}$ denotes the position of extremum point for a finite system,
and
\begin{equation}
\frac{d^{2}\ln d(J,J^{\prime })}{dJ_{x}^{2}}\sim \ln |J_{x}-J_{xc}|,
\end{equation}%
in the thermodynamic limit.

Therefore, the fidelity now is believed to be able to characterize the
topological phase transition occurring in the Kitaev honeycomb model \cite%
{JHZhao0803,SYang08}.

\subsection{Mixed-state fidelity}

\subsubsection{Reduced fidelity in quantum phase transitions}

There are several works on the role of reduced fidelity in quantum phase
transitions.

\begin{figure}[tbp]
\includegraphics[width=8cm]{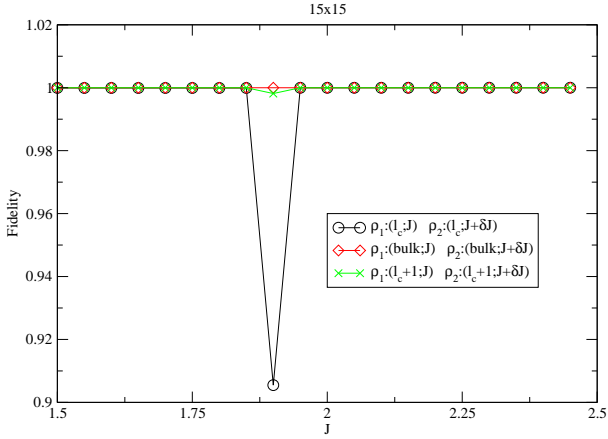}
\caption{(Color online) The single-site reduced fidelity in the
two-dimensional s-wave conventional superconductor. (From Ref. \protect\cite%
{NPaunkovic07}).}
\label{fig:impsup_fig1.eps}
\end{figure}

\textit{The one-dimensional transverse-field XY model:} The reduced fidelity
approach to the XY model was firstly touched by Zhou \cite{HQZhou07042945}
in order to confirm the relation between the fidelity per site and
renormalization group flow. He noted that the fidelity per site can capture
nontrivial information including stable and unstable fixed points in the
renormalization process in which the entanglement cannot.

Ma \textit{et al }\cite{XWang08081816} and Li \cite{YCLiPLA} studied the
two-site reduced fidelity in the ground state of the transverse-field Ising
model independently. In the work by Ma \textit{et al}, the fidelity of the
reduced state of two neighboring site in the model is%
\begin{equation}
F=\text{tr}\sqrt{\rho (\lambda )^{1/2}\rho (\lambda +\delta \lambda )\rho
(\lambda )^{1/2}},
\end{equation}%
where $\rho (\lambda )$ is the reduced-density matrix of the two spins. Then
the reduced fidelity susceptibility is defined by
\begin{equation}
\chi _{F}=\lim_{\delta h\rightarrow 0}\frac{-2\ln F}{(\delta \lambda )^{2}}.
\end{equation}%
For this model, $\rho (\lambda )$ can be expressed in terms of the two-site
correlation functions that can be calculated explicitly. This property
enables them to study the scaling and critical behavior of the reduced
fidelity easily. They found that, for a finite $N$-site system, the function
\begin{equation}
\chi _{F}(\lambda _{m},N)^{1/2}-\chi _{F}(\lambda _{m},N)^{1/2}=Q[N^{\nu
}(\lambda -\lambda _{m})]
\end{equation}%
is unversal around the critical point. The scaling relation reveals that
reduced fidelity susceptibility diverges logarithmically%
\begin{equation}
\chi _{F}(\lambda )^{1/2}\sim \ln |\lambda -\lambda _{c}|,
\end{equation}%
which is quite different from the global state fidelity in the same model.
While Li \cite{YCLiPLA} calculated the reduced fidelity directly. It is
reported that the extremum of the reduced fidelity scales like $F\propto
(\ln N)^{2.25}$ around the critical point. Later, You \emph{et al} \cite%
{WLYouRFS} also make an extension to the one-dimensional XY model. Similar
results are obtained.

\emph{The Lipkin-Meshkov-Glick model:} The quantum phase transition
occurring in the Lipkin-Meshkov-Glick model was also studied in terms of the
reduced fidelity \cite{HMKwok08,JMa08}.

It is not convenient to use the global-state fidelity to characterize those
quantum phase transitions induced the continuous level-crossing, such as the
magnetization process, because the global-state fidelity drops to zero at
each level crossing point. Kwok \textit{et al} proposed that the reduced
fidelity can overcome the difficulty. They used both the
Lipkin-Meshkov-Glick model and the one-dimensional XXX model as examples to
show that the reduced fidelity and its leading term, i.e. the reduced
fidelity susceptibility help to study scaling and critical behavior. Ma
\textit{et al }\cite{JMa08} also study the behaviors of the two-site reduced
fidelity in the Lipkin-Meshkov-Glick model model.

\emph{The one-dimensional extended Hubbard model:} Though the role of the
ground-state fidelity in the Hubbard model is still not very clear, it is
shown by Li \cite{YCLiPLA} that the two-site reduced fidelity is able to
sketch out the ground-state phase diagram of the model. Previously, it is
firstly reported by Gu \textit{et al}\cite{SJGUPRL} that the contour map of
single-site entanglement can describe the change of symmetry in the
ground-state of the system. Due to the competion betwee various correaltion,
the entanglement usually reaches a maximum at the critical point; while from
fidelity approach, it is shown that fidelity shows a minimum at the critical
point due to the abrupt change in the structure of the ground state.

\textit{Two-dimensional s-wave conventional superconductor: }The Hamiltonian
reads%
\begin{eqnarray}
H &=&-\sum_{\langle ij\rangle \sigma }tc_{i\sigma }^{\dagger }c_{j\sigma
}-\varepsilon _{F}\sum_{i\sigma }c_{i\sigma }^{\dagger }c_{i\sigma }  \notag
\\
&&+\sum_{i}\left( \Delta _{i}c_{i\uparrow }^{\dagger }c_{i\downarrow
}^{\dagger }+h.c\right)  \label{eq:Hamiltonianconvsup} \\
&&-\sum_{\sigma \sigma ^{\prime }}J\left( \cos \varphi c_{0\sigma }^{\dagger
}\sigma _{\sigma \sigma ^{\prime }}^{x}c_{0\sigma ^{\prime }}+\sin \varphi
c_{0\sigma }^{\dagger }\sigma _{\sigma \sigma ^{\prime }}^{z}c_{0\sigma
^{\prime }}\right) .  \notag
\end{eqnarray}%
The first three terms in Eq. (\ref{eq:Hamiltonianconvsup}) denote the
Hamiltonian of s-wave conventional superconductor, and the last term
represents the interaction between electrons and an classical spin placed at
the origin. Paunkovi\'{c} \textit{et al } \cite{NPaunkovic07} first studied
the reduced fidelity (or the partial-state fidelity) in the ground state of
the system. They observed that the one-site reduced fidelity shows a sudden
drop in the vicinity of the quantum phase transition. Since the one-site
reduced-density matrix depends on a single quantity, i.e. magnetization.
They interpret the drop of the reduced fidelity due to that the on-site
magnetization plays an order parameter in this model. Such a behavior is
very similar to that of entanglement \cite{PDSacramento07} in the same model.

\begin{figure}[tbp]
\includegraphics[width=8cm]{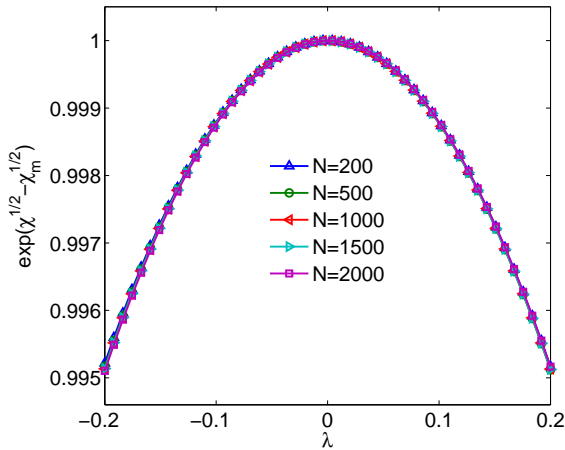}
\caption{(Color online) The scaling behavior the reduced fidelity
susceptibility of the one-dimensional transverse-field Ising model. (From
Ref. \protect\cite{XWang08081816} ).}
\label{fig:fising_fig3.eps}
\end{figure}

\subsubsection{Thermal state fidelity in strongly correlated systems}

The leading term of the fidelity, i.e. the thermal fidelity susceptibility
is simply the specific heat if we choose the temperature as the driving
parameter, and magnetic susceptibility if we choose the magnetic field as
the driving parameter. Despite of this there are still some works on
fidelity in thermal phase transitions, especially for those Hamiltonians
with non-commuting driving terms.

\begin{figure}[tbp]
\centering
\includegraphics[bb=0 0 210 150, width=7.5cm]{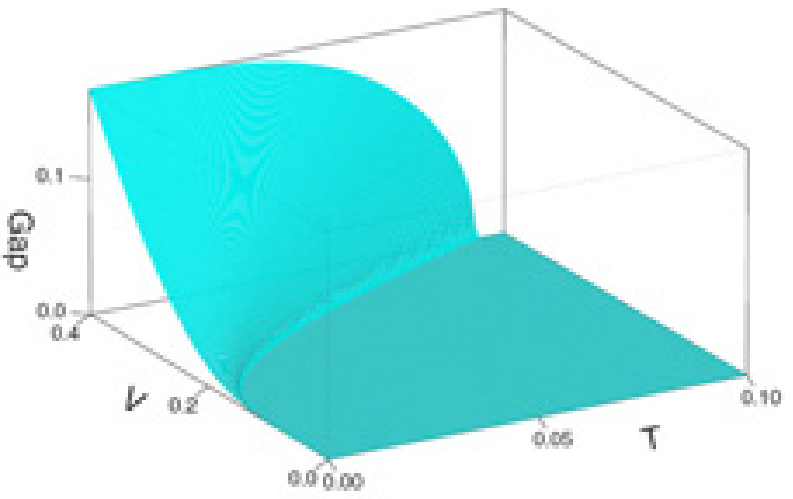}\newline
\includegraphics[bb=0 0 210 150, width=7.5cm]{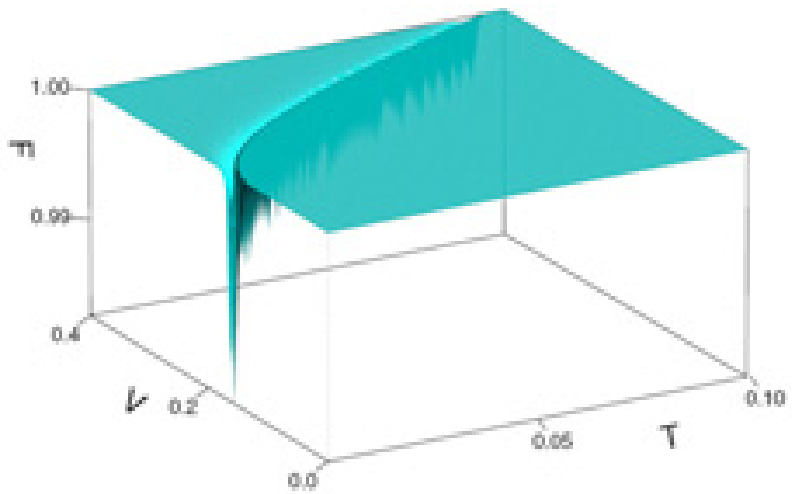}
\caption{(Color online) The gap $\Delta $ (UP) and the fidelity of $\protect%
\delta V=10^{-3},\protect\delta T=0$ (DOWN) as a function of the temperature
$T$ and the coupling constant $V$. The plot is given in rescaled quantities $%
T\rightarrow k_{B}T/(\hbar \protect\omega _{D})$, $V\rightarrow D_{F}V$ and $%
\Delta \rightarrow \Delta /(\hbar \protect\omega _{D})$ (From Ref.
\protect\cite{NPaunkovic08}).}
\label{BCS-Gap Figure}
\end{figure}

\emph{The BCS superconductor:} The effective Hamiltonian of the
Bardeen-Cooper-Schrieffer (BCS) superconductor \cite{BCS1,BCS2} can be
written as%
\begin{equation}
H_{\text{BCS}}=\sum_{k\sigma }\varepsilon _{k}c_{k\sigma }^{\dag }c_{k\sigma
}+\sum_{kk^{\prime }}V_{kk^{\prime }}c_{k^{\prime }\uparrow }^{\dag
}c_{-k^{\prime }\downarrow }^{\dag }c_{-k\downarrow }c_{k,\uparrow }.
\label{BSC_Hamiltonian}
\end{equation}%
where $\varepsilon _{k}$ is the dispersion and $V_{kk^{\prime }}$ are
coupling constant. The BCS Hamiltonian provides us with an example of a
model with a mutually non-commuting Hamiltonian. The fidelity approach to
its thermal phase transition from a normal state to a superconducting state
was done by Paunkovi\'{c} and Vieira \cite{NPaunkovic08}. Based on the
mean-field approach, the gap function and the thermal-state fidelity was
obtained analytically. \ Fig. \ref{BCS-Gap Figure} shows both the gap and
the fidelity as a function of the coupling and temperature. Clearly, when
the gap appears, the system becomes a superconducting state. A thermal phase
transition occurs at the critical point. On the $V-T$ plane, both $V$ and $T$
can be regarded as the driving parameter, therefore, the fidelity between
two thermal state separated by a small distance on the plane shows a sharp
peak around the critical point. On the other hand, since the two terms at
the right hand side of Eq. (\ref{BSC_Hamiltonian}) do not commute, the
fidelity is not easy to be calculated. However, the authors found that
distance obtained from Eq. (\ref{eq:thermalstatefidelitypf}) and the
fidelity have tha same qualitative bahavior (constant value 1 everywhere,
except for the sudden drop along the line of phase transition).

Paunkovi\'{c} and Vieira studied also the thermal phase transitions in the
Stoner-Hubbard model whose Hamiltonian reads
\begin{equation}
H_{\text{SH}}=\sum_{k\sigma }\varepsilon _{k}c_{k\sigma }^{\dag }c_{k\sigma
}+U\sum_{l}n_{l\uparrow }n_{l\downarrow }.  \label{SH_Hamiltonian-original}
\end{equation}%
where $\varepsilon _{k}=k^{2}/2m$ is the dispersion, and $U$ the on-site
interaction. Unlike the BCS superconductor, the phase transition they
addressed in the Stoner-Hubbard model is due to the existence of
Zeemann-like term for conserved quantities. In this case, the fidelity
susceptibility is simply the susceptibility of the conserved quantity\cite%
{WLYou07}. The fidelity shows a sharp peak around the transition point from
an order phase at low temperature to a disorder phase at high temperature.

\section{Numerical methods for the ground-state fidelity}

\label{sec:num}

This section is for students who are not familiar with the numerical
diagonalization and the density matrix renormalization group technique.

\subsection{Exact diagonalization}

\textit{Direct diagonalization:} One of the main goals of quantum mechanics
is to diagonalize Hamiltonian matraces. Except for a few cases, however, the
Hamiltonian of most quantum many-body systems cannot be diagonalized
explicitly. Fortunately, advance in digital computers make it possible to
diagonalize the Hamiltonian numerically. Up to now, many numerical methods
rooting in computer science have been extensively developed, among which the
most straightforward method is the exact diagonalization. Though the system
size that can be diagonalized numerical is still not large, the numerical
results actually are very instructive for studies of the quantum many-body
systems.

In the exact diagonalization, we need to express the Hamiltonian in a set of
basis, and then diagonalize it numerically. As a warm-up example, let's
consider a spin dimer system. The Hamiltonian of a spin dimer with the
Heisenberg interaction reads
\begin{eqnarray}
H &=&\sigma _{1}\cdot \sigma _{2}, \\
&=&\sigma _{1}^{x}\sigma _{2}^{x}+\sigma _{1}^{y}\sigma _{2}^{y}+\sigma
_{1}^{z}\sigma _{2}^{z}.
\end{eqnarray}%
To see its matrix form, we use the eigenstates of $\sigma ^{z}$ operators as
basis, i.e., $\{|\uparrow \uparrow \rangle ,|\uparrow \downarrow \rangle
,|\downarrow \uparrow \rangle ,|\downarrow \downarrow \rangle \}$. In this
set of basis, the Hamiltonian can be written as
\begin{equation}
H=\left(
\begin{array}{cccc}
1 & 0 & 0 & 0 \\
0 & -1 & 2 & 0 \\
0 & 2 & -1 & 0 \\
0 & 0 & 0 & 1%
\end{array}%
\right) \left(
\begin{array}{c}
|\uparrow \uparrow \rangle \\
|\uparrow \downarrow \rangle \\
|\downarrow \uparrow \rangle \\
|\downarrow \downarrow \rangle%
\end{array}%
\right) .
\end{equation}%
With some standard technique in Linear algebra, we can find that its ground
state is a spin singlet state
\begin{equation}
\Psi _{0}=\frac{1}{\sqrt{2}}\left[ |\uparrow \downarrow \rangle -|\downarrow
\uparrow \rangle \right] ,
\end{equation}%
with eigenvalue $E_{0}=-3$, while three degenerate excited states are
\begin{eqnarray}
\Psi _{1} &=&\frac{1}{\sqrt{2}}\left[ |\uparrow \downarrow \rangle
+|\downarrow \uparrow \rangle \right] ,  \notag \\
\Psi _{2} &=&|\uparrow \uparrow \rangle ,\;\;\;\Psi _{3}=|\downarrow
\downarrow \rangle .
\end{eqnarray}%
with higher eigenvalue $E_{1(2,3)}=1$.

Numerically, we can construct the basis by a set of integer which is usually
a binary digit. For the above example, we can use 0, 1, 2, 3 (that are 00,
01, 10, 11 in binary system) to denote the basis $|\uparrow \uparrow \rangle
,|\uparrow \downarrow \rangle ,|\downarrow \uparrow \rangle ,|\downarrow
\downarrow \rangle $, \ respectively. Then, we can define an array to save
the Hamiltonian. There are some standard methods \cite{NumericalRecipes} to
diagonalize a matrix, such as Householder method and QR algorithm, or
standard libraries, such as Linear Algebra Package(LAPACK) \cite{LAPACKnet}
and Intel's Math Kernel Library (MKL) \cite{MKLnet}.

\textit{Lanczos method:} For a small size sample, such as a 10 1/2-spin
chain, the dimension of the Hamiltonian matrix is not very large. We are
able to diagonalize the whole matrix. All eigenstates and eigenvalues can be
obtained. While if the dimension of the matrix size is very large, say
100000, it is almost impossible to implement the traditional diagonalization
methods. However, if we are only interested in the ground state of the
Hamiltonian, it is still possible for us to find the wavefunction via the
famous Lanczos method.

Here we would like to introduce the Lanczos method to diagonalize a large
sparse matrix and how to calculate the fidelity and its susceptibility\ via
the Lanczos method. We are not going to address the principle of the Lanczos
method, instead we show that the perturbation nature of the method
facilitates calculation of the fidelity susceptibility numerically. Now we
briefly introduce the basic steps of the Lanczos method.

The first step in the Lanczos method is to transform a large spare matrix $H$
to a tri-diagonal matrix $T$ as%
\begin{equation}
T_{m}=\left(
\begin{array}{ccccc}
\alpha _{1} & \beta _{1} & \cdots & \cdots & 0 \\
\beta _{1} & \alpha _{2} &  &  & \vdots \\
\vdots &  & \ddots &  & \vdots \\
\vdots &  &  & \alpha _{m-1} & \beta _{m-1} \\
0 & \cdots & \cdots & \beta _{m-1} & \alpha _{m}%
\end{array}%
\right) ,
\end{equation}%
where $m$ is the cut-off number. The cut-off number is determined by the
precession. For a real symmetric matrix $H$, we need an orthogonal matrix to
do such a transformation, i.e.%
\begin{equation}
V=\left( \mathbf{v}_{1},\mathbf{v}_{2},\cdots ,\mathbf{v}_{m}\right) ,
\end{equation}%
which satisfy%
\begin{equation}
\langle \mathbf{v}_{i}|\mathbf{v}_{j}\rangle =\delta _{ij},
\label{eq:lanzosorthocond}
\end{equation}%
and
\begin{equation}
HV=VT_{m}.
\end{equation}%
Therefore, starting from an initial (random) vector $\mathbf{v}$, we have
the following relations
\begin{eqnarray}
\mathbf{v}_{1} &=&\mathbf{v,}  \notag \\
\beta _{1}\mathbf{v}_{2} &=&H\mathbf{v}_{1}-\langle \mathbf{v}_{1}|H|\mathbf{%
v}_{1}\rangle \mathbf{v}_{1},  \notag \\
\beta _{2}\mathbf{v}_{3} &=&H\mathbf{v}_{2}-\langle \mathbf{v}_{2}|H|\mathbf{%
v}_{2}\rangle \mathbf{v}_{2}-\langle \mathbf{v}_{1}|H|\mathbf{v}_{2}\rangle
\mathbf{v}_{1},  \notag \\
\beta _{3}\mathbf{v}_{4} &=&H\mathbf{v}_{3}-\langle \mathbf{v}_{3}|H|\mathbf{%
v}_{3}\rangle \mathbf{v}_{3}-\langle \mathbf{v}_{2}|H|\mathbf{v}_{3}\rangle
\mathbf{v}_{2},  \notag \\
&&....  \notag \\
\beta _{m}\mathbf{v}_{m+1} &=&H\mathbf{v}_{m}-\langle \mathbf{v}_{m}|H|%
\mathbf{v}_{m}\rangle \mathbf{v}_{m},  \notag \\
&&-\langle \mathbf{v}_{m-1}|H|\mathbf{v}_{m}\rangle \mathbf{v}_{m-1}.
\end{eqnarray}%
From these relations, we can find that, using the orthogonal condition (\ref%
{eq:lanzosorthocond}),
\begin{eqnarray}
\alpha _{1} &=&\langle \mathbf{v}_{1}|H|\mathbf{v}_{1}\rangle ,  \notag \\
|\mathbf{r}_{i}\rangle &=&(H-\alpha _{i})|\mathbf{v}_{i}\rangle -\beta
_{i-1}|\mathbf{v}_{i-1}\rangle ,  \notag \\
\beta _{i} &=&\sqrt{\langle \mathbf{r}_{i}|\mathbf{r}_{i}\rangle },  \notag
\\
|\mathbf{v}_{i+1}\rangle &=&|\mathbf{r}_{i}\rangle /\beta _{i},  \notag \\
\alpha _{i+1} &=&\langle \mathbf{v}_{i+1}|H|\mathbf{v}_{i+1}\rangle ,  \notag
\\
i &=&1,2,...,m-1.
\end{eqnarray}%
The above steps are the famous Lanczos iteration, which was proposed by
Cornelius Lanczos in 1950.

After the matrix $T$ is obtained, one can easily calculate its eigenvalues $%
E_{i}$ and their corresponding eigenvectors $|\mathbf{u}_{i}\rangle $. This
procedure is simple because $T$ is already tri-diagonal. It can be proved
that the lowest eigenvalue of $T$ is the ground-state energy of $H$. Then
the ground-state wavefunction (Ritz eigenvector) can be calculated as%
\begin{equation}
|\Psi _{0}\rangle =V|\mathbf{u}_{0}\rangle ,
\end{equation}%
where $|\mathbf{u}_{0}\rangle $ is the ground state of $T$.

\begin{table}[tbp]
\caption{The fidelity susceptibility of the one-dimensional transverse-field
Ising model obtained by the exact diagonalization (middle) and exact
analytical results(right). Small difference might be caused by numerical
derivation. Here $N=20$.}
\label{tab:exactdia}
\begin{center}
\begin{ruledtabular}
\begin{tabular}{ccc}
 $h$ & FS(ED with $\delta h=0.005$) & FS of Eq. (\ref{eq:fsisingana}) \\
\hline 0.2 & 1.30206 & 1.30208\\
\hline 0.3 & 1.37360 & 1.37362\\
\hline 0.4 & 1.48807 & 1.48809\\
\hline 0.5 & 1.66672 & 1.66675\\
\hline 0.6 & 1.95535 & 1.95539\\
\hline 0.7 & 2.48564 & 2.48568\\
\hline 0.8 & 3.81093 & 3.81096\\
\hline 0.9 & 7.96763 & 7.96833\\
\hline 1.0 & 11.87198 & 11.87500\\
\hline 1.1 & 5.83999 & 5.84015\\
\hline 1.2 & 2.28066 & 2.28060\\
\hline 1.3 & 1.13390 & 1.13388\\
\hline 1.4 & 0.67720 & 0.67719\\
\hline 1.5 & 0.44735 & 0.44735\\
\hline 1.6 & 0.31372 & 0.31372\\
\hline 1.7 & 0.22904 & 0.22904\\
\end{tabular}
\end{ruledtabular}
\end{center}
\end{table}

To calculate the fidelity between two ground states in parameter space, the
Lanczos method is very promising because it is actually a numerically
perturbative method. To be concrete, if $H(\lambda )|\Psi _{0}(\lambda
)\rangle =E_{0}|\Psi _{0}(\lambda )\rangle $, then for the Hamiltonian%
\begin{equation}
H(\lambda +\delta \lambda )=H(\lambda )+\delta \lambda H_{I},
\end{equation}%
we can use $|\Psi _{0}(\lambda )\rangle $ as an initial state, then%
\begin{equation*}
\alpha _{1}=E_{0}+\delta \lambda \langle \Psi _{0}(\lambda )|H_{I}|\Psi
_{0}(\lambda )\rangle ,
\end{equation*}%
which is simply the ground-state energy up to the first-order perturbation.
To the second order, one can find that%
\begin{eqnarray}
\alpha _{1} &=&E_{0}+\delta \lambda \langle \Psi _{0}(\lambda )|H_{I}|\Psi
_{0}(\lambda )\rangle ,  \notag \\
\beta _{1} &=&\delta \lambda \sqrt{\langle \Psi _{0}(\lambda
)|H_{I}^{2}|\Psi _{0}(\lambda )\rangle -\langle \Psi _{0}(\lambda
)|H_{I}|\Psi _{0}(\lambda )\rangle ^{2}},  \notag \\
\alpha _{2} &=&\frac{\delta \lambda ^{2}}{\beta _{1}^{2}}\langle \Psi
_{0}(\lambda )|\Delta H_{I}H(\lambda +\delta \lambda )\Delta H_{I}|\Psi
_{0}(\lambda )\rangle ,  \notag \\
&&
\end{eqnarray}%
where $\Delta H_{I}=H_{I}-\langle \Psi _{0}(\lambda )|H_{I}|\Psi
_{0}(\lambda )\rangle $. Then the $T$ matrix, to the second order, becomes%
\begin{equation}
T_{2}=\left(
\begin{array}{cc}
\alpha _{1} & \beta _{1} \\
\beta _{1} & \alpha _{2}%
\end{array}%
\right) .
\end{equation}%
The eigenvalue of $T_{2}$ is
\begin{equation}
\frac{1}{2}\left[ \alpha _{1}+\alpha _{2}-\sqrt{(\alpha _{1}-\alpha
_{2})^{2}+4\beta _{1}^{2}}\right] .
\end{equation}%
\ Therefore, the Lanczos method is not only a perturbative method, but also
a variational method. The advantage is that we can use $|\Psi _{0}(\lambda
)\rangle $ as the initial state to calculate $|\Psi _{0}(\lambda +\delta
\lambda )\rangle $. Once the latter is obtained, the fidelity $\langle \Psi
_{0}(\lambda )|\Psi _{0}(\lambda +\delta \lambda )\rangle $ can be easily
calculated.

Here, we take the one-dimensional transverse-field Ising model as an
example. In section \ref{sec:fs}, we can find the fidelity susceptibility at
the point takes the form%
\begin{equation}
\chi _{F}=\sum_{k>0}\left( \frac{d\theta _{k}}{d\lambda }\right) ^{2},
\label{eq:num_isingfs}
\end{equation}%
where%
\begin{equation}
\frac{d\theta _{k}}{d\lambda }=\frac{1}{2}\frac{\sin k}{1+h^{2}-2h\cos k}.
\end{equation}%
From Eq. \ref{eq:num_isingfs}, we can calculate the fidelity susceptibility
up to a very large systems. For a comparison, we simply show both the
analytical results and numerical results for a 20-site system in Table. \ref%
{tab:exactdia}.

\subsection{Density matrix renormalization group}

\begin{figure}[tbp]
\includegraphics[bb=0 0 200 180, width=8.5 cm, clip] {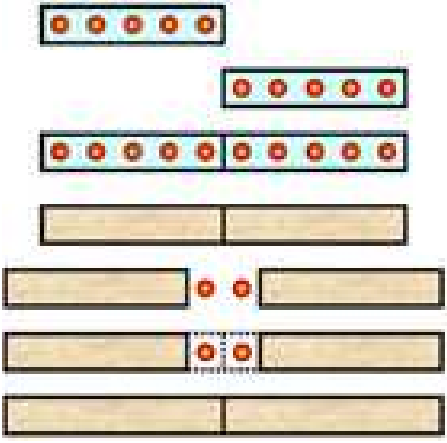}
\caption{ (Color online)The infinite-system algorithm of the DMRG technique:
The system size grows under the renormalization group transformation.}
\label{fig:dmrginfinite.eps}
\end{figure}

The density matrix renormalization group (DMRG) technique is a very
successful numerical method for one-dimensional strongly correlated systems.
The algorithm was invented by S. R. White \cite{SRWhiteDMRG} in 1992. Tzeng
and Yang \cite{YCTzeng08} firstly applied the DMRG technique to calculate
the ground-state fidelity. In this section, we briefly explain the basic
ideas of the DMRG algorithm and its application to the fidelity.

The main idea of the DMRG algorithm is a smart reduction of the number of
effective degrees of freedom and a variational and perturbative search of
the ground state within the reduced space. This idea is very important in
computational physics. For a quantum many-body system, the dimension of the
Hilbert space grows exponentially with the system size. For example, a $L$%
-site spin-1/2 system, the dimension of its Hilbert space is $2^{L}$. The
growth in the Hilbert-space dimension quickly exhausts computing resources.
This fact discourage us from doing exact diagonalization for larger systems.
Fortunately, the DMRG algorithm is able to capture the most relevant degrees
of freedom, hence allows us to reduce the effective dimension of the Hilbert
space significantly.

The DMRG algorithm consists of two fundamental sections: infinite-system
algorithm and finite-system algorithm. In both algorithms, a system of $L$
sites usually is divided into four blocks, i.e, $L^{S}$-site system block S,
two intermediate sites, and the rest environmental block E of $%
L^{E}=L-L^{S}-2$ sites. The infinite-system algorithm aims to grow the
system to the size we want to study, and the finite-system algorithm to
reduce the numerical error based on the variational principle. The DMRG
algorithm is more efficient for a system with open boundary conditions. So
we limit our discussion to this case here. Readers interested in other
details of the DMRG algorithm are recommended to refer to the review paper
by U. Schollw\"{o}ck \cite{DMRGreview} and the book by Peschel \textit{et al}
\cite{IPeschelbook} . There are also some new development in the
time-dependent DMRG method\cite{TDMRGreview}, and DMRG algorithm from the
perspective of quantum information theory \cite{FVerstraetePRL227205}.

\textit{The infinite-system algorithm }(Fig. \ref{fig:dmrginfinite.eps}):

Step 1: Start with a small system block of $l$ sites. Construct a set of
real-space basis (for instance, $\{|\uparrow \rangle ,|\downarrow \rangle \}$
for a single spin or $\{|\uparrow \uparrow \rangle ,|\uparrow \downarrow
\rangle ,|\downarrow \uparrow \rangle ,|\downarrow \downarrow \rangle \}$
for two spins). \ Then under the basis, construct the Hamiltonian matrix $%
H^{S}$ and matrices of operators responsible for interactions, say $%
O_{r}^{S} $, at the rightmost site.

Step 2: Construct the environment block of the same size (range from $l+1$
to $2l$), including the Hamiltonian matrix $H^{E}$ and interaction-operator
matrices, say $O_{l}^{E}$, at the leftmost site, say $O_{l}$, in a similar
way.

Step 3: Build the superblock by connecting the system and environment
blocks. The superblock Hamiltonian can be constructed from
\begin{equation}
H=H_{l}^{S}\otimes I^{E}+I^{S}\otimes H_{l}^{E}+O_{r}^{S}\otimes O_{l}^{E}
\end{equation}%
where $H^{I}=O_{r}^{S}\otimes O_{l}^{E}$ is the interaction Hamiltonian
between the $l$th site of the system block and $(l+1)$th of the environment
block. If there are more interaction terms, they should be included in $%
H^{I} $ too. Diagonalize the superblock Hamiltonian $H$ to obtain the ground
state $|\Psi _{0}\rangle $ by the Lanczos method or Davidson method \cite%
{ERDavidsonJCP}. Calculate the reduced-density matrices of the system block
and the environment block by%
\begin{equation}
\rho ^{S}=\text{tr}_{E}|\Psi _{0}\rangle \langle \Psi _{0}|,\rho ^{E}=\text{%
tr}_{S}|\Psi _{0}\rangle \langle \Psi _{0}|,
\end{equation}%
then diagonalize $\rho ^{S}$ by the traditional diagonalization method for
dense matrix, such as Householder-QR method,
\begin{equation}
\rho ^{S}|\mathbf{v}_{j}\rangle =w_{j}|\mathbf{v}_{j}\rangle ,
\end{equation}%
where $w_{j}$ is in a decreasing order. Form a new set of basis for the
system block by $M$ eigenstates of $\rho ^{S}$ with the largest eigenvalues,
and construct the transformation matrix
\begin{equation}
\Omega =\{\mathbf{v}_{1},\mathbf{v}_{2},\cdots \mathbf{v}_{M}\}.
\end{equation}%
Here $M$ is chosen based on the desired precision. Transform the Hamiltonian
of the system block and the interaction operators at the rightmost site into
the new basis
\begin{eqnarray}
\widetilde{H}^{S} &=&\Omega ^{\dagger }H^{S}\Omega , \\
\widetilde{O}_{r}^{S} &=&\Omega ^{\dagger }O_{r}^{S}\Omega .
\end{eqnarray}%
The environment block Hamiltonian and $O_{l}$ are transformed in a similar
way.

Step 4: Connect a new site to the rightmost site of the system block, and a
new site to the leftmost site of the environment block. The new Hamiltonian
is
\begin{eqnarray}
H^{S} &=&\widetilde{H}^{S}\otimes I+\widetilde{I}^{S}\otimes H^{N}+H^{I}, \\
H^{E} &=&I\otimes \widetilde{H}^{E}+H^{N}\otimes \widetilde{I}^{E}+H^{I},
\end{eqnarray}%
where $H^{N}$ is the single-site Hamiltonian. If the total size of the
system block, environment block, and two sites is small than the target
size, then goto the step 3, otherwise, jump to the following finite-system
algorithm.

\begin{figure}[tbp]
\includegraphics[bb=0 0 250 230, width=8.5 cm, clip] {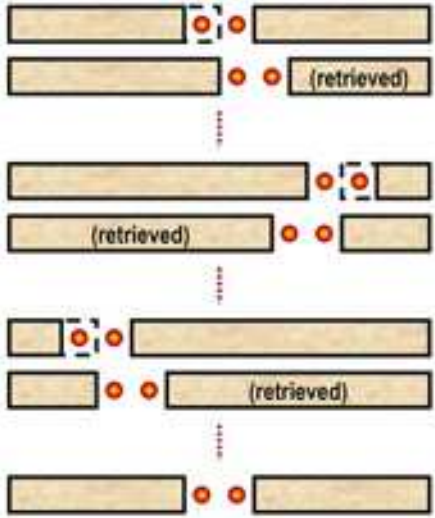}
\caption{ (Color online)The finite-system algorithm of the DMRG technique:
The system block grows under the renormalization group transformation, under
the environment block shrinks with the retrieved representation; and vice
versa.}
\label{fig:dmrgfinite.eps}
\end{figure}

\textit{The finite-system algorithm }(Fig. \ref{fig:dmrgfinite.eps}):

Once the size of the superblock in the infinite-system algorithm reaches the
target size we want to study, one can calculate the ground-state properties.
However, the results usually are not satisfactory because the error is still
very large. We need to use the finite-system algorithm to reduce the error.
The finite-system algorithm is similar to the infinite-system algorithm.
Instead of growing both blocks in the infinite-system algorithm, the growth
of one block is accompanied by shrinkage of the other block in the
finite-system algorithm. The information of the block shrunk should be
retrieved. For this purpose, one need to store the block information,
including matrices of the Hamiltonian and interaction operators in the
reduced space, obtained in the infinite-system algorithm.

Step 1: Followed from the infinite-system algorithm, now we have a system
block of $L/2-1$ sites, two independent sites, and an environment block of $%
L/2-1$ sites. Construct the superblock Hamiltonian, and compute the ground
state by the Lanczos method or Davidson method. Differ from the
infinite-system algorithm, here we do the reduced basis transformations for
the system block only. Then the new system block has $L/2$ sites.

Step 2:\ Continue to grow the system block in a similar way, while using the
stored information of the environment block, until the system block reaches
the maximum size.

Step 3:\ Once the environment block is minimized, then grow the environment
block at the expense of the system block in the same way, until the
environment is maximized.

Step 4: Grow the system block, and shrink the environment block to $L/2-1$.

A complete shrinkage and growth sequence for both blocks (From step 1-4) is
called a sweep (Fig. \ref{fig:dmrgfinite.eps}). Usually more sweeps, higher
precision the final results have. Once the desired precision is reached, the
ground state can be expressed in reduced space.

In the DMRG algorithm, the reduced basis obtained by the numerical
renormalization group strongly depend on the parameter of the Hamiltonian.
So we cannot compare two ground states at distinct points in the parameter
space directly. In order to calculate the fidelity between the two ground
states, one has to find a transformation between the two sets of reduced
basis. Such a transformation, as proposed in Ref. \cite{YCTzeng082}, can be
established in the final sweep of the finite-system algorithm.

The ground-state wavefunctions of the Hamiltonian $H_{0}+\lambda H_{I}$ at
two points $\lambda _{1}$ and $\lambda _{2}$, in their own reduced space,
can be expressed as
\begin{eqnarray}
|\Psi _{0}(\lambda _{1})\rangle &=&\sum_{i,m,n,j}\Phi _{imnj}(\lambda
)|\varphi _{i}^{S}\rangle |m\rangle |n\rangle |\varphi _{j}^{E}\rangle , \\
|\Psi _{0}(\lambda _{2})\rangle &=&\sum_{i,m,n,j}\overline{\Phi }%
_{imnj}(\lambda )|\overline{\varphi }_{i}^{S}\rangle |\overline{m}\rangle |%
\overline{n}\rangle |\overline{\varphi }_{j}^{E}\rangle ,
\end{eqnarray}%
where $|\varphi _{i}^{S}\rangle $($|\varphi _{j}^{E}\rangle $) is the
reduced basis of the system(environment) block, and $|m\rangle $, $|n\rangle
$ are the basis of the middle two sites. Since $|m\rangle $, $|n\rangle $
are the basis of local sites, we have%
\begin{equation}
\langle m|\overline{m}\rangle =\delta _{m\overline{m}},\langle n|\overline{n}%
\rangle =\delta _{n\overline{n}}.
\end{equation}%
Then the fidelity between two ground states becomes%
\begin{eqnarray}
&&\langle \Psi _{0}(\lambda _{1})|\Psi _{0}(\lambda _{2})\rangle
\label{eq:fidelityfromdmrg} \\
&=&\sum_{i,j,i^{\prime },j^{\prime },m,n}\Phi _{imnj}(\lambda )\overline{%
\Phi }_{i^{\prime }mnj^{\prime }}(\lambda )\langle \varphi _{i}^{S}|%
\overline{\varphi }_{i^{\prime }}^{S}\rangle \langle \varphi _{j}^{E}|%
\overline{\varphi }_{j^{\prime }}^{E}\rangle ,  \notag
\end{eqnarray}%
where $T_{ii^{\prime }}^{S}\equiv \langle \varphi _{i}^{S}|\overline{\varphi
}_{i^{\prime }}^{S}\rangle $ ($T_{jj^{\prime }}^{E}\equiv \langle \varphi
_{j}^{E}|\overline{\varphi }_{j^{\prime }}^{E}\rangle $) defines the
transformation matrix between the two sets of reduced basis of the system
(environment) block at $\lambda _{1}$ and $\lambda _{2}$.

Now we focus on how to obtain the $T^{S}$ of a $L^{S}$-site system block,
the $T^{E}$ of the environment block can be obtained in a similar way.

Step 1: During the final sweep of the finite-system algorithm of the two
Hamiltonians $H(\lambda _{1})$ and $H(\lambda _{2})$, if both system blocks
are minimized (to a single site), their basis are defined in the real space
and not reduced, then the transformation matrix between the two system
blocks is simply the unity matrix, i.e.,%
\begin{equation}
T^{S}(l=1)=I.
\end{equation}

Step 2: Suppose the transformation matrix of the two system blocks up to $l$
sites (including $l=1$) is obtained, i.e.
\begin{equation}
T^{S}(l)=|\overline{\varphi }_{j^{\prime }}^{E}(l)\rangle \langle \varphi
_{j}^{E}(l)|,
\end{equation}%
the basis of the system blocks together and the new site before the
renormalization group transformation are
\begin{eqnarray}
&&|\varphi _{j}^{E}(l)\rangle \otimes |m(l+1)\rangle , \\
&&|\overline{\varphi }_{j^{\prime }}^{E}(l)\rangle \otimes |m(l+1)\rangle ,
\end{eqnarray}%
respectively. The transformation matrix between the two sets of basis becomes%
\begin{equation}
T^{S}(l)\otimes I(l+1).
\end{equation}%
After the RG transformation, one can find the transformation matrix $T^{S}$
becomes
\begin{equation}
T^{S}(l+1)=\Omega ^{\dagger }[T^{S}(l)\otimes I(l+1)]\Omega ,
\end{equation}%
which gives a recursion relation of the transformation matrix $T^{S}$ in the
RG transformation.

Step 3: repeat the step 2 until both the two system block grow upto $L^{E}$
sites.

Clearly, the transformation matrix $T^{E}$ can be obtained if we start from
the minimized environment block. Finally, the fidelity of Eq. (\ref%
{eq:fidelityfromdmrg}) can be calculated. The fidelity susceptibility can be
also computed by the numerical differentiation.

\section{Summary and outlook}

\label{sec:sum}

As mentioned in the introductory section, quantum information theory
provides us opportunities to investigate quantum phenomena from new angles.
It is fair to say, besides huges of works on the role quantum entanglement
in the ground state of quantum many-body systems, the fidelity approach to
quantum phase transitions have shed new light on the critical phenomena.
Moreover, unlike the entanglement, which is somehow still mysterious in
quantum many-body system, the fidelity has, from our point of view, a
clearer physical picture. Especially, its leading term manifests scaling and
critical behaviors around the phase transition point. Therefore, the
fidelity is really a new optional method to investigate quantum phase
transitions, especially for those cases that we know nothing about order
parameters.

However, there are still some remaining problems. Firstly, the validity of
the fidelity in the Beresinskii-Kosterlitz-Thouless phase transitions is
still controversial. Secondly, the deep reason that the fidelity can signal
the topological phase transitions remains unknown. Thirdly, one of most
difficulties of the fidelity in quantum many-body systems is that it is
really difficult to find the exact ground state, except for a few cases.
Finally, it is a challenging problem to measure the fidelity in experiment
on scalable systems.

Finally, since the field is still quickly developing, we hope again that
this introductory review can offer some rough essays first, then to arouse
other people's better or more mature ideas.

\begin{acknowledgements}
We acknowledge our indebtedness to many people. We received genuine interest
and words of encouragement from Hai-Qing Lin and Chang-Pu Sun. We would also
like to thank Libin Fu, Ho-Man Kwok, Hai-Qing Lin, Li-Gang Wang, Xiaoguang
Wang, Shuo Yang, Yi-Zhuang You, Yi Zhou for many helpful discussions. We thank
N. Paunkovi\'{c} for nice feedbacks.

We thank Ho-Man Kwok for his final critical reading and efforts on textual
improvement.

We apologize if we omit acknowledging your relevant works.

This work is supported by the Direct grant of CUHK (A/C 2060344).
\end{acknowledgements}

\end{document}